\documentclass{article}

\usepackage{amssymb}
\let\UnmodifSec=\section
\renewcommand{\section}{\setcounter{equation}{0}\UnmodifSec}

\def\dateline{July 11, 2002}
\newtheorem{definition}{Definition}[section]
\newtheorem{lemma}{Lemma}[section]

\newtheorem{theorem}{Theorem}[section]
\newtheorem{remark}{Remark}[section]
\newtheorem{corollary}{Corollary}[section]

\def\l{\lambda}

\def\RR{\mathbb R}

\newcommand{\xx}{{\mathbf{x}}}\newcommand{\yy}{{\mathbf{y}}}
\newcommand{\zz}{{\mathbf{z}}}
\newcommand{\cc}{{\mathbf{c}}}

\newcommand{\z}{\underline{z}}

\def\z{{\zeta}}
\def\bC{{\bf C}}
\def\bR{{\bf R}}
\def\bN{{\bf N}}
\def\bZ{{\bf Z}}
\def\Im{{\rm Im\,}}

\def\Rp{{\bf R}_+}

\def\bCp{{\bf C}_+}

\def\tg{{\rm tg\,}}
\def\th{{\rm th\,}}
\def\ch{{\rm ch\,}}
\def\sh{{\rm sh\,}}

\def\BB{{\cal B}}
\def\CC{{\cal C}}
\def\DD{{\cal D}}
\def\EE{{\cal E}}
\def\FF{{\cal F}}
\def\GG{{\cal G}}
\def\HH{{\cal H}}

\def\JJ{{\cal J}}
\def\KK{{\cal K}}
\def\LL{{\cal L}}
\def\MM{{\cal M}}
\def\NN{{\cal N}}

\def\RR{{\cal R}}
\def\SS{{\cal S}}
\def\TT{{\cal T}}

\def\VV{{\cal V}}
\def\WW{{\cal W}}

\def\ZZ{{\cal Z}}
\def\wh{\widehat}
\def\wt{\widetilde}
\def\ovl{\overline}

\def \vhi{\varphi}
\def \veps{\varepsilon}

\def\HB{\hfill\break}
\def\interior#1{\setbox1=\hbox{$#1$}\rlap{$#1$}\kern0.4\wd1\raise1.1\ht1%
\hbox{$\scriptstyle \circ$}}
\def\bydef{\mathrel{\buildrel \hbox{\scriptsize \rm def} \over =}}
\def\boxit#1#2{\setbox1=\hbox{\kern#1{#2}\kern#1}%
\dimen1=\ht1 \advance \dimen1 by #1 \dimen2=\dp1 \advance \dimen2 by #1
\setbox1=\hbox{\vrule height\dimen1 depth\dimen2\box1\vrule}%
\setbox1=\vbox{\hrule\box1\hrule}%
\advance \dimen1 by .4pt \ht1=\dimen1 \advance \dimen2 by .4pt \dp1=\dimen2
\box1\relax}
\def\endprf{\raise .5ex\hbox{\boxit{2pt}{\ }}}

\def\Xcd{X_d^{(c)}}
\def\Xdn{X_d^n}

\def\Xcdn{X_d^{(c)n}}

\def\wXd{{\wt X_d}}
\def\wXdn{{\wt X_d^n}}

\def\wXcdn{{\wt X_d^{(c)n}}}

\def\Rep{\mbox{\boldmath $\Phi$}}
\def\amb{E_{d+1}}
\def\ambc{\amb^{(c)}}
\def\wTp{{\wt \TT_{1+}}}
\def\wTm{{\wt \TT_{1-}}}
\def\wTpm{{\wt \TT_{1\pm}}}

\def\wchi{{\wt \chi}}

\def\ifundefined#1{\expandafter\ifx\csname#1\endcsname\relax}

\title{\bf Towards a General
Theory of Quantized Fields \\ on the Anti-de Sitter Space-Time.}
\author{{Jacques Bros$^{1}$, Henri Epstein$^{2}$ and Ugo Moschella$^{3}$  }\\[20pt]
 {\small $^{1}$Service de Physique Th\'eorique, C.E. Saclay,
{91191 Gif-sur-Yvette, France}}\\
{\small $^{2}$Institut des Hautes Etudes Scientifiques,
91440 Bures-sur-Yvette,
France}\\
{\small $^{3}$Dipartimento di Scienze Matematiche Fisiche e Chimiche, Universit\`a dell'Insubria}\\
{\small Via Valleggio 11, 22100 Como,
and INFN sez. di Milano, Italy}}

\date{\dateline}

\begin{document}

\maketitle

\begin{abstract}
We propose a general framework for studying quantum field theory
on the anti-de-Sitter space-time, based on the assumption of
positivity of the spectrum of the possible energy operators. In
this framework we show that the $n$-point functions are analytic
in suitable domains of the complex AdS manifold, that it is
possible to Wick rotate to the Euclidean manifold and come back,
and that it is meaningful to restrict AdS quantum fields to
Poincar\'e branes. We give also a complete characterization of
two-point functions which are the simplest example of our theory.
Finally we prove the existence of the AdS-Unruh effect
for uniformly accelerated observers on trajectories
crossing the boundary of AdS at infinity, while
that effect does not exist for all the other
uniformly accelerated trajectories.

\end{abstract}

\section{Introduction}

   Quantum Field Theory (QFT) on  the anti-de Sitter (AdS)
space-time has come today to the general attention because of the
role that AdS geometry plays at several places in modern
theoretical physics  \cite{maldacena,RS}. AdS QFT is also
believed to provide an infrared regularization \cite{wilczek}
that can be useful for instance to understand the longstanding
problems of gauge QFT's. From a more conceptual viewpoint,
this regain of interest has led some authors to realize the need
for a deeper setting of AdS QFT and to investigate
the consequences of general principles
which, in spite of the difficulties generated by the
peculiar geometry of AdS,
might lead to a reasonable approach to
the interacting fields on this spacetime;
such an approach should of course include the case of AdS free fields,
whose preliminary versions were
given in early works \cite{Avis} \cite{Fronsdal:1974}.
In this spirit three recent works on general AdS QFT
can be mentioned.
\cite{BFS} uses the framework of observable algebras, but
without assuming local commutativity from the outset. Instead
the property of ``passivity'' (in the sense of \cite{PW})
is postulated for the vacuum, and a remarkable proportion
of the more standard properties is shown to follow.
Indeed we argue in our final remarks that, heuristically
speaking, their assumptions imply those of the present paper.
Two other works \cite{Re} and \cite{BBMS}
(motivated by \cite{maldacena}) independently exhibit explicit relationships
of a general type between AdS QFT and Conformal QFT in the Minkowskian
boundary of ADS: while \cite{Re} also pertains to the
framework of local observable algebras,
\cite{BBMS}
relies on a limited use of analytic $n$-point functions of local fields
in a Wightman-type approach (see \cite{SW}).
It is the complete setting of such a
Wightman-type framework for AdS QFT
and the derivation of a number of general results
for interacting fields belonging to that framework
which are the purpose of the present paper.

\vskip 0.1cm
It may be useful to start by recalling how AdS QFT is rendered
difficult by the lack of the global hyperbolicity property of the
underlying manifold. This manifests itself in two ways:
there exist closed timelike curves on the AdS manifold (in
particular geodesics) and there is a ``boundary'' at spacelike
infinity. The first problem is commonly avoided by considering
the covering of the manifold; the second fact implies that the
standard procedure of canonical quantization for free fields (see
e.g. \cite{BD,wald}) cannot be used, since there does not exist a
global Cauchy surface and information can flow in from spacelike
infinity.
To construct a viable QFT under these circumstances one may need
to specify suitable boundary conditions at infinity. To this end,
the nice idea in \cite{Avis} was to use the conformal embedding
of the AdS manifold in the Einstein Static Universe, which is a
globally hyperbolic space-time. It has been then possible (with
some restrictions) to produce a class of boundary conditions that
render the resulting AdS QFT well defined. Unfortunately the
procedure is very special and tricky, and can work at best for
free field theories.

Generally speaking, a well known problem when studying QFT on
gravitational backgrounds is the absence of a criterion to select
the physically meaningful states. There are indeed infinitely many
inequivalent representations of the same field algebra and it is
in general impossible to characterize the physically relevant
vacuum states as, for instance, fundamental states for the energy
operator, since the very concept of energy as a global quantity
is in general not defined.
However, an important aspect of the energy concept which
keeps its full value is the notion of energy ``relative to
an observer whose world-line is an orbit of a one-parameter
group of isometries of the spacetime'': for such observers,
the usual quantum notion of energy,
represented by the generator of time-translations,
is in fact applicable with respect to the proper-time parameter.
Therefore the properties of
energy-positivity and temperature  (i.e. the notions of
ground state and
KMS-state) are meaningful {\sl relatively to such observers}
and technically characterized by relevant analyticity properties
of the correlation functions of the fields in the corresponding
complexified orbits. Of course, this approach of the energy
concept remains particularly simple because we have to deal
with curved spacetimes of holomorphic type, such as the de Sitter
and AdS quadrics.

In a previous work dealing with
QFT on de Sitter spacetime (\cite{BEM}), we had shown that the
absence of global energy operators could
{\sl in
that case} be successfully replaced
by appropriate global analyticity
properties of the $n$-point functions of the fields
in ``natural'' tuboidal domains: the latter played the same role as
the tube domains resulting from energy-positivity in the
case of Minkowskian fields. Moreover a thermal interpretation
of these QFT for all geodesic (as well as
uniformly accelerated) observers could then be proved
as a byproduct of these global analyticity properties,
thus providing an extension of the
Bisognano-Wichmann analyticity property obtained in
the Minkowskian case \cite{BW}, or in physical terms
of the ``Unruh effect''.

Surprisingly, in the AdS case, the situation concerning
energy operators turns out to be more fortunate than in the
de Sitter case, and in fact quite favourable !
This is because there exist {\sl two classes}
of time-like
orbits of {\sl ``generic'' one-parameter isometry groups},
namely the
class of elliptic orbits and the class of hyperbolic orbits,
also supplemented by a ``boundary-type'' class of
parabolic orbits.
The hyperbolic orbits of AdS represent uniformly accelerated
motions similar to {\sl all} the geodesic or
uniformly accelerated motions of de Sitter spacetime:
the corresponding complexified orbits,
which are complex hyperbolae, are
thus also ``plagued'' by a natural periodicity in the
{\sl imaginary part} of their proper-time parameter,
which forbids the positivity of the corresponding
energy operator (but can support at best a KMS-condition
whose temperature is related to the radius of the hyperbola
and thereby to the acceleration).
On the contrary, the (complexified) elliptic
orbits, whose class contains geodesic as well as
uniformly accelerated motions, only present the
peculiarity of periodicity in the {\sl real part}
of their proper-time parameter: this  pathology already
mentioned above (namely the ``time loops'') is
cured by considering QFT on the {\sl universal covering} of AdS;
in fact
in all the following, this covering space will appear as
much more natural than
AdS itself for the setting of interacting fields.
At any rate, since there is no geometrical periodicity in the
imaginary part of the proper-time of the elliptic orbits
and since the corresponding  one-parameter groups have no
orbits of other type, their generators can be considered as genuine
global energy generators:
nothing forbids one to postulate
the positivity of the corresponding energy operators,
expressed technically by relevant analyticity
properties in
half-planes of the proper-time parameter.
So, as in the Wightman axiomatics of Minkowskian QFT,
it is still natural here to postulate the spectral condition
for all the generators of time-like elliptic orbits,
the latter playing the same role as the time-like straight-lines
(or uniform motions) in the Minkowskian spacetime.
We note that such a condition has been proposed in
Fronsdal's group-theoretical study  \cite{Fronsdal:1974} of Klein-Gordon
AdS QFT. In this connection one must also quote \cite{Deser}
whose authors point out the discrepancy between
the elliptic and hyperbolic
AdS trajectories and give arguments for
attributing to them respectively a
zero-temperature and a finite temperature specified in terms of the
acceleration.

Another postulate which will play a crucial role in our approach is
an adaptation of the property of local commutativity (or microcausality).
It still appears as natural for
the covering of AdS space-time,
in spite of its lack of global-hyperbolicity,
while it is harder
to justify on the ``pure'' AdS space-time itself, because of the time-loop
phenomenon. However we may
regard a field theory on the pure AdS space-time as just a special case of
one defined on its covering. Another justification is also
provided by \cite{BFS}.

After having set the relevant geometrical notions in Sections
\ref{PRE} and \ref{TUB}, we shall propose in
Section~\ref{HYP} a plausible set of hypotheses
for an interacting AdS QFT, among which the positivity
of the spectrum of the above mentioned energy
operators, AdS-covariance and an adaptation of
microcausality.
The spectral hypothesis
readily implies that
the $n$-point correlation functions admit
analytic continuations in tuboidal
domains in the Cartesian product of $n$ copies of
(the covering of) the complexified AdS manifold.
These analyticity properties of the correlation functions
parallel as closely as possible
what happens in Wightman QFT \cite{SW}, where the analyticity of the
correlation functions in similar tubular domains is analogously
obtained from the positivity of the spectrum of the energy.
These domains are described in Section \ref{TUB}.
Unfortunately the $n$-point tuboids are rather complicated
geometrical objects and, at present, can be given a simple description
only in the case $n = 1$, namely for the two-point functions.
Note however that for general $n$ they contain
``flat domains'' corresponding to all points moving on orbits of {\sl the
same one-parameter isometry group}, whose description is
identical with those of the corresponding
sets in complex Minkowskian spacetime;
in \cite{BBMS}, these flat domains have played a useful
role in the construction of the ``asymptotic forms'' of AdS QFT,
recognized as L\"uscher-Mack-type theories \cite{LM}
on the
asymptotic cone of AdS, in correspondence with conformal QFT
in Minkowskian spacetime of one dimension less.
At the end of our Section \ref{TUB} (subsection \ref{Comptub}), we have
tried to give a summary
of some analogies and discrepancies  between the $n$-point tuboids
of complex AdS and the corresponding ones of complex Minkowski
space.

One of the interesting points of the AdS geometry is that
there
exist families of submanifolds that can be identified with Minkowski
space-times in one dimension less (branes): as a matter of fact,
these submanifolds contain all the
two-plane sections of parabolic-type of the AdS-quadric
mentioned above as the third class of timelike orbits.
This very fact has
raised recently a large interest \cite{maldacena,RS}. Our
construction guarantees the possibility of considering
restrictions of AdS quantum fields to these ``Poincar\'e branes'' and
obtaining this way completely well-defined Minkowskian QFT's. It
is perhaps worthwhile to stress that this result, described in
Section \ref{parsec}, is not as obvious as the well-known
restrictibility of Minkowskian theories to lower dimensionality
space-times because of the more complicated geometry.
{}From a geometrical viewpoint, the conformal theories obtained
in \cite{BBMS} under asymptotic scaling assumptions
then also appear as limits of the previous Minkowskian
QFT's when the corresponding parabolic sections tend to infinity.

In any approach of QFT, the case of two-point functions
deserves a special study.
This is why we give in Section \ref{2pt} and in Appendix \ref{ap2pt}
a  complete characterization of
the AdS two-point functions; the latter are actually maximally analytic,
exactly as in the Minkowski \cite{SW} and de Sitter \cite{BM,BEM}
cases. As usual this permits the constructions of generalized
free fields and their Wick powers which fulfill all the hypotheses.
This study also provides the opportunity of displaying
some strange implications of the postulate of microcausality
in the ``pure AdS'' case. In fact, the discrepancy between the pure
AdS spacetime and the covering of AdS appears in a
characteristic way in the classification of the two-point functions;
it is revealed by the property of uniformity (or nonuniformity) of these
functions in their
analyticity domain ${\bf C}\setminus [-1,+1]$  in the complex plane of the
cosine of the AdS invariant distance.
These phenomena introduce the more general problem of characterizing
the interacting QFT's
on the pure AdS spacetime with respect to those on the covering,
which will be briefly discussed in our outlook (Section~\ref{O}).

In Section \ref{BWA} we derive from our postulates
the property of Bisognano-Wichmann analyticity in
all the hyperbolic orbits corresponding to
the class of uniformly accelerated
motions mentioned earlier, or in other words the
``AdS-Unruh effect'':
an accelerated
observer of the AdS world whose world-line belongs to that class
will perceive a thermal bath of particles with inverse temperature
equal to $2\pi$ times the radius of the corresponding
hyperbolic orbit. We note that this result, first described
in a special free-field theory in \cite{Deser},
has been also justified in the general framework of
algebraic QFT
in \cite{BFS},
by taking the principle of passivity of \cite{PW}
as a starting point. We also note that,
in spite of the peculiarities of the global geometry of AdS,
this result is similar
to the one proved in
\cite{BW} for Minkowskian QFT
and in \cite{BEM} for its extension to de Sitterian QFT.

Section~\ref{WR} is devoted to the AdS version of
the ``Euclidean'' field theory and to the
corresponding Osterwalder-Schrader reconstruction on the
covering of AdS.

Section~\ref{O} contains some final remarks among which a
brief discussion of the relationship of \cite{BFS} and the present work.

\section{Preliminaries}

\label{PRE}
We start with some notations and some well-known facts.
Let $\amb$ (resp $\amb^{(c)}$) denote $\bR^{d+1}$
(resp. $\bC^{d+1}$) equipped with the scalar product
\begin{eqnarray}
(x,y) &=& x^0 y^0 + x^d y^d - x^1 x^1 - \dots x^{d-1} x^{d-1} \nonumber\\
&=& x^0 y^0 + x^d y^d - \vec{x}\cdot \vec{y}\nonumber\\
&=& x^\mu\eta_{\mu\nu}x^\nu\ ,
\label{2.1}\end{eqnarray}
where $\vec{x}$ denotes $(x^1,\ \dots,\ x^{d-1})$.
A vector
$x$ in $\amb$ is called timelike, spacelike or lightlike
according to whether $(x,x)$ is positive,
negative or equal to zero.
We also use the notation
$||x||^2 = \sum_{\mu = 0}^d |x^\mu|^2$ for $x \in \amb$ or
$x \in \amb^{(c)}$
and we introduce the corresponding orthonormal basis
of vectors $e_\mu$ in $\amb$
($e_\mu^{\nu} = \delta_{\mu\,\nu}$).
If $A$ is a linear operator in
$\amb$ or $\amb^{(c)}$ we put
$||A|| = \sup \{||Ax||\ : ||x|| = 1\}$.

We denote $G$ (resp. $G^{(c)}$) the group of real (resp. complex) ``AdS
transformations'', i.e. the set of real (resp. complex) linear transformations
of $\amb$ (resp $\amb^{(c)}$) which preserve the scalar product (\ref{2.1}),
$G_0$ and $G_0^{(c)}$ the connected components of the identity in these
groups, $\wt G_0$ and $\wt G_0^{(c)}$ the correponding covering groups. An
element of $G_0$ (resp. $G_0^{(c)}$) will be called a proper AdS
transformation (resp. a proper complex AdS transformation).

\subsection{``Pure AdS'', complexified and ``Euclidean'' AdS,
coverings}
\label{pureetc}

\vskip 0.2cm
The real (resp. complex) {\sl ``pure'' anti-de-Sitter space-time} $X_d$
(resp. $X_d^{(c)}$) of radius $R$ is defined as the submanifold of
$\amb$ (resp $\amb^{(c)}$) consisting of the points $x$
such that $(x,x)= (x^0)^2 + (x^d)^2 - {\vec x}^2 =R^2$.
Except for the thermal considerations in Section
\ref{BWA}, we shall always take for simplicity $R=1$ throughout
this paper.
The group $G_0$ (resp. $G_0^{(c)}$) acts
transitively on $X_d$ (resp. $X_d^{(c)}$).

By changing $z^\mu$
to $iz^\mu$ for $0< \mu < d$,
the complex quadric $X_d^{(c)}$ becomes
the complex unit sphere in $\bC^{d+1}$, which has the
same homotopy type as the real unit sphere $S^d$.
In particular
$\pi_1(X_1^{(c)}) = \bZ$, $\pi_1(X_d^{(c)}) = 0$ (i.e.
$X_d^{(c)}$ is simply connected) for $d \ge 2$.
It follows that for $d \ge2$ the {\sl covering space} of $X_d^{(c)}$
is $ X_d^{(c)}$ itself .
However, as seen below, $X_d$ admits a nontrivial
{\sl covering space}
$\wt X_d$ whose ``physical'' role
is that it suppresses the time-loops
of pure AdS; its construction will also imply the existence of
nontrivial coverings of important domains of
$ X_d^{(c)}$ (although the full space
$ X_d^{(c)}$ itself has a trivial covering).

\vskip 0.2cm
It is possible to introduce in $X_d^{(c)}$ an analog of
the so-called Euclidean spacetime in complex Minkowski
space (where space is real and time purely imaginary):
we choose it to be the connected real submanifold
$X_d^{(\cal E)}$
of $X_d^{(c)}$
defined by putting
$z^0 = i y^0,\  x^1,\ldots,x^d$ real, $x^d >0$.
This sheet ($x^d >0$) of the two-sheeted hyperboloid
with equation $(x^d)^2 -(y^0)^2 - {\vec{x}}^2 = 1$,
equipped with the Riemannian metric induced by the
ambient quadratic form (\ref{2.1}), will be called
{\sl ``Euclidean'' AdS spacetime}. This choice
singles out the ``base point'' $e^d = (0,\ldots,0,1)$
as the analog of the origin in Minkowski spacetime.

\vskip 0.2cm
A concrete way of representing
the ``Euclidean'' spacetime
$X_d^{(\cal E)}$ together with $X_d$
{\sl and its covering} $\wt X_d$ is to introduce
the diffeomorphism $\chi$ of $S^1\times \bR^{d-1}$ onto $X_d$ given by
\begin{equation}
(t,\ \vec{x}) \mapsto
(\sqrt{1+\vec{x}^2}\,\sin t,\ \vec{x},\ \sqrt{1+\vec{x}^2}\,\cos t)
\label{2.11}\end{equation}
(where $S^1$ is identified to $\bR/2\pi \bZ$).

$X_d^{(\cal E)}$ is obtained by changing $t$ into $ is$ in this
representation, namely in the extension $\chi^{(c)}$ of $\chi$ to
$(S^1)^{(c)}\times \bR^{d-1}$. This yields the following
parametrization of
$X_d^{(\cal E)}$:
\begin{equation}
(s,\ \vec{x}) \mapsto
(i\sqrt{1+\vec{x}^2}\,{\rm sh}\  s,\ \vec{x},\ \sqrt{1+\vec{x}^2}\,{\rm
ch}\ s)
\label{parameucl}\end{equation}
The diffeomorphism $\wchi$, defined by lifting
$\chi$ on the covering $\bR^{d}$ of
$S^1\times \bR^{d-1}$
provides a global coordinate system on $\wt X_d$.
There also exists an
extension
$\wchi^{(c)}$ of $\wchi$ to ${\bf C}\times \bR^{d-1}$, whose image is
a partial complexification of the covering
$\wt X_d$ of AdS; this complexified covering contains
the same ``Euclidean'' spacetime
$X_d^{(\cal E)}$ as $X_d^{(c)}$.
It is clear that since $G_0$ acts transitively on $X_d$
the diffeomorphism $\chi$ can be transported by
any transformation of the group $G_0$; it follows that
$\wt G_0$ also acts transitively on
$\wt X_d$.

\vskip 0.2cm

The Schwartz space $\SS(\Xdn)$ of test-functions on $\Xdn$ will be
defined as the space of functions on $\Xdn$ which admit
extensions in $\SS(\amb^n)$. A $\CC^\infty$ function $f$ on $\wXdn$
belongs to $\SS(\wXdn)$ if every derivative of $f$ with respect to
the ambient coordinates decreases faster than any power of the geodesic
distance in the Riemannian geometry induced by $||x||$. Equivalently
$f\circ \wchi \in \SS(\bR^{d})$.

\subsection{The Lorentzian structure of AdS}

The restriction to $X_d$ of the pseudo-Riemannian metric
$\eta_{\mu\nu}dx^\mu dx^\nu$ is locally Lorentzian with signature
$(+,\ -,\ \ldots,\ -)$.
An elementary description follows from the fact that
$G_0$ acts transitively on $X_d$: it is sufficient to
look at the situation in the tangent hyperplane to the
base point $x= e_d$ (i.e. $ \{x; \ x_d =1\}$) whose intersection with
$X_d$ is the light-cone $\{e_d+y\ :\ (y^0)^2 - {\vec y}^2 =0,\ y^d=0\}$.
At the base point $e_d$, the future (resp. past) cone is then defined
by $(y^0)^2 - {\vec y}^2 >0,\ y^0>0$ (resp. $y^0 <0$).
At any point $x \in X_d$, the tangent hyperplane
to $X_d$ is $\{x+y\ :\ (x,y) = 0\}$. Its intersection with $X_d$
is the light-cone with apex at $x$, $\{x+y \in X_d\ :\ (y,y) =
0\}$. We say that a tangent vector $y$ at any point $x$
is time-like, light-like or space-like according to whether
$(y,y) >0$, $(y,y) =0$ or $(y,y) < 0$, which is consistent
with the situation at the base point.
Defining the local future (resp. past) cone at each point $x$
can also be done by using the transitive action of $G_0$
(or simply by continuity)
starting from $e_d$, but a more explicit
characterization can be
given by using the Lie algebra of $G$ (see below).
It can be easily seen
that the circular sections of $X_d$ by the planes
parallel to $(e_0, e_d)$, parametrized by $t$ in the representation
(\ref{2.11}) have time-like tangents whose future is
in the direction of increasing t, for all points $x= \chi (t,\vec x)$;
this clearly exhibits the phenomenon of closed
time-loops announced earlier and its suppression by going to
the covering of AdS.

\vskip 0.2cm
{}From a global viewpoint, two events  $x_1$, $x_2$ of $X_d$ are space-like
separated if $(x_1-x_2)^2<0$, i.e. $(x_1,x_2)>1$. The acausal set of the
base point $e_d$, i.e. the set of points $x$ which are space-like with respect
to $e_d$, is then given by
\begin{equation}
\Gamma_{a}(e_d) = \{x\in X_d; x^d >1 \}.
\end{equation}
Let us introduce the causal set of $e_d$ as the complement in $X_d$ of the
closure of $\Gamma_a(e_d)$:
\begin{equation}
\Gamma_c(e_d)= \{x\in X_d \ ; \ x^d\leq 1\}.
\end{equation}
$\Gamma_c (e_d)$ can be decomposed into three sets $\Gamma_c
(e_d)=\Gamma_+(e_d)\cup\Gamma_- (e_d) \cup\Gamma_{ex} (e_d)$:
\begin{eqnarray}
&&\Gamma_+ (e_d) = \{x\in X_d\ ;\ -1<x^d < 1,\, x^0 >0\},\nonumber\\
&&\Gamma_- (e_d) =  \{x\in X_d\ ;\ -1<x^d <1,\, x^0 <0\},\nonumber\\
&&\Gamma_{ex} (e_d) =  \{x\in X_d\ ; \ x^d \leq - 1,\}.
\end{eqnarray}
The regions $\Gamma_+ (e_d)$  and  $\Gamma_- (e_d)$ can be
conventionally called future and past of $e_d$.
Similar regions $\Gamma_a (x)$,
$\Gamma_c (x)$,
$\Gamma_{\pm} (x)$,
$\Gamma_{ex} (x)$ can be associated with any point $x$ of $X_d$
(again by the transitive action of $G_0$).

The geodesics of $X_d$ are conic sections by 2-planes containing the origin of
the ambient space; if $x_1$ and $x_2$ are two points on a connected branch of
geodesic $\gamma$ the distance $d(x_1,x_2)$ is defined by $d(x_1,x_2) =
\theta(x_1,x_2)$, where $\theta$ is the angle under which the arc $(x_1, x_2)$
of $\gamma$ is seen from the origin; otherwise stated, it is the parameter of
the isometry which transforms $x_1$ into $x_2$ in the minimal subgroup of $G$
admitting $\gamma$ as an orbit. $\theta$ is an ordinary angle if $\gamma$ is
an ellipse, and a hyperbolic angle if $\gamma$ is a branch of hyperbola. The
former case is interpreted as ``normal'' time-like separation if $|\theta| \le
\pi $ (future and past being distinguished by the sign of $\theta$) and one
has $(x_1,x_2) = \cos{d(x_1,x_2)}$; $d(x_1,x_2)$ is interpreted as the
interval of proper time elapsed between $x_1$ and $x_2$ for the geodesic
observer sitting on $\gamma$. This case is typically realized by the
geodesic with equations $(x^0)^2 + (x^d)^2 =1, \vec x =0$
(i.e. the curve $x= \chi(t,\vec 0)$ in (\ref{2.11})).
The second case corresponds to space-like
separation ($(x_1- x_2)^2 <0$) and one has
$(x_1,x_2) =  \ch {d(x_1,x_2)} > 1 $.
It is typically realized by taking
the section of $X_d$ by the plane $(e_0, e_1)$
(or $(e_d, e_1)$).

The AdS space-time is not geodesically convex: indeed for any
point $x$, all the temporal geodesics which contain $x$ also
contain the antipodal point $-x$, and the proper time interval
beetwen $x$ and $-x$ is $\pi $ on each of these geodesics.
Furthermore, the temporal geodesics emerging from an event $x$ do
not cover the full causal region $\Gamma_c(x)$ but only
$\Gamma_+(x)\cup\Gamma_- (x) $. In order to go from $x$ to a
point $y$ in the region $\Gamma_{ex}(x)$, one needs to follow at
least two arcs of temporal geodesics, which implies a ''boost''
(i.e. some ``interaction'') at the junction of these two arcs.
Similar remarks apply as regards causality and geodesics on the
covering $\wt X_d$.

\subsection{Three types of planar trajectories in AdS and its covering}

Two-planes of $\amb$ containing the origin
can always be spanned by a pair of linearly independent orthogonal
vectors $(a,b)$ and classified by considering
the possibility for each vector $a$ and $b$
to be timelike, spacelike or lightlike,
which gives six different cases. To each pair $(a,b)$ there corresponds
an element $M$ of the Lie algebra of $G$,
an isometry group $e^{t M}$ which is a one-parameter subgroup of $G_0$ and
a family of parallel two-planes whose sections by AdS are conics
(possibly degenerated into straight lines)
invariant under this subgroup; it follows that each connected
component of these sections is either a timelike or a spacelike
curve (or a lightlike straightline in the degenerate case).
Being interested by the classification of the {\sl timelike} curves
(or trajectories), we retain three possible cases which
reduce to simple models in terms of the basis $\{e_{\mu}\}$,
by using again the fact that $G_0$ acts transitively on $X_d$.

\vskip 0.2cm
i) {\sl The elliptic trajectories}:
this is the case when $a^2 >0$ and $b^2 >0 $. Since
this entails that (for all $\alpha,\beta$)\
$(\alpha a + \beta b)^2 > 0$, all the corresponding
parallel sections of AdS (in a family specified by $(a,b)$)
are timelike. Their models are the circular sections by all two-planes
parallel to $(e_0,e_d)$, described above in
(\ref{2.11}): the isometry subgroup associated with them is the
rotation group with parameter $t$.
There is one and only one geodesic in each such family
(in the unique plane of the family containing the origin).

\vskip 0.2cm
ii){\sl The hyperbolic trajectories}:
this is the case when $a^2 >0$ and $b^2 <0 $.
In each family of parallel sections of AdS specified
by $(a,b)$, there are two subfamilies.
One of them is composed of spacelike branches
of hyperbolae and contains
the unique geodesic of the family. The other one is composed
of timelike branches of hyperbolae, interpreted as
uniformly accelerated motions:
there is no timelike geodesic of that type.
The isometry subgroup associated with such a family is
a group of pure Lorentz transformation.
A model for this family is given by the sections parallel to
the $(e_0,e_1)-$plane. Since it will be used repeatedly
(in particular in Section \ref{BWA}) we introduce the
following notations.
For each $\lambda \in \bC \setminus \{0\}$, we denote $[\lambda]$
the special Lorentz transformation such that
\begin{equation}
([\lambda]x)^0 = {\lambda + \lambda^{-1} \over 2} x^0 +
{\lambda - \lambda^{-1} \over 2} x^1,\ \ \
([\lambda]x)^1 = {\lambda - \lambda^{-1} \over 2} x^0 +
{\lambda + \lambda^{-1} \over 2} x^1\ ,
\label{2.2}\end{equation}
the other components of $x$ remaining unchanged. In other words
\begin{equation}
([e^s]x)^0 = x^0\ch s + x^1\sh s,\ \ \ \
([e^s]x)^1 = x^0\sh s + x^1\ch s\ .
\label{2.3}\end{equation}
The corresponding subfamily of timelike orbits on AdS
is characterized by the following condition to be
satisfied by the other components:
$\rho(x)^2= (x^d)^2 - (x^2)^2 - \cdots -(x^{d-1})^2 -1 > 0$.
The corresponding orbits are then branches of
``hyperbolae with radius $|\rho|$'', namely with
equations
\begin{equation}
x^0 = \rho {\rm sh }t,\
x^1 = \rho {\rm ch} t.
\label{hyperbol}
\end{equation}

\vskip 0.2cm
iii){\sl  The parabolic trajectories}:
this is the case when $a^2 >0$ and $b^2 =0 $: since
this entails that (for all $\alpha,\beta$)
$(\alpha a + \beta b)^2 \ge 0$, all the corresponding
parallel sections of AdS (in a family specified by $(a,b)$)
are timelike or exceptionally lightlike.
Their models are the parabolic sections by all two-planes
parallel to $(e_0,\ e_{d-1} - e_d)$, admitting the following
representation in terms of a translation group parameter $t$:
\begin{equation}
x^0 = \sigma e^v  t,\ \
x^{d-1} = \sigma ({\rm sh} v + {1\over 2}e^v t^2),\ \
x^d = \sigma ({\rm ch} v - {1\over 2}e^v t^2),
\label{parabol}
\end{equation}
where $\sigma^2 = 1 + (x^1)^2 + \cdots + (x^{d-2})^2$.
The complete description of the corresponding subgroup of $G_0$
admitting these parabolic orbits
will be given and used in Sec \ref{parsec} (see Eq. (\ref{6.4})).
These timelike sections admit as a limiting case (for
$v$ tending to $-\infty$) the lightlike section by the plane
with equations $x^{d-1} + x^d = 0, x^1 =\cdots = x^{d-2} =0$.
This lightlike section gives the only geodesics in the family.

\vskip 0.2cm
We note that only the elliptic trajectories have nontrivial
liftings into the covering of $X_d$.
The hyperbolic and parabolic
trajectories belong to ``a single sheet of $\wt X_d$''.
This is connected with the following difference between these families
of trajectories:
while each family of elliptic trajectories
(resp. their liftings) covers the
{\sl whole space} $X_d$ (resp. $\wt X_d$),
each family of the two other types is decomposed into
two subfamilies which cover disjoint domains of $X_d$,
namely {\sl wedge-shaped regions} of the form $X_d \cap \{x ;\
\pm x^1 > |x^0|\}$ for the hyperbolic case and {\sl halves of  AdS}
of the form
$X_d \cap \{x ;\ \pm (x^{d-1} + x^d) >0\}$ for the
parabolic case.

It is also interesting to note that by putting $t=i\tau$
in the representations of the {\sl three families of trajectories}
(\ref{2.11}),(\ref{hyperbol}),(\ref{parabol}), one obtains curves
inside the ``Euclidean'' AdS spacetime $X_d^{({\cal E})}$,
namely respectively, branches of hyperbolae, circles and
parabolae. The orbits of the second case,
which are associated with purely imaginary
Lorentz transfomations and exhibit periodicity
in the corresponding imaginary time
parameter are to be compared with those occurring
as well in the Euclidean space of complex Minkowski space
as in the (``Euclidean'') hypersphere of the
complex de Sitter spacetime (see e.g. (\cite{BEM})):
as shown below in Sec \ref{BWA}, they correspond to
the existence of an Unruh effect as in these other two spacetimes.
On the other hand,
the other two classes of orbits
are rather to be compared with those of the time-translation
groups of Minkowskian space (although not corresponding to
geodesical motions on $X_d$ generically),
as far as they allow a topologically equivalent  ``Wick-rotation''
procedure to be performed (see Sec \ref{WR} and \ref{parsec}).

\vskip 0.3cm
\noindent
{\sl Interpretation of the planar trajectories as
uniformly accelerated motions:}

\vskip 0.2cm
In this paragraph we reintroduce the radius $R$ of the AdS spacetime.
If $x=x(t)$ denotes an arbitrary AdS trajectory parametrized by its
{\sl proper time} $t$, the corresponding velocity-vector
$u(t) = {d\over dt} x(t)$ at the point $x(t)$
satisfies the following relations
(in terms of scalar products in the ambient space $E_{d+1}$):
$ (u(t),u(t)) =1$, $ (x(t),u(t)) =0$. It follows that the  {\sl ambient
acceleration-vector}
$w(t) = {d\over dt} u(t)$
satisfies the relation $(w(t),x(t))  = -1$.
The latter together with the AdS equation $(x(t),x(t)) = R^2$ then
imply that the {\sl AdS acceleration-vector} $\gamma(t)$,
defined as the
$E_{d+1}-$orthogonal projection of $w(t)$ onto the tangent hyperplane
to AdS at the point $x(t)$, is given by the following formula:
$\gamma(t) = w(t) +{1\over R^2} x(t)$.
In view of the previous relations, this entails that
$ (\gamma(t),\gamma(t)) =
(w(t),w(t)) -{1\over R^2}$.
{\sl Motions with constant acceleration} are those for which
$ (\gamma(t),\gamma(t))$ or equivalently
$(w(t),w(t))$ is independent of $t$.

Consider any planar trajectory of AdS; if it is either elliptic
or hyperbolic, we can represent it by the equation
$x(t) = \hat x(t) + R c$, where $\hat x(t)$ varies in a two-plane $\Pi$
and the vector $c$ can always be chosen orthogonal to $\Pi$.
If $c^2 <0$, the trajectory is
elliptic (this is the case described by (\ref{2.11}), with
$c= (0,\vec x,0)$).  If $c^2 >1$, the trajectory is hyperbolic
(this is the case described by (\ref{hyperbol}), with
$c=   (0,0,x^2,\ldots, x^d)$, $c^2= (1+\rho^2)$).
Since $w(t)$ as well as $u(t)$ remain in $\Pi$, and since
$(w(t),u(t))= (\hat x(t),u(t)) =0$ , there holds the colinearity
condition $w(t)= \lambda (t) \hat x(t)$ and the condition
$(w(t),x(t))= (w(t),\hat x(t)) =-1$ then yields
$\lambda (t) = -{1\over (\hat x(t),\hat x(t))} = {1\over R^2(c^2-1)}$.
It then follows that
$(w(t),w(t))=
- {1\over R^2(c^2-1)}$.
is independent of $t$. This motion is therefore uniformly
accelerated with the following value of the AdS acceleration:
$ {\rm a}= {\sqrt {-(\gamma(t),\gamma(t))}} =
{1\over R}\  \sqrt {c^2\over c^2 -1}$.
This leads us to state

\begin{lemma}
All the planar trajectories of the AdS spacetime correspond to
uniformly accelerated motions, with the following specifications:

i) If the trajectory is elliptic, the corresponding acceleration
${\rm a} =
{1\over R}\  \sqrt {|c^2|\over |c^2| +1}$
takes any value between $0$
(i.e.the geodesic case obtained for $c=0$)
and $1\over R$ (i.e. the case
$c \to \infty$ corresponding to
ellipses in far-away two-planes).

ii) If the trajectory is hyperbolic, the corresponding acceleration
${\rm a} =
{1\over R}\  \sqrt {c^2\over c^2 -1}$
takes any value between
$1\over R$ (i.e. the case
$c \to \infty$ corresponding to
hyperbolae in far-away two-planes)
and $+\infty$
(i.e. the case $c^2 \to 1$ corresponding to
degenerate hyperbolae (or ``bifurcate horizons'')).

iii) If the trajectory is parabolic, the corresponding acceleration is
${\rm a} = {1\over R}$
\label{accel}
\end{lemma}

To complete the proof,
we just have to treat the case of parabolic trajectories,
whose prototype is described by Eqs (\ref{parabol}).
Introducing the proper time parameter and the radius
$R$ of AdS, the latter can be rewritten:

$x^0 =t,\
x^{d-1} = \sigma {\rm sh} v + {t^2\over 2\sigma e^v },\ \
x^d = \sigma {\rm ch} v - {t^2\over 2\sigma e^v }\ $,
with $\ \sigma^2 = R^2 + (x^1)^2 + \cdots + (x^{d-2})^2$.

The ambient acceleration-vector is $w(t)= (0,\ldots,
{1\over \sigma e^v},
-{1\over \sigma e^v})$, which is such that
$(w(t),w(t))=0$ and therefore the AdS acceleration-vector
$\gamma(t) = w(t) +{1\over R^2} x(t)$ is such that
$ (\gamma(t),\gamma(t)) =
 -{1\over R^2}$.

\subsection{The Lie algebra of $G$, time-like and isotropic two-planes}

The Lie algebra $\GG$ of $G$ can be identified with the real vector
space of the linear operators $A$ on $\amb$
such that ${A^\mu}_\nu = A^{\mu \rho}\eta_{\rho \nu}$ with
$A^{\mu \rho} = -A^{\rho \mu}$.
Hence there is a canonical linear bijection $\ell$ of the
space of antisymmetric 2-tensors over $\amb$ onto $\GG$.
A basis of $\GG$ is
provided by $\{M_{\mu \nu}\ :\ 0 \le \mu < \nu \le d\}$, with
${(M_{\mu \nu})^\rho}_\sigma = e_\mu^\rho e_{\nu \sigma}
-e_\nu^\rho e_{\mu \sigma}$;
using the standard notation
$a \wedge b = a \otimes b
- b \otimes a$, we write:
$M_{\mu \nu} = \ell(e_\mu \wedge e_\nu)$.

In particular
\begin{equation}
M_{0d} = \left (
\begin{array}{ccc}
0 & \ldots & 1 \\
\vdots & 0 & \vdots \\
-1 & \ldots & 0
\end{array}
\right ),\ \ \ \ \ e^{tM_{0d}} = \left (
\begin{array}{ccc}
\cos t & \ldots & \sin t\\
\vdots & 1 & \vdots \\
-\sin t & \ldots & \cos t
\end{array}
\right )\
\label{2.4}\end{equation}
and the diffeomorphism $\chi$ (see Eq. (\ref{2.11})) can be
rewritten as follows
\begin{equation}
(t,\vec x) \mapsto {\rm exp}(t M_{0d})(0,\ \vec{x},\
\sqrt {1+ {\vec{x}}^2}),
\label{diffLie}
\label{2.4.a}\end{equation}
with $t \in S^1$.
The same formula can also be used for describing the lifting $\wchi$
of $\chi$ on the covering spaces, provided $t$ is allowed to vary in $\bR$,
and $A \mapsto \exp(A)$ is understood as the exponential
map of $\GG$ into $\wt G_0$, and $(0,\ \vec{x},\
\sqrt {1+ {\vec{x}}^2})$ is identified with a point of one
of the fundamental domains of $\wXd$. The same remark applies to
the formulae
\begin{equation}
\chi(t+s,\ \vec{x}) =
\exp(s M_{0d})\chi(t,\ \vec{x}),\ \ \
\wchi(t+s,\ \vec{x}) =
\exp(s M_{0d})\wchi(t,\ \vec{x}).
\label{2.4.b}\end{equation}
For simplicity, we keep the
same notation, and let the context decide on its interpretation.

It is also easy to verify that, with the
notations of (\ref{2.3}), $[e^s] = \exp sM_{10}$.

The scalar product (\ref{2.1}) naturally induces a
$G$-invariant scalar product
in the space of the contravariant tensors of order $p$, namely
$(A,B) = A^{\mu_1\dots\mu_p}g_{\mu_1 \nu_1}\dots g_{\mu_p \nu_p}
B^{\nu_1\dots\nu_p}$. This satisfies
\begin{equation}
(a_1\otimes\ldots\otimes a_p,\ b_1\otimes\ldots\otimes b_p) =
(a_1,b_1)\ldots(a_p,b_p)\ .
\end{equation}
In particular, given $a, b \in \amb$,
\begin{equation}
{1\over 2}(a \wedge b,\ a \wedge b) = (a,a)(b,b) - (a,b)^2\ ,
\label{2.4.1}\end{equation}
\begin{equation}
{1\over 2}(e_0 \wedge e_d,\ a \wedge b) = a_0 b_d - a_d b_0\ .
\label{2.5}\end{equation}
The square of this 2-dimensional determinant is a 2-dimensional
Gramian :
\begin{eqnarray}
(a^0 b^d - a^d b^0)^2 &=&
\left[(a^0)^2 + (a^d)^2\right]\left[(b^0)^2 + (b^d)^2\right]
- (a^0 b^0 + a^d b^d)^2 \nonumber\\
&=& ((a,a) + \vec{a}^2)((b,b) + \vec{b}^2) -
((a,b) + \vec{a}\cdot \vec{b})^2\nonumber\\
&=& (a,a)(b,b)-(a,b)^2 + \vec{a}^2 \vec{b}^2 - (\vec{a}\cdot \vec{b})^2
\nonumber\\
&&+\  (a,a)\vec{b}^2 + (b,b)\vec{a}^2 - 2(a,b)\vec{a}\cdot \vec{b}\ .
\label{2.5.1}\end{eqnarray}
Supposing $(a,a)> 0$, $(b,b)>0$, and $(a,a)(b,b) - (a,b)^2 >0$, we find
$(a^0 b^d - a^d b^0)^2 > 0$.
If $a$ and $b$ are continuously varied while $(a,a)$,
$(b,b)$, $(a,b)$ are kept constant with the above inequalities
satisfied, the sign of $(e_0 \wedge e_1,a \wedge b)$
remains constant. Under the same conditions, $a$ can be brought to
the form $a^0 e_0$, $a^0 >0$, by a transformation in $G_0$.
Then, denoting $b' = (0,\ b^1,\ldots, b^d)$,
we have $(a^0)^2 (b',b')>0$, i.e. $b'$ is time-like in the Minkowski
space $\{x\ :\ x^0 =0\}$. It can be brought to the form $b'^d e_d$ by a
Lorentz transformation acting in the same space, after which
$a = a^0 e_0$, $b = b^0e_0 + b^d e_d$, $a^0 b^d - a^d b^0 = a^0 b^d$.
In particular if $a$ and $b$ are orthonormal,
the necessary and sufficient condition
for $a = \Lambda e_0$, $b =\Lambda e_d$,
$\Lambda \in G_0$, is that the above
scalar product (\ref{2.5}) be positive.
Suppose $a$ and $b$ are as above and $(e_0\wedge
e_d,a \wedge b) > 0$. Then the two dimensional real vector subspace spanned by
$a$ and $b$, and any parallel 2-plane, equipped with the metric induced by
(\ref{2.1}), is a Euclidean space (with positive metric). Conversely given a
2-plane with strictly positive induced metric, the parallel two dimensional
real vector subspace has an orthonormal base $(a,\ b)$ and the scalar product
$(e_0\wedge e_d,\ a \wedge b)$ can be made positive by changing $b$ to
$-b$ if necessary. We call such a 2-plane time-like, and always regard it as
oriented by the 2-form $d(a,x)\wedge d(b,x)$. The one-parameter subgroup of
$G_0$ defined by $t \mapsto \exp t\ell(a\wedge b)$ leaves this 2-plane
invariant, and every orbit of this subgroup is contained in a parallel
2-plane. This subgroup is conjugated in $G_0$ to $\exp tM_{0d}$. It follows
that
\begin{eqnarray}
\exp t\ell(a\wedge b)\,a &=& \cos(t)\,a - \sin(t)\,b\ ,\nonumber\\
\exp t\ell(a\wedge b)\,b &=& \sin(t)\,a + \cos(t)\,b\ ,\nonumber\\
\exp t\ell(a\wedge b)\,x &=& x \ \ {\rm if}\ \ (a,x) = (b,x) =0\ .
\label{2.7}\end{eqnarray}
We denote $\CC_1$ the subset of $\GG$ consisting of
all elements of the form $\ell(a\wedge b)$ with
$(a,a) = (b,b) = 1$, $(a,b) = 0$,
and $(e_0\wedge e_d,\ a \wedge b) > 0$. Equivalently,
\begin{equation}
\CC_1 = \{ \Lambda\, M_{0d}\,\Lambda^{-1}\ :\ \Lambda \in G_0\}\ .
\label{2.8}
\end{equation} We denote $\CC_+$ the cone generated in
$\GG$ by $\CC_1$, i.e. $\CC_+ = \bigcup_{\rho >0} \rho \CC_1$.

We note that for all the elements $M$ of $\CC_1$
the corresponding group elements $\exp t M$ can be
considered as belonging either to $G_0$ or to
$\wt G_0$ according to whether $\exp$ is regarded as the exponential
map of one or the other group. As mentioned at the beginning
of this section,
we always identify $G_0$ (resp. $G_0^{(c)}$)
with a subgroup of $SL(d+1,\ \bR)$ (resp. $SL(d+1,\ \bC)$), and
its exponential map with the matrix exponential.

\vskip 0.2cm
While in the real Minkowski space the isotropic subspaces are
one-dimen\-sional, the maximal isotropic subspaces in $\amb$ are
two-dimensional when $d > 2$.

\begin{lemma}
Let $a,\ b \in \amb$ be linearly independent and
satisfy $(a,a) = (b,b) = (a,b) = 0$. Then $d+1 \ge 4$ and there
is a $\Lambda \in G_0$ such that $\Lambda a = e_0+e_1$ and
$\Lambda b = \pm(\veps e_{d-1}+e_d)$, where $\veps = \pm 1$.
Any $x \in \amb$ such that
$(x,a) = (x,b) = (x,x) = 0$ is a linear combination of $a$ and $b$.
\end{lemma}
\noindent {\it Proof.}
If $d =2$, $\amb$ has a metric opposite to a Minkowskian metric
and the isotropic subspaces are one-dimensional. We assume that
$d \ge 3$. The vector $a$ must have a non-vanishing projection
in the 2-plane spanned by $e_0$ and $e_d$, and can be brought
by some $\exp(\theta M_{0d})$ to have $a^d = 0$, $a^0 > 0$.
In the Minkowski space $\{x\ :\ x^d =0\}$, there is a Lorentz
transformation in the connected component of the identity
which brings $a$ to be equal to $e_0+e_1$. After these
transformations $b$ satisfies $b^0 = b^1 = c$. Hence the projection
of $b$ into the Minkowski space orthogonal to $e_0$ and $e_1$ is
light-like and cannot vanish, since then $b$ would be colinear
to $a$. In particular $b^d \not= 0$. Assuming e.g. $b^d >0$,
there exists, in the Minkowski space generated by $\{e_2,\ldots,e_d\}$,
a Lorentz transformation in the connected component of the identity
which brings $b'$ to the form $\veps e_{d-1}+e_d$ ($\veps = \pm 1$)
without affecting the
subspace generated by $e_0$ and $e_1$. At this point,
$a = e_0+e_1$, $b = c(e_0+e_1) +\veps e_{d-1}+e_d$. We now apply the
transformation $A(\veps c) \in G_0$ given, in the space generated
by $\{e_0,\ e_1,\ e_{d-1},\ e_d\}$, by
\begin{equation}
A(\veps c) = \left (
\matrix{
\strut 1+ {c^2 \over 2} & -{c^2 \over 2} & -\veps c & 0\cr
\strut {c^2 \over 2} & 1- {c^2 \over 2} & -\veps c & 0\cr
\strut -\veps c & \veps c & 1 & 0\cr
\strut 0 & 0 & 0 & 1\cr
} \right )\ \ =
\ \ \exp \veps c \left (
\matrix{
\strut 0 & 0 & -1 & 0\cr
\strut 0 & 0 & -1 & 0\cr
\strut -1 & 1 & 0 & 0\cr
\strut 0 & 0 & 0 & 0\cr
} \right )\ .
\label{2.8.1}\end{equation}
This brings $a$ and $b$ to the forms $e_0+e_1$ and $\veps e_{d-1}+e_d$,
respectively. Suppose now that $x \in \amb$ satisfies
$(x,x) = (x,a) = (x,b) = 0$. Then $x^0 = x^1$, $x^{d-1} = \veps x^d$
and the projection of $x$ into the subspace orthogonal to
$\{e_0,\ e_1,\ e_{d-1},\ e_d\}$ has to vanish. Thus
$x = x^0 a + x^d b$.
\endprf
\vskip15pt

We finally give an explicit characterization of the local future cone
$V^+_x$ at any point $x$ of $X_d$.
We define
$V^+_x$ as the connected component of $\{y\ : (y,x) =0,\ \
(y,y) >0\}$ which contains the timelike vector $M_{0d}\,x$.
In the case when
$x= e_d$, $M_{0d}\,e_d = e_0$ and any $y \in V^+_{e_d}$ can be
written as $\Lambda \rho e_0$, where $\Lambda \in G_0$ belongs to
the stabilizer of $e_d$: $\Lambda e_d = e_d$ and $\rho >0$. Hence
\begin{eqnarray}
V^+_{e_d} &=& \{\Lambda\,\rho M_{0d}\,\Lambda^{-1}\,e_d\ :\
\Lambda \in G_0,\ \ \ \Lambda e_d = e_d,\ \ \ \rho > 0\}\ ,
\nonumber \\
&=& \{\rho\,Me_d\ :\ M = \ell(a\wedge e_d),\ \ \ (a,a) = 1,\ \
(a,e_d) = 0,\nonumber \\
&& (e_0\wedge e_d,\ a\wedge e_d) >0,\ \ \ \rho>0\}\ ,
\label{2.9}
\end{eqnarray}
and therefore, for any $x \in X_d$
\begin{eqnarray}
V^+_x &=& \{\rho\,Mx\ :\ M = \ell(a\wedge x),\ \ \ (a,a) = 1,\ \
(a,x) = 0,\nonumber \\
&& (e_0\wedge e_d,\ a\wedge x) >0,\ \ \ \rho>0\}\ .
\label{2.10}\end{eqnarray}

\section{Tuboids}

\label{TUB}
\subsection{Future and past tuboids of $X_d^{(c)}$}
\label{TFP}

We denote
\begin{equation}
\bC_+ = \{\z \in \bC\ :\ \Im \z > 0\} = -\bC_-\ .
\label{3.1}\end{equation}
For $z = x+iy \in \ambc$, we define
\begin{equation}
\epsilon(z) = {1\over 2}(e_0\wedge e_d,\ y\wedge x) = y^0 x^d - x^0 y^d\ ,
\label{3.1.0}\end{equation}

\begin{definition}
\label{deftub1}
The future tuboid $\ZZ_{1+}$ and the past tuboid $\ZZ_{1-}$
of $X_d^{(c)}$ are defined
by
\begin{equation}
\ZZ_{1+} = \{\exp(\tau M)\,c\ :\ M \in \CC_1,\ \ c \in X_d,\ \ \
\tau \in \bC_+\} = \ZZ_{1-}^*
\label{3.1.1}\end{equation}
($\CC_1$ being defined by Eq. (\ref{2.8})).
The tuboids $\wt \ZZ_{1\pm}$ are defined as
the universal coverings of $\ZZ_{1\pm}$.
They are given  by the same formula
where
$\exp(\tau M)$ is now understood as an element of $\wt G_0^{(c)}$,
$\tau$ varies in $\bC_+$
and $c_1$ as an arbitrary element of $\wXd$.
\end{definition}

\begin{lemma}
\label{pseuclid}
The ``Euclidean'' AdS spacetime $X_d^{(\cal E)}$
is contained in $X_d \cup
\ZZ_{1+}
\cup \ZZ_{1-}$
\end{lemma}
\noindent{\it Proof.}
Since $X_d^{(\cal E)}$ is represented by
(\ref{diffLie})
where $t$ is changed into $it$,
all the complex points of this manifold are
seen to be obtained as points of
$\ZZ_{1+}
\cup \ZZ_{1-}$
by taking $M=M_{0d}$, $\tau = it$ and $c= (0,\vec x, \sqrt {1+{\vec x}^2})$
in (\ref{3.1.1}).

\begin{lemma}
\label{tub1}
$\ZZ_{1+}$ has the following properties:
\begin{eqnarray}
&\makebox{\it (i)}\ &\ \ \ZZ_{1+} = \{z=x+iy \in \amb^{(c)}\ :\
(x,x) - (y,y) =
1,\ \ (x,y) = 0,\nonumber\\
&& \hbox to 6 cm{\hfill} (y,y) > 0, \ \ \epsilon(z) > 0\}.
\label{3.2}\\
&\makebox{\it (ii)}&\ \ \ZZ_{1+} = \{\Lambda\, \exp(it
M_{0d})\,e_d\ :\
\Lambda \in G_0,\ \ \ t > 0\}\ .\label{3.2.1}\\
&\makebox{\it (iii)}& \strut\hbox{\it If}\ z = x+iy \in \ZZ_{1+}\
\hbox{\it then}\ z^0 \not= 0,\ \ z^d \not=0,\ \ \hbox{\it and}\
\Im (z^0/z^d) >0.\nonumber
\end{eqnarray}
\end{lemma}
\noindent{\it Proof.} Denote $A_1$ the rhs of (\ref{3.2}), and
$A_2$ the rhs of (\ref{3.2.1}). It is clear that $\ZZ_{1+}$,
$A_1$, $A_2$ are invariant under $G_0$, and that $A_2 \subset
\ZZ_{1+}$. We first prove that $\ZZ_{1+} \subset A_1$. Suppose
that $z = \exp(\tau M)\,c$ with $c \in X_d$, $M \in \CC_1$, and
$t = \Im\tau > 0$. Let $\Lambda_1 \in G_0$ be such that
$\Lambda_1 M \Lambda_1^{-1} = M_{0d}$. Then $\Lambda_1 z =
\exp((s+ it)M_{0d}) c'$, where $c' \in X_d$, ${c'}^0 = 0$,
${c'}^d > 0$. Thus $z = \Lambda z'$ with $\Lambda =
\Lambda_1^{-1}\exp(sM_{0d})$ and $z' = x'+iy' =
\exp(itM_{0d})c'$, i.e. ${x'}^0 = 0$, ${x'}^d = {c'}^d\ch(t)$,
$y' = {c'}^d\sh(t)e_0$, hence $\epsilon(z') > 0$. It follows that
$z'$, and therefore also $z$, belong to $A_1$.\HB
We now show that $A_1 \subset A_2$.
Suppose that a point $z=x+iy$ belongs to
$A_1$. There exists a $\Lambda \in G_0$ and a real $t>0$ such
that $y = \Lambda\,\sh(t)\,e_0$ and $x = \Lambda\,\ch(t)\,e_d$, i.e.
$\Lambda^{-1}(x+iy) = \sin(it)\,e_0 + \cos(it)\,e_d = \exp(itM_{0d})\,e_d$.
This can be rewritten as
\begin{eqnarray}
z &=& \exp(is\,\ell(y\wedge x))\,x/\sqrt{(x,x)},
\nonumber\\
s &=& \log(\sqrt{(x,x)} + \sqrt{(y,y)})/ \sqrt{(x,x)(y,y)}\ .
\label{3.2.2}\end{eqnarray}
To prove {\em (iii)}, we suppose that $x+iy \in A_1$. Then
$x^d = y^d =0$ implies that both $x$ and $y$
belong to a ``usual'' Minkowski space, hence they cannot be both time-like as
well as orthogonal. The same argument excludes $z^0=0$.
Finally $\epsilon(z) = |z^d|^2 \Im (z^0/ z^d)$.
\endprf
\vskip15pt

The form (\ref{3.2}) makes it obvious that $\ZZ_{1+}$
and therefore
$\ZZ_{1-}$ are open
subsets of $\Xcd$ while the definition makes it obvious that they are
connected. $\ZZ_{1+}$ and $\ZZ_{1-}$ are disjoint since every point
$z$ of $\ZZ_{1-}$ satisfies $\epsilon(z) < 0$.

\begin{lemma}
\label{z2}
Let $a+ib \in \ovl{\ZZ_{1+}}$.
Then for every $M \in \CC_1$ and $\tau \in \bC_+$,
the point $\exp(\tau M)(a+ib)$ is in $ \ZZ_{1+}$.
\end{lemma}

\noindent{\it Proof.} By the invariance of $\ZZ_{1+}$ under $G_0$, it suffices
to prove the statement in the case when $M = M_{0d}$ and $\tau = it$ with real
$t > 0$. We suppose that $(b,b)\ge 0$, $(a,a)-(b,b) =1$, $(a,b) =0$, and
$\epsilon(a+ib) \ge 0$. This implies
$(a^0)^2 + (a^d)^2 \ge (a,a) \ge 1$.
Let $x+iy = \exp(it M_{0d})(a+ib)$. A simple calculation shows that
\begin{eqnarray}
(y,y) - (b,b)&=&
\left[(a^0)^2 + (a^d)^2 + (b^0)^2 + (b^d)^2\right]\sh^2 t \nonumber\\
&+&  2(b^0 a^d - b^d a^0)\sh t\, \ch t > 0,
\end{eqnarray}
\begin{eqnarray}
y^0 x^d - y^d x^0 &=&
\left[(a^0)^2 + (a^d)^2 +(b^0)^2 + (b^d)^2\right]\sh t\, \ch t\nonumber\\
&+& (b^0 a^d - b^d a^0)(\ch^2 t + \sh^2 t) >0.
\end{eqnarray}

\vskip15pt

\begin{lemma}
\label{cutplane}
\begin{description}
\item (i) The image of the domain $\ZZ_{1+}$  (or $\ZZ_{1-})$
under the coordinate map $z \mapsto z^d = (e_d,z)$ is the cut-plane
$\Delta = {\bf C} \setminus[-1,1]$.
\item (ii)
The image of the domain $\ZZ_{1-} \times \ZZ_{1+}$ (or $\ZZ_{1+} \times
\ZZ_{1-}$) by the (scalar product)
mapping $z_1,z_2 \mapsto (z_1,z_2)$ is the cut-plane
$\Delta$.
\item (iii)
The map $z_1, z_2 \mapsto (z_1,z_2)$ of $\ZZ_{1-} \times \ZZ_{1+}$
onto $\Delta$ can be lifted to a map of
$\widetilde\ZZ_{1-} \times \widetilde\ZZ_{1+}$ onto
the covering $\widetilde\Delta$ of the cut-plane
$\Delta$.
\end{description}
\end{lemma}

\noindent{\it Proof.}\HB
{\em (i)}
If $z \in \ZZ_{1+}$, then, by Lemma \ref{tub1}, it can be written
as $z = \Lambda \exp(it M_{0d})e_d$, with $\Lambda \in G_0$
and $t > 0$. Hence $(e_d,\ z) = (a,\ \exp(it M_{0d})e_d)$
where $a \in X_d$ can be written as
\begin{equation}
a = (\sqrt{1+\vec{a}^2}\,\sin s,\ \vec{a},\ \sqrt{1+\vec{a}^2}\,\cos s) =
\exp(s M_{0d})\,(0,\ \vec{a},\ \sqrt{1+\vec{a}^2})\ ,
\end{equation}
hence $(e_d,\ z) = \sqrt{1+\vec{a}^2}\,\cos(it -s) \in \Delta$.
Conversely any $\zeta \in \Delta$ can be written as
$\zeta = \cos(u+iv)$, $v>0$, i.e. $\zeta = (e_d,\ \exp((u+iv)M_{0d})e_d)$
is in the image of $\ZZ_{1+}$.

\vskip10pt

\noindent {\em (ii)}
Let $z_1 \in \ZZ_{1-}$, $z_2 \in \ZZ_{1+}$. Then
$z_1 = \exp(-\tau_1M_1)c_1$, $z_2 = \exp(\tau_2M_2)c_2$, with
$M_j \in \CC_1$, $\tau_j \in \bC_+$, $c_j \in X_d$, $j = 1,\ 2$.
Hence
\begin{equation}
(z_1,\ z_2) = (c_1,\ \exp(\tau_1M_1)\exp(\tau_2M_2)c_2) =
(e_d,\ z'),
\end{equation}
where $z' = \Lambda \exp(\tau_1M_1)\exp(\tau_2M_2)c_2$ (for some
$\Lambda \in G_0$) belongs to $\ZZ_{1+}$ by Lemma \ref{z2}.
Therefore $(z_1,\ z_2) \in \Delta$ by {\em (i)}. Conversely any
$\zeta \in \Delta$ can be written as $\zeta = \cos(u+iv)$, $v>0$,
i.e. $\zeta = (\exp(-(u+iv)M_{0d}/2)e_d,\
\exp((u+iv)M_{0d}/2)e_d)$ is in the image of
$\ZZ_{1-}\times\ZZ_{1+}$.

\vskip10pt

\noindent {\em (iii)}
It is easy to see that elements of
$\ZZ_{1-} \times \ZZ_{1+}$
such that $z_1 = {\rm exp}(t_1+is_1) M_{0d} e_d$, $s_1 >0$,
and $z_2 = {\rm exp}(t_2+is_2) M_{0d} e_d$, $s_2 <0$,
have a scalar product $(z_1, z_2) = \cos (t_1-t_2 +i (s_1-s_2))$
which runs on an elliptic path with foci $-1,+1$ in the
infinite-sheeted domain $\wt \Delta$
when $t_1$ and $t_2$ vary in $\bf R$ at fixed $s_1,s_2$.
These elliptic paths
remain in
$\wt \Delta$, which they fully recover, when $s_1$ and $s_2$ vary in
their ranges. Similar elliptic paths homotopic to the latter would
be generated by starting from more general elements of
$\ZZ_{1-} \times \ZZ_{1+}$.
See also Appendix \ref{ap2pt} for an alternative
argument based on a more global representation of the tuboids. \endprf

\begin{lemma}
\label{natural1} The domain $\ZZ_{1+}$ is a domain of holomorphy.
\end{lemma}

\noindent{\it Proof.}
As proved in Appendix \ref{ap2pt}, there exists a biholomorphic map of
$\ZZ_{1+}$ onto the domain $\TT_+ \setminus \{z\ :\ z^1 = 0\}$
in the usual $d$-dimensional complex Minkowski space. This also
exhibits the topoplogy of $\ZZ_{1+}$.
\endprf

\subsection{The minimal $n$-point tuboids}

We define the minimal $n$-point future tuboid
$\ZZ_{n+}$ as the subset of $\Xcdn$ consisting of
all points $(z_1,\ \ldots,\ z_n)$ such that
\begin{equation}
z_1 = e^{\tau_1 M_1}\,c_1,\ \
z_2 = e^{\tau_1 M_1}\,e^{\tau_2 M_2}\,c_2,\ \ \ \ldots,\ \ \
z_n = e^{\tau_1 M_1}\,\ldots e^{\tau_n M_n}\,c_n,
\label{3.3}\end{equation}
where, for $1 \le j \le n$,
$\tau_j \in \bC_+$,
$M_j \in \CC_1$,
and $c_j \in X_d$. Equivalently
\begin{eqnarray}
\ZZ_{n+} &=& \{(z_1,\ \ldots,\ z_n) \in \Xcdn\ :\ \forall j = 1,\ldots,n,
\ \ \ z_j = e^{it_1 M_1}\,\ldots e^{it_j M_j}\,c_j\ ,\nonumber\\
&& t_j > 0,\ \ \ M_j \in \CC_1,\ \ \ c_j \in X_d\}\ .
\label{3.4}\end{eqnarray}
Indeed if $z$ is of the form (\ref{3.3}) with $\tau_j = s_j + it_j$
then
\begin{eqnarray}
z_j &=& e^{it_1 M'_1}\,\ldots e^{it_j M'_j}\,c'_j\ , \nonumber\\
M'_j &=& e^{s_1 M_1}\ldots e^{s_{j-1} M_{j-1}}\, M_j\,
e^{-s_{j-1} M_{j-1}}\ldots e^{-s_1 M_1}\ ,\nonumber\\
c'_j &=& e^{s_1 M_1}\ldots e^{s_{j} M_{j}}\,c_j\ .
\label{3.5}\end{eqnarray}
We define the minimal $n$-point past tuboid as
$\ZZ_{n-} = \ZZ_{n+}^*$. We denote $\wt \ZZ_{n\pm}$
the universal coverings of these sets.
However, we shall write simply
$\ZZ_{n\pm}$ when no confusion arises.

\begin{lemma}
\label{open}
The set $\ZZ_{n+}$ is open in $\Xcdn$.
\end{lemma}

\noindent{\it Proof.}
We prove, by induction on $n$, the following more detailed
statement (St$_n$):
{\it
Let $z = (z_1,\ \ldots,\ z_n) \in \Xcdn$ be such that,
for $0 \le j \le n$,
\begin{equation}
z_j = e^{it_1 M_1}\,\ldots e^{it_j M_j}\,c_j\ ,
\ \ \ t_j > 0,\ \ \ M_j \in \CC_1\ ,\ \ \ c_j \in X_d\ .
\label{3.6}\end{equation}
For every neighborhood $W$ of
$(t_1,\ldots,t_n,\ M_1,\ldots,M_n,\ c_1,\ldots,c_n)$
in $(0,\ \infty)^n\times \CC_1^n \times \Xdn$ there is a neighborhood $V$
of $z$ in $\Xcdn$ such that for every $z' \in V$ there exists
a $(t',\ M',\ c') \in W$ such that
$z'_j = e^{it'_1 M'_1}\,\ldots e^{it'_j M'_j}\,c'_j$ for
every $j = 1,\ldots, n$.
}

We start with the case $n=1$, for which we give a detailed
proof. The cases $n >1$ will then be more sketchily
treated.
Since the statement St$_1$ is invariant under $G_0$,
it suffices to consider the case when
$z = z_1 = x+iy = \exp(itM_{0d})\,c$ with $t >0$ and $c_0 = 0$,
$c_d > 0$. Thus
\begin{equation}
x+iy = (ic_d \sh(t),\ \vec{c},\ c_d \ch(t)),\ \ \ t>0,\ \ c_d >0\ .
\label{3.7}\end{equation}
Let $R = 1 +||z||$.
If $z'' \in \Xcd$ is of the form $(iy''_0,\ \vec{x''},\ x''_d)$,
with $x''_d > 0$, then
\begin{eqnarray}
&z'' = x'' + i y'' = (ic''_d \sh(t''),\ \vec{c''},\ c''_d \ch(t''))
= \exp(it''M_{0d})c''\ ,&
\nonumber\\
&\vec{c''} = \vec{x''},\ \ \ c''_d = ((x''_d)^2 - (y''_0)^2)^{1/2},\ \ \
\th(t'') = y''_0 /x''_d\ ,&
\label{3.8.1}\end{eqnarray}
where $c''$ and $t''$ are continuous functions of $z''$. Hence
given $\veps \in (0,\ 1)$, there is a $\delta_1 > 0$ such that
if $z''$ is of the form $(iy''_0,\ \vec{x''},\ x''_d)$,
and $||z''-z|| < \delta_1$, then (\ref{3.8.1}) holds with
$|t''-t| < \veps$ and $||c''-c|| < \veps/4$, and
$||z''|| < R(1+\veps/4)$. There is a $\delta > 0$ such that,
for every $z' \in \Xcd$ such that $||z'-z|| < \delta$, there exists
a $\Lambda \in G_0$, satisfying $||\Lambda-1|| < \veps/4R$,
$||\Lambda^{-1}-1|| < \veps/4R$, and such that $z'' = \Lambda z'$
satisfies $||z''-z|| < \delta_1$ and $z'' = (iy''_0,\ \vec{x''},\ x''_d)$,
$x''_d >0$. Then (\ref{3.8.1}) holds, and $z' = \exp(it''M')c'$,
with $M' = \Lambda^{-1}M_{0d}\Lambda$, $c' = \Lambda^{-1}c''$.
Since $||M_{0d}||=1$, $||c|| \le ||z||$, this implies
$||M'-M_{0d}|| < \veps$ and $||c'-c|| < \veps$.

We now assume that the statement St$_m$ has been proved for all
$m \le n-1 \ge 1$. Let $z = (z_1,\ \ldots,\ z_n)$ satisfy (\ref{3.6})
and let $z' = (z'_1,\ \ldots,\ z'_n)$ be sufficiently close to $z$.
By St$_1$, $z'_1 = \exp(it'_1 M'_1)c'_1$ with $t'_1$, $M'_1$, $c'_1$
respectively close to $t_1$, $M_1$, $c_1$.
The point
\begin{equation}
(z''_2, \ldots, z''_n) = (\exp(-it'_1 M'_1)z'_2, \ldots,
\exp(-it'_1 M'_1)z'_n)
\label{3.9}\end{equation}
is close, in $X_d^{(c)(n-1)}$, to the point
\begin{equation}
(e^{it_2 M_2}c_2\ ,\ldots,\ e^{it_2 M_2}\ldots e^{it_n M_n}c_n)\ .
\label{3.10}\end{equation}
By St$_{n-1}$ the point (\ref{3.9}) can be rewritten as
\begin{equation}
(e^{it,_2 M'_2}c'_2\ ,\ldots,\ e^{it'_2 M'_2}\ldots e^{it'_n M'_n}c'_n)\ ,
\label{3.11}\end{equation}
where, for $2 \le j \le n$, $t'_j$, $M'_j$, $c'_j$ are respectively
close to $t_j$, $M_j$, $c_j$. This proves St$_n$. \endprf

\subsection{The $n$-point future and past tuboids}

\label{TBIG}
\begin{definition}
\label{bigs}
We denote $G_0^+$ (resp. $\wt G_0^+$)
the subset of $G_0^{(c)}$ (resp. $\wt G_0^{(c)}$) consisting
of all elements of the form $\exp(\tau_1 M_1)\ldots \exp(\tau_N M_N)$,
where $N \in \bN$, $\tau_j \in \bC_+$, $M_j \in \CC_+$ for all
$j = 1,\ldots,N$.
We denote $G_0^- = \{\Lambda\ :\ \Lambda^{-1} \in G_0^+\}$.
This is also the complex conjugate of $G_0^+$.
We define the $n$-point future and past tuboids $\TT_{n+}$ and $\TT_{n-}$
by
\begin{eqnarray}
\TT_{n+} &=& \{(z_1,\ \ldots,\ z_n) \in \Xcdn\
\ :\ \forall j = 1,\ldots,n,\nonumber\\
&& z_j = \Lambda_1\ldots \Lambda_j\,c_j\ ,\ \ \
\Lambda_j \in G_0^+\ , \ \ \
c_j \in X_d\  \}\ ,\nonumber\\
\TT_{n-} &=& \TT_{n+}^*\ .
\label{3.12}\end{eqnarray}
\end{definition}
Note that if
$\Lambda_1,\ \Lambda_2 \in G_0^+$, then $\Lambda_1\Lambda_2 \in G_0^+$.
If $\Lambda \in G_0$ then $\Lambda G_0^+ \Lambda^{-1} = G_0^+$. Similar
properties hold for $\wt G_0^+$.

\begin{lemma}
\label{g+open}\HB
(i) $G_0^+$ is open in $G_0^{(c)}$.\HB
(ii) $G_0^+ = G_0 G_0^+ = G_0^+  G_0 $.
\end{lemma}
This lemma is proved in Appendix \ref{g+pf}. Similar properties hold for
$\wt G_0^+$.

The tuboids $\TT_{n\pm}$ are invariant under $G_0$, i.e.
if $\Lambda \in G_0$ and $(z_1,\ldots,z_n) \in \TT_{n+}$,
then $(\Lambda z_1,\ldots,\Lambda z_n) \in \TT_{n+}$.
Obviously $\ZZ_{n\pm} \subset \TT_{n\pm}$. As a consequence of
Lemma \ref{z2}, $\ZZ_{1+} = \TT_{1+}$.

\begin{lemma}
\label{bopen}
For any $n \in \bN$, $\TT_{n+}$ is open.
\end{lemma}

\noindent {\it Proof.}
As in the proof of Lemma \ref{open}, we prove, by induction on $n$,\HB
(P$_n$):
{\it Let $z = (z_1,\ \ldots,\ z_n) \in \Xcdn$ be such that,
for $0 \le j \le n$,
\begin{equation}
z_j = \Lambda_1\,\ldots\Lambda_j\,c_j,
\ \ \ \Lambda_j \in G_0^+\ ,\ \ \ c_j \in X_d\ .
\label{3.14}\end{equation}
Then, for any $z' \in \Xcdn$ sufficiently close to $z$, there exist
$\Lambda'_1,\ldots,\Lambda'_n \in G_0^+$ and $c'_1,\ldots,c'_n \in X_d$,
respectively
close to $\Lambda_1,\ldots,\Lambda_n$ and $c_1,\ldots,c_n$, such that
$z'_j = \Lambda'_1\,\ldots\Lambda'_j\,c'_j$.}

We start with $n=1$ and suppose, without loss of generality in view
of $G_0$-invar\-ian\-ce, that $z = \Lambda\,c$ with $c\in X_d$, and
$\Lambda = \exp(it_1 M_1)\ldots\exp(it_L M_L)$, $t_k > 0$ for
all $k=1,\ldots,L$. Assume e.g. $L>1$ and let $z'\in \Xcd$ be sufficiently
close to $z$. Then $z'' = \exp(-it_{L-1} M_{L-1})\ldots\exp(-it_1 M_1) z'$
is close to $\exp(it_L M_L) c$ and it can, by the proof of
Lemma \ref{open}, be written as $z'' = \exp(it'_L M'_L)c'$,
where $t'_L >0$, $M'_L \in \CC_1$ and $c' \in X_d$ are respectively close to
$t_L >0$, $M_L$, and $c$. Thus $z' = \Lambda'\,c'$ with
$\Lambda' = \exp(it_1 M_1)\ldots\exp(it_{L-1} M_{L-1})\exp(it'_L M'_L)$.

The inductive proof of P$_n$ for all $n > 1$ follows the same
line as in the proof of Lemma \ref{open}. \endprf \vskip15pt

Note that $\TT_{n+} \subset (\TT_{1+})^n$. Another proof of
Lemma \ref{bopen} can be based on Lemma \ref{g+open}.

\begin{definition}
The tuboids $\wt \TT_{n\pm}$ are defined as the universal covering spaces of
$\TT_{n\pm}$.
\end{definition}

\begin{remark}\rm
The transformation $[-1] = e^{i\pi M_{10}}$ belongs to $G$ and
$G_0^{(c)}$ but not to $G_0$. Indeed $[-1] M_{0d} [-1]\linebreak[0]
= -M_{0d}$.
If $L = \ell(u \wedge v) \in \CC_1$, and
$L' = [-1] L [-1] = \ell(u' \wedge v')$
we find
$(M_{0d},\ L') = -(M_{0d},\ L)$, i.e.
$u'_0 v'_d - u'_d v'_0 < 0$ hence $L' \in -\CC_1$.
As a consequence if $(z_1,\ldots,\ z_n) \in \TT_{n+}$ (resp. $\ZZ_{n+}$)
then $([-1]z_1,\ldots,\ [-1]z_n) \in \TT_{n+}^*$
(resp. $\ZZ_{n+}^*$).
\end{remark}

\subsection{Permuted tuboids}
\label{permtub}
For $n \ge 2$ and for any permutation $\pi$ of $\{1,\ldots,\ n\}$,
the permuted tuboid $\TT_{n,\pi}$ is defined by
\begin{equation}
\TT_{n,\pi} = \{z \in \Xcdn\ :\ (z_{\pi(1)},\ldots,\ z_{\pi(n)})
\in \TT_{n+} \}\ .
\label{3.14.1}\end{equation}
An analogous definition is used for $\wt\TT_{n,\pi}$.
Two permuted tubes $\TT_{n,\pi}$ and $\TT_{n,\pi'}$
are called adjacent if $\pi'\pi^{-1}$ is the transposition
of two consecutive indices.
An interesting peculiarity of the AdS space-time is that
(in contrast to the Minkowskian situation) adjacent
permuted tuboids are not disjoint. The simplest example
is

\begin{lemma}
\label{inters}
Let $c_1$ and $c_2$ be
real and $|(c_1,c_2)| > 1$. Then,\HB
(i) for each $A \in G_0^+$,
$(A c_1,\ A c_2) \in \TT_{2+} $.\HB
(ii) for each $A \in G_0^+$,
$(A c_1,\ A c_2) \in \TT_{2+} \cap \TT_{2,(2,1)}$.
\end{lemma}
Here $\TT_{2,(2,1)} = \{z\ :\ (z_2,\ z_1) \in \TT_{2+}\}$.

\noindent {\it Proof.}
It is obvious that (ii) follows from (i), since the hypotheses are
invariant under the exchange of $c_1$ and $c_2$.
To prove (i), we may assume
that, in the coordinates $0$, $1$, $d$ (all others kept equal to 0),
\begin{equation}
c_1 = e_d = (0,\ 0,\ 1),\ \ \
c_2 = (0,\ \sh u,\ \veps\ch u),\ \ \
\veps = \pm 1,\ \ \ u \not= 0.
\label{T2.23}\end{equation}
Let $s$ be real and satisfy $|s| \in (0,\ \pi)$ and $\veps u s > 0$. Then
\begin{eqnarray}
e^{isM_{10}}\,c_2 &=&
(i\sh u \sin s,\ \sh u \cos s,\  \veps\ch u)\\
&=& (ib^d \sh t,\ b^1,\ b^d \ch t)\ \ =\ \
e^{it M_{0d}}\,b
\label{T2.24}\end{eqnarray}
\begin{equation}
\th t = \veps \sin s\,\th u \in (0,\ 1),\ \ \ b^0 =0,\ \ \
b^1 = \cos s\,\sh u,\ \ \
b^d = \veps \ch u/\ch t\ .
\label{T2.25}\end{equation}
Hence $e^{isM_{10}}\,c_2 \in \TT_{1+}$. Let $A \in G_0^+$ and
$z_1 = A c_1$, $z_2 = A c_2$. For sufficiently
small $|s| > 0$, since $G_0^+$ is open,
we have $A e^{-isM_{10}} \in G_0^+$, and
\begin{equation}
z_1 = (A e^{-isM_{10}})\, c_1,\ \ \
z_2 = (A e^{-isM_{10}})\, (e^{isM_{10}}\,c_2)
= (A e^{-isM_{10}})\, e^{it M_{0d}}\,b,
\label{T2.26}\end{equation}
so that $(z_1,\ z_2) \in \TT_{2+}$. Similarly
$(z_2,\ z_1) \in \TT_{2+}$. \endprf

\begin{remark}\rm
\label{disj}
Denote
\begin{equation}
\RR_2 = \{(c_1,\ c_2) \in X_d^2 \ :\ (c_1,c_2) > 1\},\ \ \ \
\RR'_2 = \{(c_1,\ c_2) \in X_d^2 \ :\ (c_1,c_2) < -1\}\ .
\label{T2.26.1}\end{equation}
then the sets $G_0^+ \RR_2 =
\{(A c_1,\ A c_2)\ :\ (c_1,\ c_2) \in \RR_2, \ \
A \in G_0^+\}$ and $G_0^+ \RR'_2$ are disjoint. If $d >2$,
each of them is connected.
\end{remark}

Lemma \ref{inters} can be generalized to $n$-point tubes as
follows:

\begin{lemma}
\label{intersn}
Let $(c_1,\ldots,\ c_j,\ c_{j+1},\ldots,\ c_n) \in \Xdn$ be such
that $|(c_j,\  c_{j+1})| > 1$.
Then for any choice of $\Lambda_k \in G_0^+$, $1\le k < j$ or
$j+1 < k \le n$, and of $A \in G_0^+$, the point $z$ such that
\begin{eqnarray}
z_k &=& \Lambda_1\ldots\Lambda_k\,c_k\ \ {\rm for}\ k <j,
\nonumber\\
z_j &=& \Lambda_1\ldots\Lambda_{j-1}A\,c_j,\ \
z_{j+1}\ =\ \Lambda_1\ldots\Lambda_{j-1}A\,c_{j+1},
\nonumber\\
z_k &=&  \Lambda_1\ldots\Lambda_{j-1}A\Lambda_{j+2}
\ldots \Lambda_k\,c_k\ \ {\rm for}\ k > j+1,
\label{T2.27}\end{eqnarray}
belongs to $\TT_{n+}$.
\end{lemma}

\vskip 0.25cm
\noindent Therefore points of the form (\ref{T2.27}) belong to the
intersection of $\TT_{n+}$ with the permuted tuboid $\TT_{n,(j+1,j)}$
obtained by exchanging the indices $j$ and $j+1$. Let
$\RR_{j,k} = \{x \in \Xdn\ :\ (x_j,x_k) >1\}$. This is an open
subset of $\Xdn$ which has two connected components if $d=2$,
only one otherwise. As a result of Lemma~\ref{intersn}, the intersection
$\TT_{n+} \cap \TT_{n,(j+1,j)}$ is an open tuboid which has connected
components bordered by those of $\RR_{j,j+1}$.

\vskip 0.25cm
\noindent {\it Proof of Lemma \ref{intersn}.}
We may again assume that
\begin{equation}
c_j = e_d = (0,\ 0,\ 1),\ \ \
c_{j+1} = (0,\ \sh u,\ \veps\ch u),\ \ \
\veps = \pm 1,\ \ \ u \not= 0.
\label{T2.28}\end{equation}
For sufficiently small $|s| >0$, and $\veps u s >0$,
let $t$ and $b = (0,\ b^1,\ b^d)$ be
given by (\ref{T2.25}), and let
$\Lambda_j = Ae^{-isM_{01}} \in G_0^+$,
$\Lambda_{j+1} = e^{it M_{0d}} \in G_0^+$. Then
\begin{equation}
z_j = \Lambda_1\ldots\Lambda_{j-1}\Lambda_j  c_j,\ \ \
z_{j+1} = \Lambda_1\ldots\Lambda_{j-1}\Lambda_j \Lambda_{j+1} b.
\label{T2.29}\end{equation}
Furthermore
\begin{equation}
A\Lambda_{j+2} =
\Lambda_j \Lambda_{j+1} \Lambda'_{j+2},\ \ \
\Lambda'_{j+2} =
e^{-it M_{0d}} e^{isM_{01}} \Lambda_{j+2} \in G_0^+\ .\ \hbox{\endprf}
\label{T2.30}\end{equation}
On the other hand ``opposite'' tuboids such as $\TT_{n+}$ and
$\TT_{n-}$ do not intersect since they are respectively contained
in $\TT_{1+}^n$ and $\TT_{1-}^n$.

\subsection{Comparing the $n$-point tuboids of $X_d^{(c)}$
and their coverings with the tubes of complex Minkowski space}
\label{Comptub}
The family of one-parameter subgroups $e^{\tau M}$ of $\wt G_0$
whose generator $M$ belongs to the set $\CC_1$ of $\cal G$
can be considered as the analog of the family of timelike
translation groups $e^{\tau a}$ acting on Minkowskian spacetime,
whose generator $a$ belongs to the unit hyperboloid shell
$H_1 =\{a\in V^+; a^2=1 \}$, but of course there is a major difference:
while in the latter case, this family of one-parameter groups form
a commutative subgroup $T_+$ of the group of spacetime translations,
the groups of the former family are all mutually noncommutative.
Nevertheless, the analogy between the two families has provided us
with the basic idea for defining the $n$-point tuboids in the AdS case.
In fact, if in the definitions of $\ZZ_{n+}$ (resp. $\TT_{n+}$)
one replaces the Lie elements $M_j \in \CC_1$ by $a_j \in H_1$ (resp.
$\Lambda_j \in G_0^+$ by $g_j \in T_+$) and the points $c_j$ on AdS by
points in Minkowskian spacetime, one exactly reobtains the usual
$n$-point tubes of complex Minkowski space as they are defined in
\cite{SW}. The most obvious (and unpleasant) effect of noncommutativity
in the AdS case is that for $n \ge 2$ the description of the
tuboids remains very implicit, since  the defining conditions
involve group elements in a heavy way
instead of being characterized directly by equations on the
AdS manifold; in particular it is not clear whether the ``minimal
tuboids'' $\ZZ_{n+}$ are really smaller than the (``complete'')
$n$-point tuboids $\TT_{n+}$, which will be used
in a natural way for expressing
the spectral condition of AdS quantum fields
(see below in Sec. \ref{HYP}).
We wish however to emphasize a simple result which displays
a reassuring analogy between the AdS tuboids and the Minkowskian
tubes, namely the inclusion of regions of the
corresponding ``Euclidean'' spacetimes in these domains. In
fact, one has
the following property, in which $\wt \chi^{(c)}$
denotes the extension to ${\bf C} \times {\bf R}^{d-1}$ of
the diffeomorphism $\wt \chi$ (see subsection \ref{pureetc}
after formulae (\ref{2.11}) and (\ref{parameucl})).

\begin{lemma}
\label{tubeucl}
For each n, the tuboid $\wt \ZZ_{n+}$
contains the image by $\wt \chi^{(c) n}$
of the following ``flat tube'':

\noindent
$\{(\tau_1,\vec x_1)\ldots,(\tau_n,\vec x_n) \in
{\bf C}^n \times \bR^{(d-1)n};\ 0< \Im \tau_1 < \cdots <\Im \tau_n\}.$
In particular
$\wt \ZZ_{n+}$
contains the following open subset of
$X_d^{(\EE) n}$:
\begin{eqnarray}
\lefteqn{
\{(z_1,\ldots,z_n) \in X_d^{(c) n}\ :}\nonumber\\
&&z_j = (i\sqrt {1+ \vec x_j^2}\  {\rm sh} \ s_j,\  \vec x_j,\
\sqrt {1+ \vec x_j^2}\  {\rm ch} \ s_j),\nonumber\\
&&\ 1 \le j \le n;\
0< s_1<\cdots < s_n \}
\label{final3} \end{eqnarray}
\end{lemma}
The proof is readily obtained by putting
$M_1 = \cdots = M_n = M_{0d}$ and $c_j =
(0,  \vec x_j, \linebreak[0]
\sqrt {1+ \vec x_j^2})$,\ \ ($\ 1 \le j \le n$) in (\ref{3.3})
and changing the sequence
$(\tau_1, \tau_1 +\tau_2,\ldots, \tau_1 +\cdots +\tau_n),\ \tau_j \in
\bC_+$
into  $(\tau_1, \tau_2,\ldots, \tau_n) $.
The subset (\ref{final3}) of $\wt \ZZ_{n+}$ which is then exhibited
is clearly (in view of (\ref{parameucl})) a subset of
$X_d^{(\EE) n}$.

This result will  entail the validity of a ``Wick-rotation procedure''
for the {\sl universal covering } of AdS spacetime
(see Sec. \ref{WR}). In fact, we note that
the flat tubes of lemma \ref{tubeucl} represent
domains in the complexification (in the time variable) of
Minkowski space as well as $\wt X_d$: these domains are isomorphic.
In the pure AdS spacetime, this isomorphism does not exist since
one has to consider quotients of the
previous flat tubes corresponding to  the geometric periodicity
conditions under the transformations $\tau_j \to \tau_j +2\pi$
as described in (\ref {3.3}).

\vskip 0.2cm
Another close analogy between the tuboids $\ZZ_{n\pm}$ and
the corresponding Minkow\-skian $n$-point tubes will be displayed
in subsection \ref{DBW}: it will be proved that there exists
a special subset of real points in $X_d^n$ enjoying the same property
as the Jost points of Minkowski space, namely all the complex points
obtained by the action of appropriate {\sl one-parameter complex subgroups}
of $G_0^{(c)}$ on these ``Jost points'' are  contained in
$\ZZ_{n+}
\cup \ZZ_{n-}$ and this is the starting point of analytic completions
(of the type of Glaser-Streater's theorem) which are crucial for
proving the Bisognano-Wichmann property (see Section \ref{BWA}).

\vskip 0.3cm
We shall now mention two interesting discrepancies between the
tuboids $\ZZ_{n+}$ and the corresponding $n$-point tubes of
complex Minkowski space.
The first one, which is developed in Section \ref{parsec},
concerns the parabolic trajectories already introduced in Section
\ref{PRE}; the corresponding complexified curves
turn out to exhibit sections of the
tuboids $\ZZ_{n+}$ which lie in
the complexified ``Poincar\'e sections of AdS''.
These peculiarities are at the origin of the fact that
QFT's in these Poincar\'e sections can be generated as
restrictions of QFT's on AdS.
No analogs of such families of complexified trajectories
in isotropic (or ``lightlike'')
hyperplanes exist in complex Minkowski space.

\vskip 0.2cm
Finally, we have exhibited above in subsection \ref{permtub}
a property of pairs of permuted tuboids in $X_d$ which is definitely
new with respect to the corresponding pairs in complex Minkowski space.
In the latter case, such tubes always have an empty intersection
and the property of a common analytic continuation
for pairs of functions analytic in ``adjacent permuted
tubes''
necessitates the application of the edge-of-the-wedge theorem
to the
boundary values from these two domains
through an appropriate ``coincidence region''.
In the present case, the situation is somewhat simpler,
since according to lemma \ref{inters}
it is a general fact that {\sl adjacent pairs of permuted tuboids in
$X_d^{(c)}$ have nonempty intersections}.

\section{QFT on $X_d$ and $\wt X_d$}

\label{HYP}
As usual, it is possible to formulate the main assumptions in terms
of distributions (the test-functions having then compact supports)
or tempered distributions. We denote $\BB_n$ the space of test-functions on
$\Xdn$ or $\wXdn$. This may be either $\DD(\Xdn)$ (resp. $\DD(\wXdn)$)
or $\SS(\Xdn)$ (resp. $\SS(\wXdn)$).
The Borchers algebra $\BB$ on $X_d$ (resp. $\wt X_d$) is the
complex vector space of terminating sequences
of test-functions $f = (f_0, f_1(x_1),\ldots,
f_n(x_1,\ldots,x_n),\linebreak[0]\ \ldots)$, where $f_0 \in {\bC}$
and $f_n \in \BB_n$ for all $n \ge 1$,
the product and $\star$ operations being given by
\begin{equation}
(fg)_n =
\sum_{p,\ q \in \bN \atop p+q = n}\ f_p \otimes g_q,\ \ \ \
(f^\star)_n (x_1,\ldots,\ x_n) =
\ovl{f_n (x_n,\ldots,\ x_1)},
\label{4.1}\end{equation}
The action of $\Lambda \in G_0$ (resp. $\wt G_0$) on $\BB$ is defined by
$ f \mapsto f_{\{\Lambda\}}$, where
\begin{eqnarray}
f_{\{\Lambda\}} = (f_0, f_{1\{\Lambda\}},\ldots,f_{n\{\Lambda\}},\ldots),
\nonumber \\
f_{n\{\Lambda\}} ({x_{1}},\ldots,x_{n})
=f_{n} ({\Lambda}
^{-1}{x_{1}},\ldots,{\Lambda}^{-1}x_{n}).
\label{4.2}\end{eqnarray}
It will also be useful to denote
\begin{equation}
f_{n\{\Lambda_1,\ldots,\Lambda_n\}} (x_1,\ldots,x_n)
=f_{n} (\Lambda_1^{-1}x_1,\ldots,\Lambda_n^{-1}x_n),
\label{4.2.1}\end{equation}
where $f_n \in \BB_n$ and $\Lambda_1,\ldots,\Lambda_n$
belong to $G_0$ or $\wt G_0$.
If $\pi$ is a permutation of $(1,\ldots,n)$, and $f_n \in \BB_n$
we define the function $_\pi f_n \in \BB_n$ by
\begin{equation}
_\pi f_n(x_{\pi(1)},\ldots,x_{\pi(n)} =
f_n(x_1,\ldots,x_n)\ .
\label{4.2.2}\end{equation}
The theory of a single scalar quantum field theory on $X_d$
(resp. $\wt X_d$) is specified by a continuous linear functional
$\WW$ on $\BB$,
i.e. by a sequence $\{\WW_n \in \BB'_n\}_{n \in \bN}$
(resp. $\{\WW_n \in \BB'_n\}_{n \in \bN}$),
called Wightman functions, with the
following properties:

\begin{description}
\item{1.} {\bf Covariance:} Each ${\cal W}_{n}$ is invariant under
the group $G_0$ (resp. $\wt G_0$), i.e. for all
\begin{equation}
\langle {\cal W}_{n},\ f_{n\{\Lambda\}}\rangle =
\langle {\cal W}_{n},\ f_{n} \rangle
\label{4.3}
\end{equation}
for all $\Lambda \in G_0$ (resp. $\wt G_0$).

\item{2.} {\bf Locality:}
\begin{equation}
{\cal W}_{n}({x_{1}},\ldots,x_{j},x_{j+1},\ldots,x_{n})
={\cal W}_{n}({x_{1}},\ldots,x_{j+1},x_{j},\ldots,x_{n})
\label{4.4}\end{equation}
if $x_{j}$ and $x_{j+1}$ are space-like separated.

\item{3.} {\bf Positive Definiteness:} For each $f \in \BB$,
$\WW(f^\star f) \ge 0$. Explicitly,
given $f_{0} \in {{\bC}},
f_{1} \in \BB_1,\ldots,$
$f_{k} \in \BB_k$,
then
\begin{equation}
\sum_{n,m=0}^{k}\langle
{\cal W}_{n+m},\ f_{n}^\star\otimes f_{m}\rangle\geq 0.
\label{4.5}\end{equation}
\end{description}

If these conditions are satisfied the GNS construction (see
\cite{Bo,J}) provides a Hilbert space $\HH$, a continuous unitary
representation $\Lambda \mapsto U(\Lambda)$ of $G_0$ (resp. $\wt
G_0$) and a representation $f \mapsto \Rep({f})$ (by unbounded
operators) as well as a unit vector $\Omega \in \HH$, invariant
under $U$, such that $\WW(f) = (\Omega,\ \Rep(f)\Omega)$ for all
$f \in \BB$. As a special case the field operator $\phi$ is the
operator valued distribution over $X_d$ (resp. $\wt X_d$) such
that $\phi(f_1) = \Rep(f)$ where $f = (0,\ f_1,\ 0,\ \dots)$. In
addition the construction provides vector valued distributions
$\Phi_n^{(b)}$ such that
\begin{eqnarray}
\langle \Phi_n^{(b)},\ f_n \rangle &=&
\Rep(f)\Omega\nonumber\\
&=& \int f_n(x_1,\ldots,\ x_n)\,\phi(x_1)\ldots\phi(x_n)\,\Omega\,
d\sigma(x_1)\ldots d\sigma(x_n)
\label{4.6}\end{eqnarray}
where $f = (0,\ \ldots,\ 0,\ f_n,\ 0,\ \ldots )$. As usual, for every
$\Lambda \in G_0$ (resp. $\wt G_0$),
\begin{equation}
U(\Lambda)\,\Rep(f)\,U(\Lambda)^{-1} = \Rep(f_{\{\Lambda\}}),\ \ \ \ \
U(\Lambda)\,\Omega = \Omega\ .
\label{4.7}\end{equation}
The details of this construction are completely analogous to those
of the Min\-kow\-skian case. To every element $M$ of the Lie algebra $\GG$
we can associate the one-parameter subgroup $t \mapsto \exp tM$ of
$G_0$ (resp. $\wt G_0$) and a self-adjoint operator $\wh M$ acting in
$\HH$ such that $\exp it\wh M = U(\exp tM)$ for all $t \in \bR$.
With these notations, we postulate the following

\begin{description}
\item{4.} {\bf Strong Spectral Condition:}
For every $M \in \CC_+$, every $\Psi \in \HH$, and every $\CC^\infty$
function $\wt \vhi$ with compact support contained in
$(-\infty,\ 0)$,
\begin{equation}
\int_{\bR} \left ( \int \wt \vhi(p)\,e^{-itp}\,dp \right )\,
U(\exp tM)\,\Psi\,dt = 0\ .
\label{4.8}\end{equation}
Equivalently $\wh M$ has its spectrum contained in $\Rp$.
\end{description}

Using the spectral decomposition of $\wh M$, this implies that,
for every $\Psi \in \HH$,
$t \mapsto \exp(it \wh M)\,\Psi = U(\exp tM)\,\Psi$ extends to
a function, again denoted $z \mapsto \exp(iz \wh M)\,\Psi$,
continuous on $\ovl{\bC_+}$ and holomorphic in $\bC_+$,
and bounded in norm by $||\Psi||$.
For any finite sequence $\{M_j\}_{1 \le j \le N}$ of elements of
$\CC_+$, the function
\begin{equation}
(z_1,\ldots,z_N) \mapsto
e^{iz_1 \wh M_1}\ldots e^{iz_N \wh M_N}\, \Psi
\label{4.9}\end{equation}
is therefore continuous and bounded in norm by $||\Psi||$
on the ``flattened tube''
\begin{equation}
\bigcup_{j=1}^N \{(z_1,\ldots,z_N) \in \bC^N\ :\
z_j \in \ovl{\bC_+},\ \ \ z_k \in \bR\ \ \ \forall k \not= j\}\ .
\label{4.10}\end{equation}
It is holomorphic in $z_j$ in $\bC_+$ when the other $z_k$ are kept
real.
By the flattened tube (Malgrange-Zerner) theorem, this function
extends to a continuous function on $\ovl{\bC_+}^N$,
holomorphic in $\bC_+^N$, and bounded in norm by $||\Psi||$. Thus
the function
$\Lambda \mapsto U(\Lambda)\Psi$ extends to a bounded holomorphic
function on $G_0^+$ (resp. $\wt G_0^+$ ) with continuous
boundary value on $G_0$ (resp. $\wt G_0$ ).

We now consider
\begin{eqnarray}
\lefteqn{(\tau_1,\ldots,\tau_n) \mapsto} \nonumber\\
&&\int (\Omega,\ e^{i\tau_1 \wh M_1}\phi(x_1) \ldots
e^{i\tau_n \wh M_n}\phi(x_n)\,\Omega)\,f_n(x_1,\ldots,x_n)\,
d\sigma(x_1) \ldots d\sigma(x_n)\nonumber\\
&=& \int
(\Omega,\ \phi(e^{\tau_1 M_1}x_1)\,
\phi(e^{\tau_1 M_1}e^{\tau_2 M_2}x_2) \ldots
\phi(e^{\tau_1 M_1}\ldots e^{\tau_n M_n}x_n) ,\Omega)\nonumber\\
&& \hbox to 4cm{\hfill} f_n(x_1,\ldots,x_n)\,
d\sigma(x_1) \ldots d\sigma(x_n)\nonumber\\
&\bydef& \langle \WW_n,\ f_{n\{\Lambda_1,\ldots , \Lambda_n\}}
\rangle\ ,
\label{4.11}\end{eqnarray}
where, for $1 \le j \le n$, $\tau_j \in \bR$, $M_j \in \CC_+$, and
$\Lambda_j = e^{\tau_1 M_1}\ldots e^{\tau_j M_j}$.
Suppose that $f_n = g_1\otimes \ldots \otimes g_n$ where $g_j \in \BB_1$.
Then (\ref{4.11}) extends to a function of $\tau_1,\ldots,\tau_n$
which is $\CC^\infty$ on the flattened tube
\begin{equation}
\bigcup_{j=1}^n\,
\{(\tau_1,\ldots,\tau_n)\ :\
\Im \tau_j \ge 0,\ \ \ \Im \tau_k = 0\ \ \ \forall k \not= j\}\ ,
\label{4.12}\end{equation}
and holomorphic in $\tau_j$ in $\bC_+$ when the other $\tau_k$ are
kept real.
For every $K > 0$, the restriction of this function to
\begin{equation}
\bigcup_{j=1}^n\,
\{(\tau_1,\ldots,\tau_n)\ :\ |\tau_j| < K,\ \ \
\Im \tau_j \ge 0,\ \ \ |\tau_k| <K,\ \
\Im \tau_k = 0\ \ \ \forall k \not= j\}
\label{4.13}\end{equation}
is bounded in modulus by
\begin{equation}
C(K) \prod_{j=1}^n ||g_j||_{m(K)}\ ,
\label{4.14}\end{equation}
where $||g_j||_{m(K)}$ is one of the seminorms defining the topology
of $\BB_1$. The envelope of holomorphy of the set (\ref {4.13})
contains the topological product
\begin{equation}
H_n(K) = \prod_{j=1}^n \{\tau_j \in \bC_+\ :\ |\tau_j| < K\tg(\pi/4n)\}.
\label{4.15}\end{equation}
Therefore the function (\ref{4.11}) extends to a $\CC^\infty$ function
on $(\ovl{\bCp})^n$, holomorphic in $(\bCp)^n$ and bounded in modulus by
(\ref{4.14}) on the set $H_n(K)$ for every $K > 0$. For every
$\tau \in H_n(K)$, the value of (\ref{4.11}) defines a continuous
$n$-linear form on $\BB_1^n$, hence, by the nuclear theorem, a unique
continuous linear functional on $\BB_n$. We conclude that, for a general
$f_n \in \BB_n$, the function (\ref{4.11}) extends to a $\CC^\infty$ function
on $(\ovl{\bCp})^n$, holomorphic in $(\bCp)^n$. By standard arguments
it follows that there exists a function $W_n$, holomorphic in
$\ZZ_{n+}$, having $\WW_n$ as its boundary value in the sense of
distributions. The same proof (but with a more cumbersome notation)
shows that $W_n$ is holomorphic in $\TT_{n+}$.

In the remainder of this paper we will require the Wightman functions $\WW_n$
to be tempered distributions on $\Xdn$ (resp. $\wXdn$), i.e.
we will take $\BB_n = \SS(\Xdn)$ (resp. $\SS(\wXdn)$), and we will assume,
instead of the ``strong spectral condition'', that
the following holds:

\begin{description}
\item{5.} {\bf Tempered Spectral Condition:}
For each pair of integers $m \ge 0$ and $n \ge 0$,\hfill\break
$\WW_{m+n}(w_m,\ldots,w_1,\ z_1,\ldots,z_n)$ is the boundary value,
in the sense of tempered distributions,
of a function $W_{m,n}$ of $(w,\ z)$ holomorphic and of tempered growth in
$\TT_{m+}^*\times\TT_{n+} = \TT_{m-}\times\TT_{n+}$.
(In particular
$\WW_{n}(z_1,\ldots,z_n)$ is the boundary value
of a function $W_n$ holomorphic and of tempered growth in $\TT_{n+}$.)
Moreover for any
$f_n \in \BB_n$ and every choice of $M_1,\ldots,M_n \in \CC_1$ the function
defined by (\ref{4.11}) is $\CC^\infty$ and at most of polynomial
growth in $\ovl{\bCp^n}$.
\end{description}

If the positive definiteness condition holds, the tempered spectral
condition implies the strong spectral condition. However the
tempered spectral condition makes sense even if the positive definiteness
condition does not hold.

The following lemma follows from arguments given in \cite{BEM} (Sect. 5),
using as the main tool a theorem of V.~Glaser \cite{G1}.

\begin{lemma}
\label{locpos}
Let $\WW$ be a Wightman functional satisfying the conditions of
covariance and locality and the tempered spectral condition.
Suppose in addition that there is a real open set $V \in X_d$ such that
$\WW(f^\star f) \ge 0$ for all $f \in \BB$ with support in
$V$ (i.e. such that $f_n$ has support in $V^n$ for all $n\in \bN$).
Then there exists, for each $n \in \bN$, a vector valued
function $\Phi_n$, holomorphic in $\TT_{n+}$, and with tempered
growth at infinity and near the boundaries, having
$\Phi_n^{(b)}$ as a boundary value in the sense of tempered
distributions. In particular $\WW$ satisfies the unrestricted
positive definiteness condition, and the Reeh-SchliederTheorem
holds.
\end{lemma}

The permuted Wightman functions $\WW_{n,\pi}$ defined, as usual, by
$\langle \WW_{n,\pi}\,,\ f_n \rangle =\linebreak[0]
\langle \WW_n\,,\linebreak[0]\ _\pi f_n\rangle$,
are boundary values of functions $W_{n,\pi}$ holomorphic in the
permuted tuboids
$\TT_{n,\pi} = \{z\ :\ (z_{\pi(1)},\ldots, z_{\pi(n)}) \in \TT_{n+}\}$.
Owing to local commutativity, they are branches of a single
holomorphic function.

\begin{remark}
\label{extloc}\rm
If a set of Wightman functions satisfies the tempered spectral condition
but only a part of the locality condition, i.e. if it is
assumed that $\WW_n$ and $\WW_{n,(j+1,j)}$ coincide in an open
subset of $\RR_{j,j+1}$
(necessarily symmetric under the exchange of $j$ and $j+1$),
then it follows from Lemma~\ref{intersn} that
they coincide in the whole of $\RR_{j,j+1}$. Similar extension theorems
are well-known in the Minkowskian case (see e.g. \cite{SW}) due
to phenomena of analytic completion. It is remarkable that no
completion is needed in the AdS case.
\end{remark}

If the conditions 1-5 hold, one may want to assume

\begin{description}
\item{6.} {\bf Uniqueness of the vacuum:}
The invariant subspace of $\HH$,
$\{\Psi \in \HH\ :\ U(\Lambda)\Psi = \Psi\ \forall \Lambda \in G_0
\ ({\rm resp.\ }\wt G_0)\}$
is one-dimensional, i.e. it is equal to $\bC\Omega$.
\end{description}

In the Minkowskian case this condition is equivalent to a clustering
property of the Wightman functions, namely the truncated
Wightman functions tend to zero when a proper subset of their
arguments tend to space-like infinity while the others remain bounded.
(The truncated Wightman functions have the same inductive definition
as in the Minkowskian case (\cite{J} p. 66). They have the same linear
properties as the Wightman functions.)
A similar equivalence holds in the anti-de Sitter case.
In Sect.~\ref{parsec}, we discuss some properties of this type,
which are equivalent to the uniqueness of the vacuum in the presence
of positivity.

\section{Parabolic (Poincar\'e) sections}

\label{parsec} A convenient chart of a part of $X_d$ (resp.
$\Xcd$) is provided by the parabolic coordinates $\zz,\ v \mapsto
z(\zz,\ v)$ given by
\begin{equation}
\zz,v \rightarrow z(\zz,v)= \left\{\begin{tabular}{lcll}
 $z^{\mu} $&=& $e^{ {v}} \zz^\mu  $ &  \cr
 $z^{d-1} $&=& $\sh  {v} + \frac 12 e^{ {v}} \zz^2 $  &
\cr
 $z^{d}$&=&$ \ch  {v} - \frac 12 e^{ {v}}\zz^2$ & \label{6.1}
\end{tabular}\right. .
\label{poinc}\end{equation} In this equation ${
\mu=0,1,...,d-2}$, \ $\zz^0,...,\zz^{d-2}$ are the coordinates of
an arbitrary event in a real (resp. complex) $(d-1)$-dimensional
Minkowski space-time with metric\footnote{Here and in the
following where it appears, an index {\em {\small M}} stands for
Minkowski.} $ds^2_{{M}}= d{\zz^0}^{\,2}- d{\zz^1}^{\,2} - \ldots
-d{\zz^{d-2}}^{\,2}$, \ $\zz^2 = \zz^0\zz^0 - \sum_{j=1}^{d-2}
\zz^j \zz^j$ and $v \in \bR$ (resp. $\bC$).

This explains why  the coordinates $\zz,v$ of the parametrization (\ref{poinc})
are  also called Poincar\'e coordinates. As $\zz,\ v$ vary in $\bR^d$, the
image of this map is $\{x \in X_d\ :\ x^{d-1} +x^d >0\}$. The scalar product
and the AdS metric can then be rewritten as follows:
\begin{eqnarray}
&& (z,z') = \ch( {v}- {v} ')  - \frac 12 e^{ {v}+ {v}'}
\left(\zz-\zz'\right)^2,
\label{7}\\
&& {\mathrm d}s^2_{AdS} = e^{2 {v}} {\mathrm d}s^2_{{M}}-{\mathrm d} {v}^2.
\label{metric1}
\end{eqnarray}
 Eq. (\ref{7}) implies that
\begin{equation}
(z({\zz},\ v) - z'({\zz'},\ v'))^2 = e^{v+v'}(\zz-\zz')^2
- 2\ch(v-v') +2\ .
\label{lll}
\end{equation}

For a given real $v$ we denote $\MM_v$ the parabolic section
\begin{equation}
\MM_v = \{z \in X_d\ ({\rm resp.}\ \Xcd)\ :\ (z,\ e_d - e_{d-1}) = z^{d-1} +
z^d = e^v\}. \label{7a}
\end{equation}
The subgroup $G_{(d-1)d}$ (resp. $G_{(d-1)d}^{(c)}$) of
$G_0$ (resp. $G_0^{(c)}$) which fixes $e_d - e_{d-1}$, and therefore
leaves $\MM_v$ globally invariant, is isomorphic to
the real (resp. complex) Poincar\'e group operating on the
$(d-1)$-Minkowski space. In particular, for $b = (b^0,\ldots,b^{d-2})$,
the transformation
\begin{equation}
\exp (b^\mu L_\mu) =
\left (
\matrix{
\strut 1 & \ldots & 0 & b^0 & b^0 \cr
\strut 0 & \ldots & 0 & b^1 & b^1 \cr
\vdots   &        &\vdots & \vdots & \vdots \cr
\strut 0 & \ldots & 1 & b^{d-2} & b^{d-2} \cr
\strut b_0 & \ldots & b_{d-2} & 1 + {(b,b)\over 2} & {(b,b)\over 2}\cr
\strut -b_0 & \ldots & -b_{d-2} & -{(b,b)\over 2} & 1 - {(b,b)\over 2}\cr
} \right )
\label{6.3}\end{equation}
operates in the Minkowski leaf as $\zz \mapsto \zz + b$.
If $b = (\tau,\ \vec{0},\ 0)$, this transformation leaves the
coordinates $z^1,\ldots,z^{d-2}$ unchanged and, in the 3-dimensional
space of the coordinates $z^0,\ z^{d-1},\ z^d$, is given by
\begin{equation}
e^{\tau L_0} =
\left (
\matrix{
\strut 1 & \tau & \tau \cr
\strut \tau & 1 +{\tau^2 \over 2} & {\tau^2 \over 2} \cr
\strut -\tau & -{\tau^2 \over 2} & 1 -{\tau^2 \over 2} \cr
} \right )\ .
\label{6.4}\end{equation}
Here
\begin{equation}
L_0 = \ell (e_0 \wedge (e_d - e_{d-1})) = M_{0d} - M_{0(d-1)} =
\left (
\matrix{
0 & 1 & 1 \cr
1 & 0 & 0\cr
-1 & 0 & 0\cr
} \right )\ ,
\label{6.5}\end{equation}
Similarly $L_\mu = \ell(e_\mu \wedge (e_d - e_{d-1}))$
where $0 \le \mu \le d-2$.
With these notations,
\begin{equation}
z(\zz,\ v) = \exp(\zz^\mu L_\mu)\exp(v M_{(d-1)d})\,e_d\ .
\label{6.5.1}\end{equation}
We denote
\begin{eqnarray}
\Gamma_{(d,d-1)} &=& \{b^\mu L_\mu \in \GG\ :\ b^0 > |\vec{b}|\}\nonumber\\
&=& \{\Lambda\,tL_0\,\Lambda^{-1}\ :\ t > 0,\ \ \Lambda \in G_0,\ \
\Lambda e_d = e_d,\ \ \Lambda e_{d-1} = e_{d-1}\}\ .\ \ \ \
\label{6.6}\end{eqnarray}
If $c$ is a real point of $\MM_v$, i.e.
$c = z(\cc,\ v)$ for some real $\cc$ and $v$,
and $Q = b^\mu L_\mu  \in \Gamma_{(d,d-1)}$
then $\exp( iQc) = z(\cc + ib,\ v)$ is a point of the future tube in
the complexified $\MM_v$ considered as a Minkowski space. Conversely
any point of the future tube is of this form.

\vskip 0.25 cm
More generally, we define the $n$-point forward tuboid $\FF_{d,d-1,n}$ as
\begin{eqnarray}
\lefteqn{
\FF_{d,d-1,n} =
\{(z_1,\ldots,z_n) \in \Xcdn \ :\
\forall j = 1,\ldots,n}\nonumber\\
&&z_j = \exp(iQ_1)\cdots \exp(iQ_j)\,c_j\,,\ \ \
Q_j \in \Gamma_{(d,d-1)}\,,\nonumber\\
&&c_j \in X_d,\ \ c_j^{d-1}+c_j^d > 0\}\ .
\label{6.12}\end{eqnarray}
The intersection of this set with $\MM_v^n$ is the set obtained by
restricting the $c_j$ to lie in $\MM_v$ in the above definition.
This intersection is just the $n$-point forward tube in the
Minkowskian variables $z_1,\ldots,z_n$. The main point of this section
is

\begin{lemma}
\label{partube}
For all $n$, $\FF_{d,d-1,n} \subset \ZZ_{n+}$.
\end{lemma}

The proof of this lemma consists of Lemmas \ref{1tube} and \ref{ntube}
below. We begin with the following remark.

\begin{remark}
\label{ulim}\rm
Let
\begin{equation}
\Lambda_u = \ \exp(u\,M_{d(d-1)})\ \ =\
\left ( \matrix{
1 & 0 & 0 \cr
0 & \ch u & -\sh u\cr
0 & -\sh u & \ch u \cr
} \right )\ .
\label{6.6.1}\end{equation}
Then $\Lambda_u e_\mu = e_\mu$ for $0 \le \mu \le d-2$, and
\begin{equation}
\Lambda_u e_{d-1} = (\ch u)e_{d-1} - (\sh u)e_d,\ \ \ \
\Lambda_u e_d = (\ch u)e_d - (\sh u)e_{d-1}\ .
\label{6.6.2}\end{equation}
Therefore
\begin{equation}
\lim_{u \rightarrow \pm\infty} 2e^{-|u|}\Lambda_u e_{d-1}
= e_{d-1} \mp e_d\ ,\ \ \
\lim_{u \rightarrow \pm\infty} 2e^{-|u|}\Lambda_u e_d
= e_d \mp e_{d-1}\ ,
\label{6.6.3}\end{equation}
and, for $0 \le \mu \le d-2$,
\begin{equation}
\Lambda_u M_{\mu d}\Lambda_u^{-1} = \ell(e_\mu \wedge \Lambda_u e_d)
= (\ch u) M_{\mu d} - (\sh u)M_{\mu (d-1)}
\label{6.6.4}\end{equation}
\begin{equation}
\lim_{u \rightarrow \pm\infty} 2e^{-|u|}\Lambda_u M_{\mu d}\Lambda_u^{-1} =
M_{\mu d} \mp M_{\mu (d-1)}\ .
\label{6.6.5}\end{equation}
\begin{equation}
\lim_{u \rightarrow \pm\infty} 2e^{-|u|}\Lambda_u M_{\mu (d-1)}\Lambda_u^{-1}
= M_{\mu (d-1)} \mp M_{\mu d}\ .
\label{6.6.6}\end{equation}
\end{remark}

\begin{lemma}
\label{1tube}
Let $c \in \MM_v$ and $L \in \Gamma_{(d,d-1)}$, and let $z=\exp(iL)c$. Then\HB
(i) There is an $M \in \CC_+$, and
a $c' \in X_d$, arbitrarily close to $L$ and $c$, respectively, such that
$z = \exp(iL)c = \exp(iM)c'$.\HB
(ii)
For every neighborhood $W$ of $(L,\ c)$ in $\GG \times X_d$, there is a
neighborhood $V$ of $z = \exp(iL)c$ in $\Xcd$ such that every
$z'' \in V$ can be written as $z'' = \exp(iM'')\,c''$, with
$M'' \in \CC_+$ and $(M'',\ c'') \in W$.
\end{lemma}
\noindent{\it Proof.}
(i) It suffices to prove the statement in case $L = s L_0$ for some $s > 0$,
and $c^0 =0$.
In the coordinates $(z^0,\ z^{d-1},\ z^d)$, for any $\tau \in \bC$,
$c \in \MM_v$,
\begin{equation}
\exp(\tau L_0)\,c = \left (
\begin{array}{c}
\strut c^0 + e^v \tau\\
\strut c^0\tau  + c^{d-1} + e^v {\tau^2 \over 2}\\
\strut -c^0\tau  + c^d - e^v {\tau^2 \over 2}
\end{array}
\right )\ .
\label{6.7}\end{equation}
As expected the two last components add up to $e^v$.
We now set $c^0 =0$ and $\tau = is$ ($s > 0$) in (\ref{6.7}).
We look for $M$ in the form
\begin{equation}
M = \Lambda_u\,tM_{0d}\,\Lambda_u^{-1},
\label{6.8}\end{equation}
with $\Lambda_u$ as in (\ref{6.6.1}) and $u > 0$ large. Then
\begin{eqnarray}
\lefteqn{
\Lambda(u,\ it) = \Lambda_u\,\exp(itM_{0d})\,\Lambda_u^{-1} = }\nonumber \\
&&\left ( \matrix{
\strut \ch t & i\sh u\, \sh t & i\ch u\, \sh t \cr
\strut i\sh u\, \sh t & \ch^2 u\, -\sh^2 u\, \ch t &
\ch u\, \sh u\, (1-\ch t)\cr
\strut -i\ch u\, \sh t & - \ch u\, \sh u\, (1-\ch t) &
-\sh^2 u\, + \ch^2 u\, \ch t\cr
} \right )\ .
\label{6.9}\end{eqnarray}
The set of all vectors $z = (iy^0,\ \vec{x},\ x^{d-1},\ x^d)$ with pure
imaginary 0-component and all other components real is mapped into itself
by $\Lambda(u,\ it)$ for all real $u$ and $t$.
As $u$ tends to $+\infty$,
\begin{equation}
\Lambda_u\,tM_{0d}\,\Lambda_u^{-1} = (\ch u) tM_{0d} - (\sh u) tM_{0(d-1)}
\label{6.10}\end{equation}
tends to $L_0$ provided $t \approx 2e^{-u}$, and $\Lambda(u,\ it)$ tends to
$\exp (is L_0)$ provided $t \approx 2se^{-u}$.
For fixed real $c = (0,\ \vec{c},\ c^{d-1},\ c^d)$,
satisfying $c^{d-1} + c^d = e^v$, and real $s >0$,
we wish to find a real $c' = (0,\ \vec{c},\ c^{\prime d-1},\ c^{\prime d})$
and $t > 0$ such that $\Lambda(u,\ it) c' = \exp(isL_0) c$, i.e.
$c' = \Lambda(u,\ -it) \exp(isL_0) c$. The condition that $c^{\prime 0} = 0$
gives
\begin{equation}
\th t = {se^v \over
c^{d-1}\sh u + c^d \ch u + (s^2/2) e^v e^{-u}}
\approx 2se^{-u}\ .
\label{6.11}\end{equation}
The other components of $c'$ are then real and tend to those of $c$
as $u \rightarrow \infty$. This proves (i).

\noindent (ii) Using (i), $z = \exp(iM)c'$, where $M \in \CC_+$ and
$c' \in X_d$ can be chosen arbitrarily close to $L$ and $c$ respectively.
If $z'' \in \Xcd$ is sufficiently close to $z$, we may apply
St$_1$ of the proof of Lemma \ref{open}, which yields the conclusion of (ii).
\endprf
\vskip15pt
 The condition that $c\in \MM_v$ in Lemma \ref{1tube}
can be replaced by the condition $c^d + c^{d-1} >0$, and in fact
by $c^d + c^{d-1} \not= 0$ since the case $c^d + c^{d-1} <0$ can
be dealt with by changing $c$ to $-c$. We note also that $(c^d +
c^{d-1})^2 = (L_0 c,\ L_0 c)$. We define
\begin{eqnarray}
\Gamma_+ &=& \{\Lambda\rho L_0\Lambda^{-1}\ :\
\Lambda \in G_0,\ \ \rho > 0\}\nonumber\\
&=& \{a\wedge b\ :\ (a,a) = 1,\ \ (b,b) = (a,b) = 0,\ \ a^0b^d -
a^db^0 >0\}
 \label{6.106}\end{eqnarray} (the same calculations as
in (\ref{2.5.1}) show that if $(a,a)=1$,
$(b,b)=(a,b)=0$, then $(a^0b^d - a^db^0)^2 \ge \vec{b}^2 = b_0^2
+ b_d^2$). The following lemma extends Lemma \ref{1tube} to the
case of $n$ points. It may be useful to spell it out in some
detail.

\begin{lemma}
\label{ntube}
let $z = (z_1,\ \ldots,\ z_n) \in \Xcdn$ be such that,
for $1 \le j \le n$,
\begin{equation}
z_j = e^{iQ_1}\,\ldots e^{iQ_j}\,c_j\ ,
\ \ \ Q_j \in \Gamma_{(d,d-1)}\ ,\ \ \ c_j \in X_d,\ \ (Q_jc_j,Q_jc_j) > 0\ .
\label{6.13}\end{equation}
For each $\veps > 0$, there exists a $\delta(n,\ \veps,\ Q,\ c) > 0$
such that, for any
$z' = (z'_1,\ldots,z'_n) \in \Xcdn$ satisfying
$||z_j-z'_j|| < \delta(n,\ \veps,\ Q,\ c)$ for all $j= 1,\ldots,n$,
there exist
$M_1,\ldots,\linebreak[0]M_n \in \CC_+$
and $c'_1,\ldots,c'_n \in X_d$ such that
$||M_j - Q_j|| < \veps$, $||c_j - c'_j|| < \veps$, and
$z'_j = \exp(iM_1)\,c'_1\,\ldots \exp(iM_j)\,c'_j$
for all $j= 1,\ldots,n$.
\end{lemma}

\noindent {\it Proof.}
Let St$'_n$ denote the statement of the lemma. St$'_1$ follows from
Lemma \ref{1tube} in view of the preceding remarks.
Assume that St$'_m$ has been proved for all $m \le n-1$
and let $z = (z_1,\ldots,z_n)$ be as
in (\ref{6.13}). Let $\veps \in (0,\ 1)$, $R = 1+\sup_{1\le j\le n}||z_j||$.
Denote $\wt z = (z_2,\ldots,z_n) \in X_d^{(c)(n-1)}$,
$\wt Q = (Q_2,\ldots,Q_n)$, $\wt c = (c_2,\ldots,c_n)$, and
\begin{equation}
\delta_2 = {1 \over 4R}
\delta\left(n-1,\ \veps,\
\wt Q,\ \wt c \right).
\end{equation}
By St$'_1$, there exists a $\delta_1 > 0$ such that
for any $z'_1$ with
$z'_1-z_1  < \delta_1$, there exist $M_1 \in \CC_+$ and $c'_1 \in X_d$
such that $z'_1 = \exp(iM_1)c'_1$, $||c'_1-c_1|| < \veps$,
$||M_1 -Q_1|| < \veps$,
and $||\exp(iM_1)-\exp(iQ_1)|| <\delta_2$.
Let $z'_1,\ldots,z'_n \in \Xcdn$ satisfy
$||z'_j-z_j|| < \min\{R/4,\ \delta_1,\
\delta_2/(1+||\exp(-iQ_1)||)\}$, and let $M_1$, $c'_1$ be as above.
Then, for $2\le j\le n$,
\begin{eqnarray}
||\exp(-iM_1)z'_j - \exp(-iQ_1)z_j|| &\le&
||\exp(-iM_1)-\exp(-iQ_1)||\,||z'_j||\nonumber\\
 &+& ||\exp(-iQ_1)||\,||z'_j-z_j||
\nonumber\\
&\le& \delta(n-1,\ \veps,\ \wt Q,\ \wt c).
\end{eqnarray}
By St$'_{n-1}$, there exist $M_2,\ldots,M_n\in \CC_+$ and
$c'_2,\ldots,c'_n\in X_d$ such that, for $2 \le j\le n$, $||M_2
-Q_2|| < \veps$, $||c'_2-c_2|| < \veps$, and $\exp(-iM_1) z'_j =
\exp(iM_2)\ldots\exp(iM_j)c'_j$. This proves St$'_n$.
\endprf\vskip15pt

We digress at this point and take advantage of the above calculations
to recall the proof of the following lemma (which appears in \cite{BB}).

\begin{lemma}[Borchers-Buchholz]
\label{bobu}
Let $U$ be a continuous unitary representation of $G_0$ in
a Hilbert space $\HH$. Let $\Psi \in \HH$
be such that $U(\exp(u\,M_{d(d-1)}))\Psi\linebreak[0] = \Psi$ for all real $u$.
Then $U(\Lambda)\Psi = \Psi$ for all $\Lambda \in G_0$.
\end{lemma}
\noindent Obviously $M_{d(d-1)}$ can be replaced, in the statement of
Lemma \ref{bobu}, by any of its conjugates under $G_0$, e.g. $M_{01}$ etc.

\vskip 0.25cm
\noindent {\it Proof of Lemma \ref{bobu}.}
By Remark \ref{ulim}
\begin{eqnarray}
\lim_{u \rightarrow \pm\infty}&&
\exp(uM_{d(d-1)})\,\exp(2re^{-|u|}M_{\mu d})\,\exp(-uM_{d(d-1)})\nonumber\\
&&= \exp(r(M_{\mu d} \mp M_{\mu(d-1)})),\ \ 0\le \mu \le d-2.
\label{G.4}\end{eqnarray}
Here $r$ is any real. Suppose that $\Psi \in \HH$ is such that
$U(\exp(uM_{d(d-1)}))\Psi = \Psi$ for all real $u$. Then for all
real $u$ and $t$, $0 \le \mu \le d-2$,
\begin{equation}
||U(e^{uM_{d(d-1)}}\,e^{tM_{\mu d}}\,e^{-uM_{d(d-1)}})\,\Psi - \Psi|| =
||U(e^{tM_{\mu d}})\,\Psi - \Psi||\ .
\label{G.5}\end{equation}
We set $t = 2re^{-|u|}$
for some fixed real $r$. As $u \rightarrow \pm\infty$,
$t$ tends to 0, so that the rhs of this equation tends to 0.
By the continuity of $U$ and (\ref{G.4}), the lhs tends to
$||U(\exp(r(M_{\mu d} \mp M_{\mu(d-1)})))\Psi -\Psi||$ hence this
quantity is equal to 0.
By the continuity of $U$, the subgroup
$N_\Psi = \{\Lambda \in G_0\ :\ U(\Lambda)\Psi =\Psi\}$ of $G_0$ is closed,
hence it is a Lie subgroup (see e.g. \cite{St}, pp.~228 ff).
Its Lie algebra contains $M_{d(d-1)}$ and
$M_{\mu d} \mp M_{\mu(d-1)}$ for all $\mu = 0,\ldots,d-2$, hence all
$M_{\mu d}$ for $\mu = 0,\ldots,d-1$.
These elements generate the whole of $\GG$
since $[M_{\mu d},\ M_{\nu d}] = M_{\nu\mu}$ for $0 \le \mu < \nu \le d-1$.
Hence $N_\Psi = G_0$. \endprf

\vskip15pt
 Lemma \ref{partube} follows, as announced, from Lemmas
\ref{1tube} and \ref{ntube}. If $\{\WW_n\}_{n \in \bN}$ is a
sequence of Wightman functions satisfying the conditions of Sect.
\ref{HYP}, but not necessarily the positive definiteness
condition, the tempered distribution $\WW_n$ can be restricted to
$\MM_v^n$, and more generally to $\MM_{v_1}\times \cdots \times
\MM_{v_n}$. The distributions $\WW_n(z(\zz_1,\ v_1),\ \ldots,\
\linebreak[0] z(\zz_n,\ v_n))$ have the linear properties of the $n$-point
Wightman functions for a set of Minkowskian fields on $\bR^{d-1}$,
$A(\zz,\ v) = \phi(z(\zz,\ v))$, labelled by a real parameter $v$
and depending in a $\CC^\infty$ manner on $v$. The usual
Minkowskian covariance and analyticity properties are satisfied
by virtue of Lemma \ref{partube}. The local commutativity is
inherited from that postulated for the $\WW_n$ in view of the
formula (\ref{lll}). If the $A(.\,,\ v)$ satisfy the positivity
condition for all $v$ in a non-empty open interval $(a,\ b)$,
then by Lemma \ref{locpos} (Sect.~\ref{HYP}), the original
$\WW_n$ satisfy the positivity condition on the whole of $X_d$.

In case the positive definiteness condition holds, we also have

\begin{lemma}
\label{limspec}
Let $Q \in \Gamma_+$ and let $\wh Q$ denote the self-adjoint operator
on $\HH$ such that $\exp(it\wh Q) = U(\exp(tQ))$ for all $t \in \bR$. Then
the spectrum of $\wh Q$ is contained in $\bR_+$. More precisely,
for every $\Psi \in \HH$, and every $\CC^\infty$
function $\wt \vhi$ on $\bR$ with compact support contained in
$(-\infty,\ 0)$,
\begin{equation}
\int_{\bR} \left ( \int \wt \vhi(p)\,e^{-itp}\,dp \right )\,
U(\exp tQ)\,\Psi\,dt = 0\ .
\label{6.14}\end{equation}
\end{lemma}

\noindent {\it Proof.}
Let $\vhi(t) = \int \wt \vhi(p)\,e^{-itp}\,dp$. We may assume that
$\int |\vhi(t)| dt \le 1$ and $||\Psi|| = 1$.
Given $\veps >0$, let
$T> 0$ be such that $\int_{|t| > T} |\vhi(t)| dt < \veps/3$.
Let $V$ be a neighborhood of the identity in $G_0$ (or $\wt G_0$)
such that $||(U(\Lambda) -1)\Psi|| < \veps/3$ for all $\Lambda)\in V$.
It is possible to choose $M \in \CC_+$ such that
$\exp(-tQ)\exp(tM) \in V$ for all $t \in [-T,\ T]$. Then
\begin{eqnarray}
\lefteqn{
||\int \vhi(t) U(e^{tQ})\Psi\,dt|| \le ||\int \vhi(t)U(e^{tM})\Psi\,dt||
+ ||\int_{|t| > T} \vhi(t)U(e^{tM})\Psi\,dt||}\nonumber\\
&&+||\int_{|t| > T} \vhi(t) U(e^{tQ})\Psi\,dt||
+ ||\int_{|t| \le T} \vhi(t) U(e^{tQ})
\left(1-U(e^{-tQ}e^{tM})\right)\,\Psi\,dt||\nonumber\\
&&< \veps \label{6.15}\end{eqnarray} since the first term in the
rhs is zero. \endprf\vskip15pt

Under the same hypotheses, we note that if $\Psi \in \HH$ is invariant
under $U(G_{(d-1)d})$ then it is in particular invariant under
$U(\exp(\bR M_{01}))$, so that, by Lemma \ref{bobu}, it is invariant
under $U(G_0)$.
Let again
$A(\zz,\ v) = \phi(z(\zz,\ v))$, which we consider
as an operator valued distribution on the Minkowski space $\bR^{d-1}$
depending smoothly on the real parameter $v$.
By the Reeh-Schlieder theorem for the field $\phi$, the vacuum is cyclic
for the set of fields $\{A(.,\ v)\ :\ v \in (a,\ b)\}$, where
$(a,\ b)$ is any non-empty open interval in $\bR$. As a consequence, the
uniqueness of the vacuum for the original theory is equivalent to
the uniqueness of the vacuum for the fields $A$, hence to
the cluster property for their Wightman functions, which is also
a certain cluster property for the Wightman functions of $\phi$.
We note that the Borchers-Buchholz Lemma also provides the
following characterization of the uniqueness
of the vacuum in terms of the Wightman functions.

\begin{lemma}
Assume that the conditions 1-5 of Sect. \ref{HYP} hold
(this includes the positivity condition). Then the condition
of uniqueness of the vacuum is equivalent to
\begin{equation}
\forall f,g \in \BB,\ \ \
\lim_{T \rightarrow +\infty}\ {1\over T} \int_0^T \left(
\langle \WW,\ f^\star\,g_{\{\exp(t M_{d(d-1)})\}} \rangle
- \langle \WW,\ f^\star \rangle \langle \WW,\ g \rangle \right ) dt = 0
\label{U.1}\end{equation}
\end{lemma}

\noindent {\it Proof}.
If all the conditions 1-5 of Sect. \ref{HYP} hold,
(\ref{U.1}) is equivalent to
\begin{eqnarray}
\forall f,g \in \BB, &&
\lim_{T \rightarrow +\infty}\ {1\over T} \int_0^T
(\Rep(f)\Omega,\ \exp(it \wh M_{d(d-1)}) \Rep(g)\Omega)\,dt
\nonumber\\
&=&
(\Rep(f)\Omega,\ \Omega)(\Omega,\ \Rep(g)\Omega)\ .
\label{U.2}\end{eqnarray}
By the mean ergodic theorem, the limit in the lhs is equal to
$(\Rep(f),\ E\,\Rep(g))$, where $E$ is the projector on the subspace
of all vectors invariant under $\exp(it \wh M_{d(d-1)})$.
By Lemma \ref{bobu} this is the subspace of all vectors invariant
under $U(G_0)$. Therefore
(by the density of $\{\Rep(f)\Omega\ :\ f \in \BB\}$) the condition
(\ref{U.1}) is equivalent to the fact that the subspace of all vectors
invariant under $U(G_0)$ is $\bC\Omega$.
\endprf

\section{Two-point functions}

\label{2pt}
We now consider the two-point function of a scalar field theory
on $\wXd$ which satisfies the general requirements described in the
previous sections:
\begin{equation}
\WW_2(x_1,x_2)=\WW(x_1,x_2) = (\Omega, \phi(x_1)
\phi(x_2)\Omega)\ . \label{2pt.1}\end{equation} By the tempered
spectral condition, this is the boundary value on $\wXd^2$, in
the sense of tempered distributions, of a function $W_+(z_1,z_2)$,
holomorphic and of tempered growth in $\wTm \times \wTp$. The
permuted Wightman function $\WW(x_2,x_1)$ is the boundary value
of $W_-(z_1,z_2) = W_+(z_2,z_1)$, holomorphic in $\wTp \times
\wTm$. The two permuted Wightman functions are invariant under
$\wt G_0$ and coincide in the real open subset of $\wXd^2$ in
which $x_1$ and $x_2$ are space-like separated. Therefore $W_\pm$
are branches of a single holomorphic function $W(z_1,z_2)$.
Extensions of standard arguments (see Appendix \ref{ap2pt}) show that there
exists a function $w$, holomorphic on the universal covering $\wt
\Delta$ of $\Delta = \bC \setminus [-1,\ 1]$, such that
$W_+(z_1,z_2) = w((z_1,z_2))$ when $z_1$ and $z_2$ belong to
$\wTm$ and $\wTp$, respectively. Theories on the {\sl covering}
of Anti-de Sitter spacetime
are thus in this respect closely similar to
Minkowski \cite{SW} and de Sitter  \cite{BM} field theories.

In the case of a field theory on $\wXd$, the commutator function
can be written (non uniquely) as the difference of
a retarded and an advanced ``function'' with supports in
$\{(x_1,\ x_2) \in \wXd^2\ :\ x_j = \wt \chi(s_j,\ \vec{x_j}),\ \
\pm(s_1-s_2) \ge 0\}$, respectively (see (\ref{2.11})).

In the case of a field theory on
the {\sl pure} AdS spacetime $X_d$, $w$ is actually a function
holomorphic on $\Delta$, and, in particular,
$W_+(x_1,x_2)$ and
$W_-(x_1,x_2)$
coincide not only on
$\RR_2 = \{x \in X_d^2\ :\ (x_1,x_2) > 1\}$,
but also on the ``exotic region''
$\RR'_2 = \{x \in X_d^2\ :\ (x_1,x_2) < -1\}$.
In this case the support of
the commutator function $\WW(x_1,x_2) - \WW(x_2,x_1)$ is
contained in $X_d^2 \setminus \RR_2\cup\RR'_2$.
This is the union of the two closed sets
$\{x \in X_d^2\ :\ |(x_1,x_2)| \le 1,\
(x_2\wedge x_1, e_0\wedge e_d) \ge 0\}$
and $\{x \in X_d^2\ :\ |(x_1,x_2)| \le 1,\
(x_2\wedge x_1, e_0\wedge e_d) \le 0\}$,
and the commutator function can be written as the difference
of an advanced and retarded function with supports in these two
sets (respectively). This splitting is, as usual, not unique.

Any two-point function with the above mentioned properties
is the two-point function of a generalized free field on $\wXd$ or $X_d$,
which satisfies the positive definiteness condition if and only
if, for any $\CC^\infty$ function $\vhi$ with compact support in $\wXd$
(resp. $X_d$),
\begin{equation}
\langle \WW,\ \ovl\vhi \otimes \vhi \rangle \ge 0.
\label{2pt.2}\end{equation}
Wick powers
of such a field are well-defined: their Wightman functions
can be obtained by the standard formulae, using $W$, as
boundary values of holomorphic functions in the relevant tuboids.
They satisfy all the requirements of Sect. \ref{HYP} with the possible
exception of positive definiteness, which holds if and only if
(\ref{2pt.2}) holds. Note that two Wick powers of generalized free
fields on $X_d$ share the property of commuting when their arguments
are in $\RR_2\cup\RR'_2$.

\subsection{The simplest example: Klein-Gordon fields.}

Let us consider fields satisfying the AdS Klein-Gordon equation on $\widetilde
X_{d}$:
\begin{equation}
\square \phi + m^2 \phi = 0. \label{kg}
\end{equation}
There is a preferred choice of solutions of that equation that display the
simplest example of the previous analytic structure. The corresponding
two-point functions are expressed in terms of generalized Legendre functions
\cite{BV-2}
\begin{equation}
Q^{(d+1)}_{\lambda}(\zeta)= \frac{\Gamma\left(\frac d2\right)}{\sqrt \pi
\Gamma\left(\frac {d-1}{2}\right)}\int_1^\infty (\zeta + it \sqrt{\zeta^2
-1})^{-\lambda-d+1}(t^2-1)^{\frac{d-3}{2}}dt \label{glf}
\end{equation} by the
following formula:
\begin{eqnarray}
\lefteqn{
W_{\lambda + \frac{d-1}{2}}(z_1,z_2) = w_{\lambda + \frac{d-1}{2}}(\zeta)}
\nonumber\\
&&=\frac{e^{-i\pi d}}{\pi^{\frac{d-1}2}} \Gamma\left(\frac{d+1}{2}\right)
h_{d+1}(\lambda) Q^{(d+1)}_{\lambda}(\zeta), \ \ \zeta = (z_1,z_2),
\label{kgtp1}\end{eqnarray}
where the parameter $\lambda$ is related to the mass by the formula
\begin{equation}
m^2=\lambda(\lambda + d-1).
\end{equation}
The normalization can be obtained by imposing the local Hadamard condition,
that gives
\begin{equation}
h_{d+1}(\lambda) = \frac{(2\lambda + d
-1)}{\Gamma(d)}\frac{\Gamma(\lambda + d-1)}{\Gamma(\lambda +1)}.
\label{hd+1}
\end{equation}
It can be checked directly that the functions
$w_{\lambda + \frac{d-1}{2}}((z_1,z_2))$
defined by Eqs (\ref{kgtp1}), (\ref{glf}), (\ref{hd+1})
are holomorphic in $\Delta$ or $\wt\Delta$
according to whether $\lambda$ is an integer or not.

\vskip 0.2cm
It is useful to display also an equivalent expression for the
Wightman functions (\ref{kgtp1}) in terms of the usual associate
Legendre's function of the second-kind\footnote{Note the slightly
different notation with the generalized Legendre's function
defined in Eq. (\ref{glf}) with the upper index in parentheses.
This is the way these Wightman functions were first written in
\cite{Fronsdal:1974} for the four dimensional case ($d=4$). Their
identification with second--kind Legendre functions is worth
being emphasized again, in place of their less specific (although
exact) introduction under the general label of hypergeometric
functions, used in recent papers. In fact Legendre functions are
basically linked to the geometry of the dS and AdS quadrics from
both group--theoretical and complex analysis viewpoints
\cite{BV-2,BM,vilenkin}}  \cite{Bateman}:
\begin{equation}W_{\lambda + \frac{d-1}{2}}(z_1,z_2)=
 W_{\nu}(z_1,z_2) = w_\nu(\zeta) = \frac
{e^{-i\pi\frac {d-2}2}}{(2\pi)^{\frac{d}2}} (\zeta^2-1)^{-\frac
{d-2}4} Q^{\frac {d-2}2}_{\nu-\frac 1 2}(\zeta), \label{kgtp}
\end{equation}
where we introduced the parameter $\nu = \lambda + \frac{d-1}{2}$
such that
\begin{equation}
\nu^2  = \frac {({d-1})^2 }{4} + m^2. \label{nu}
\end{equation}
Theories with $\nu > -1$ are acceptable in the sense that they
satisfy all the axioms  of Section \ref{HYP} including the
positive-definiteness property, which we shall prove below.
There are in fact two regimes \cite{Breitenlohner:1982jf}:
\begin{description}
\item{i)}
for $\nu > 1$ the axioms uniquely select one field theory for each given mass;
\item{ii)} for $|\nu| < 1$ there are two acceptable theories for each given
mass.
\end{description}
The case $\nu = 1$ is a limiting case.
Eq. (\ref{kgtp}) shows that the difference
between the theories parametrized by opposite values of $\nu$ is in their
large distance behavior. More precisely, in view of Eq. (3.3.1.4) of
\cite{Bateman}, we can write:
\begin{equation}
w_{-\nu} (\zeta)  = w_{\nu} (\zeta)  + \frac{\sin
\pi\nu}{(2\pi)^{\frac{d+1}{2}}} \, \Gamma\left(\frac d 2-\nu\right) \Gamma
\left(\frac d 2 +\nu\right) (\zeta^2-1)^{-\frac {d-1}4}P^{-\frac
{d-1}2}_{-\frac 1 2 -\nu}(\zeta) .
\end{equation}
The last term in this relation is {\em regular on the cut}
$\zeta \in [-1,1]$ and therefore does not contribute to the
commutator. By consequence the two theories represent the same
algebra of local observables at short distances. But since the
last term in the latter relation grows the faster the larger is
$|\nu|$ (see \cite{Bateman} Eqs. (3.9.2)), the two theories
drastically differ by their long range behaviors.

Let us discuss for now the positive definiteness of the Wightman
function (\ref{kgtp}). To this end, following the remarks in
Section \ref{parsec} we can consider the restriction
\begin{equation}
W_{M,v,v'}(\zz,\zz')  = W_\nu\left( \ch(v-v') - \frac 1 2
\exp(v+v')({\zz} -{\zz'})^2\right) \label{restriction}
\end{equation}
of the two-point function $W_\nu$ to $\MM_v\times \MM_{v'}$
defined for $v,v'$ real and $\zz = \xx + i \yy$, $\zz' = \xx' + i
\yy'$ such that $\yy^2 > 0$, $\yy^0 <0$, ${\yy'}^2 > 0$ and
${\yy'}^0>0$ (see eq. (\ref{7a})). The results of Section
\ref{parsec} imply that this restriction defines  a local and
(Poincar\'e) covariant two-point function which satisfies the
condition of positivity of the energy spectrum \cite{SW}.

Let us consider now the following Hankel-type transform of
$W_{M,v,v'}(\zz,\zz')$  w.r.t. the brane coordinate $v'$:
\begin{eqnarray} && H(\l',v,(\zz-\zz')^2)  = \int {\rm d}v'\, e^{(d-3)v'}  \theta_{\l'}(v')
W_{M,v,v'}(\zz,\zz') \label{htransform}
\end{eqnarray}
with \begin{equation} \theta_\l(v)=\frac 1 {\sqrt{2}} e^{-\frac
{d-1}2 v} J_\nu\left(\sqrt{\l} \, e^{-v }\right)\ .
\label{mode}\end{equation}

Inversion gives
\begin{eqnarray}
W_{\nu}(z( v,\zz),z'(v',\zz)) &=& \frac{1}{2(2\pi)^{\frac {d-2} 2}}
\int_0^\infty \frac{{\rm d}\lambda} 2 \, \lambda^{\frac {d-3} 4}
e^{-\frac {d-1} 2(v+v')}\nonumber\\
&& J_\nu(\sqrt{\lambda} \,e^{-v}) J_\nu
(\sqrt{\lambda}\,e^{-v'}) \ \delta^\frac{d-3}{2} K_{\frac {d-3}2}
(\sqrt{\lambda }\ \delta) \label{aaa}
\label{MAdS22}\end{eqnarray}
where $\delta^2 = - (\zz - \zz')^2$ (see \cite{B2}, p. 64, formula
(12), and \cite{BBGMS}
 for more details).

\vskip 0.2cm
Eq. (\ref{aaa}) together with Lemma \ref{locpos} show that the
Wightman functions \ref{kgtp} are positive-definite.

\subsection{K\"all\'en-Lehmann-type representations}
We postpone to a further paper the proof of
K\"all\'en-Lehmann-type representations which will be based
on an appropriate Laplace-type transformation, in a
spirit similar to that given in \cite{BM} for the case
of two-point functions on de Sitter spacetime.

\vskip 0.2cm
For two-point functions of general fields on $\wt X_d$:
\begin{equation}
W(z_1,z_2) = \int_{-{d+1\over 2}}^{=\infty} \rho(\lambda)
\ W_{\lambda + \frac{d-1}{2}}(z_1,z_2)\  d\lambda,
\end{equation}

For two-point functions of general fields on $X_d$:
\begin{equation}
W(z_1,z_2) = \sum_{\ell > -{d+1\over 2}}\ \rho_{\ell}
\ W_{\ell + \frac{d-1}{2}}(z_1,z_2).
\end{equation}

In these equations, $\rho(\lambda)$ (resp. $\{\rho_{\ell}\}$) represents
a positive measure (resp. sequence), as a consequence of the
positive-definiteness property.

\section{Bisognano-Wichmann analyticity and ``AdS-Unruh effect''}

\label{BWA}
In this section we prove that our assumptions of locality
and tempered spectral condition imply
a KMS-type analyticity property
of the Wightman functions in the complexified orbits of
one-parameter Lorentz subgroups of
$G_0$: the physical interpretation of this
property can be called
an ``AdS-Unruh effect'', since the relevant
hyperbolic-type orbits we are considering are
trajectories of uniformly accelerated motions.
In fact, this property is similar to the
analyticity property  in the complexified orbits
of the Lorentz boosts proved by Bisognano
and Wichmann \cite{BW} in Minkowskian theory.
However, in contrast with the method used in \cite{BW},
our method does not rely at all on the positivity properties
of the theory but only makes use of an analytic completion procedure of
geometrical type which has been presented in \cite{BEM}
for the case of de Sitter spacetime (noting there that it applied
similarly to the Minkowskian case) and which can be adapted in a
straightforward way to the AdS case as we show below.
Throughout this section, the distinction between the cases of pure AdS
and covering of AdS will be irrelevant because all the regions
and corresponding analyticity properties considered
will always take place in a {\sl single sheet} of $\wt X_d$; so
we can speak here of AdS
without caring about that distinction;
in particular the notation $\ZZ_{n\pm}$ will denote here as well
the corresponding tuboids
$\wt \ZZ_{n\pm}$ if the spacetime considered is $\wt X_d$.

\vskip 0.2cm
To be specific, let us consider once for all the Lorentz subgroup
of $G_0$ whose orbits are parallel to the two-plane of
coordinates $x_0, x_1$, with the notations of Eqs
(\ref{2.2}),(\ref{2.3}) (all the other Lorentz subgroups of $G_0$
acting on AdS being conjugate of the latter by the action of $G_0$
itself). The following ``wedge-shaped'' region of AdS, which is
invariant under that subgroup is then distinguished:
\begin{equation}
W_R(e_d) = \{x \in X_d\ : x^1 > |x^0|, \ x^d >0\}\ .
\label{wedgeshape}\end{equation}
(Note that this region of AdS admits the base point
$e_d$ as an exposed boundary point).
Considering $n$-point test-functions $f_n \in \BB_n$ and
Lorentz transformations $[\lambda] = [e^s] \in {\bf R}\setminus \{0\}$
we abbreviate
$f_{n\{[\lambda]\}}$ into ${f_{n}}_\lambda$, i.e.
\begin{equation}
{f_n}_\lambda(x_1,\ldots,\ x_n) =
f_n([\lambda^{-1}]x_1,\ldots,\ [\lambda^{-1}]x_n).
\label{5.9}\end{equation}
Then the main result of this section is the following
\begin{theorem}
\label{bw}
If a set of Wightman functions satisfies the locality and tempered
spectral conditions, then for all $m,\ n\in \bN$,
$f_m\in \DD(W_R(e_d)^m)$ and $g_n\in \DD(W_R(e_d)^n)$,
there is a function $G_{(f_m,g_n)}$ holomorphic on $\bC\setminus \bR_+$
with continuous boundary values $G_{(f_m,g_n)}^\pm$
on $\bR_+\setminus\{0\}$ from the upper and lower half-planes such that,
for all $\lambda \in \bR_+$,
\begin{equation}
G_{(f_m,g_n)}^+(\lambda) =
\langle\WW_{m+n},\ f_m\otimes {g_n}_\lambda\rangle,\ \ \ \
G_{(f_m,g_n)}^-(\lambda) =
\langle\WW_{m+n},\ {g_n}_\lambda\otimes f_m\rangle\ .
\label{5.11}\end{equation}
\end{theorem}
The thermal interpretation of this theorem could
be presented with all the details of Theorem 3 of \cite{BEM},
but here we shall only dwell on the following remarks,
specific to the geometry of AdS.

Consider an AdS spacetime with radius $R$ (i.e.
with equation $(x,x) = R^2$ and base point $R\ e_d =(0,\ldots,0,\ R)$)
and assume that all the test-functions
$f_m,\ g_n$ considered in the statement of Theorem \ref{bw}
have their supports contained respectively in sets of the form
${\cal V}_a^m$,
${\cal V}_a^n$, where ${\cal V}_a$ is a neighbourhood in $X_d$
of a certain point $x(a)$ in the wedge $W_R(e_d)$
which we take of the following form:
\begin{equation}
x(a) = (0,\ R a,\ 0\ldots,0,\ R \sqrt {a^2+1}),
\label{x(a)} \end{equation}
$a$ being a positive number.
The Lorentzian orbits followed by the points in the supports of
$f_m$ and $g_n$ are in a neighbourhood of the orbit of $x(a)$
which is the branch of hyperbola with equations
\begin{equation}
[e^{t\over Ra}]x(a) = (Ra\,\sh {t\over Ra},\
Ra\,\ch {t\over Ra},\
0\ldots,0,\ R \sqrt {a^2+1})\ .
\label{groupx(a)}\end{equation}
In the latter, the normalization of the group parameter $t\over Ra$
has been chosen such that $t$ is the
{\sl proper time} along the trajectory of $x(a)$ with respect
to the metric of the AdS spacetime of radius $R$.
Expressed in this parameter $t$ such that
$\lambda = e^{t\over Ra}$,
the analyticity property stated in the theorem corresponds
to the analyticity in a KMS-type domain, namely a
$\beta$-periodic cut-plane generated by
the strip $\{t; \ 0<\Im t< \beta\}$,
with the period $i\beta = 2i\pi \ aR$ (accompanied by the
relevant conditions of KMS-type for the
boundary values at $t$ real and $t-i\beta$ real
in terms of the products of field-observables).
Thus an
observer living on the trajectory (\ref{groupx(a)}),
and whose measurements are therefore supposed to be performed
in terms of field observables whose support
at $t=0$ lies in the
neighbourhood ${\cal V}_a$ of $x(a)$,
will perceive a thermal bath of particles
at the temperature $T= \beta^{-1} = {1\over 2\pi \ Ra}$.
In view of lemma \ref{accel}, the motion of such an observer is
uniformly accelerated and its acceleration $\rm a$ is related
to the ``radius'' $Ra$ of the hyperbolic trajectory
by the following formula (noting that the parameter $c^2$ of
lemma \ref{accel} is $c^2= a^2 +1$):
${\rm a} =
{1\over R}\  \sqrt {c^2\over c^2 -1}
= {1\over R}\  \sqrt {1 + {1\over a^2}}$.
Therefore one has:
\begin{lemma}
\label{temperature}
the temperature $T$ perceived by an AdS-Unruh observer living
on the trajectory described by  (\ref{groupx(a)}) is given
in terms of his (or her) acceleration $\rm a$
by
\begin{equation}
T={1\over 2\pi} \sqrt {{\rm a}^2 -{1\over R^2}}.
\label{tempeq}\end{equation}
\end{lemma}

\vskip 0.2cm
The proof of Theorem \ref{bw} contains two steps (as the similar
Theorem 2 of \cite{BEM}). In subsection \ref{Jostpt} we introduce
special sets of points in $X_d^n$ which play the role of the
Jost points of the Minkowskian case, because
all the complex Lorentz transformations $[\lambda]$
transport them into $\ZZ_{n+}\cup\ZZ_{n-}$.
This is the starting point of a standard analytic
completion procedure which generates the analyticity of the
Wightman functions in domains obtained by the action of all
complex Lorentz transformations on the tuboids
$\ZZ_{n+}$
(or $\ZZ_{n-}$) as in Lemma 2 of \cite{BEM}.
Here also, there is (for each pair $(m,n)$ considered in the statement of
Theorem \ref{bw}) a quartet of Wightman functions analytic
respectively in the tuboids
$\ZZ_{n\pm}$
and in two other tuboids
$\ZZ'_{n\pm}$
which are involved in that procedure. This is
fully explained in subsection \ref{DBW}.

\subsection{A special set of Jost points}

\label{Jostpt}
Let $a \in X_d$ be given by
\begin{equation}
a = (0,\ \sh u,\ \vec{0},\ \ch u),\ \ \ u > 0\ .
\label{5.1}\end{equation}
With the notations of (\ref{2.2}) and (\ref{2.3}), and real $s \in (0,\ \pi)$,
\begin{equation}
[e^{is}]\,a = e^{isM_{10}}\,a =
(i\sh u \sin s,\ \sh u \cos s,\ \vec{0},\ \ch u)\ .
\label{5.2}\end{equation}
This can be reexpressed as $\exp(itM_{0d})b$ with
$b = (0,\ \sh u \cos s,\ \vec{0},\ r)$, i.e.
\begin{eqnarray}
[e^{is}]\,a &=& (ir\sh t,\ \sh u \cos s,\ \vec{0},\ r\ch t),\nonumber\\
\th t &=& \th u \sin s,\nonumber\\
r &=& \sqrt{(\ch u)^2 -(\sh u)^2 (\sin s)^2}
\label{5.3}\end{eqnarray}
Let $\JJ_{n0}$ be the set of points $(a_1,\ldots,a_n) \in \Xdn$
such that
\begin{eqnarray}
&& a_j = (0,\ \sh u_j,\ \vec{0},\ \ch u_j),\ \ \ j = 1,\ldots, n,\nonumber\\
&& 0 < u_1 <\cdots < u_n\ .
\label{5.4}\end{eqnarray}
Let $(a_1,\ldots,a_n) \in \Xdn$ be of the form (\ref{5.4}). Then for
$0<s <\pi$,
\begin{eqnarray}
&& [e^{is}]\ a_j = \exp(it_j M_{0d})b_j,\nonumber\\
&& \th t_j = \th u_j \sin s,\ \ \ j = 1,\ldots, n,\nonumber\\
&& 0 < t_1 < \cdots < t_n\ ,
\label{5.5}\end{eqnarray}
with $b_j \in X_d$.
Since $t_j$ can be rewritten as $\theta_1 + \cdots + \theta_j$ with
$\theta_k >0$, the point $[e^{is}]a$ is in $\ZZ_{n+}$.
By the invariance of $\ZZ_{n+}$ under $G_0$ (resp. $\wt G_0$) it follows
that $[\lambda] a \in \ZZ_{n+}$ for all $\lambda \in \bC_+$. Obviously
$[\lambda] a \in \ZZ_{n-}$ for all $\lambda \in \bC_-$, since
$\ZZ_{n-} = \ZZ_{n+}^*$.

\subsection{Derivation of Bisognano-Wichmann analyticity}

\label{DBW}

This subsection closely follows the treatment given in \cite{BEM}.
We refer the reader to that reference for more details.
Let
\begin{equation}
W_R = \{x \in \amb\ :\ x^1 > |x^0|\}\ .
\label{5.6}\end{equation}
If $x \in W_R \cap X_d$ and $x^d >0$, we denote
(consistently with the notation $W_R(e_d)$ introduced in
(\ref{wedgeshape}))
\begin{equation}
W_R(x) = \{y \in X_d\ : y-x \in W_R,\ \ y^d >0\}\ .
\label{5.7}\end{equation}
Let
\begin{equation}
\KK_{n+} = \{(x_1,\ldots,x_n)\in \Xdn\ :\ x_1 \in W_R(e_d),\ \ \
x_j \in W_R(x_{j-1})\ \ \ \forall j = 2,\ldots,n\}\ .
\label{5.8}\end{equation}
$\KK_{n-}$ is defined by reflecting $\KK_{n+}$ across the
hyperplane $\{x \in \amb\ :\ x^1 = 0\}$ or equivalently
$\KK_{n-} = [-1]\KK_{n+}$ with $[-1] = [e^{i\pi}]$ as in (\ref{2.3}).
If $x \in \KK_{n+}$, then $x_j$ and $x_k$ are space-like separated
whenever $1 \le j < k \le n$.
Note that $\JJ_{n0} \subset \KK_n$.
If $(z_1,\ldots,z_n)$ belongs to $\Xdn$ or $\Xcdn$,
we denote $z_\leftarrow = (z_n,\ldots,z_1)$.

Besides $\ZZ_{n+}$ and $\ZZ_{n-}$ we shall also use the
other two tuboids
\begin{eqnarray}
\ZZ'_{n+} &=& \{z \in \Xcdn\ :\
z_\leftarrow \in \ZZ_{n-}\},\nonumber\\
\ZZ'_{n-} &=& \{z \in \Xcdn\ :\
z_\leftarrow \in \ZZ_{n+}\} = {\ZZ'_{n+}}^*\ .
\label{5.10}\end{eqnarray}

We fix $m \ge 0$, $n \ge 1$ and a function ${f_m}\in \DD(X_d^m)$
with support in $W_R(e_d)^m$. There exist two functions
$z \mapsto F_+({f_m};\ z)$ and $z \mapsto F_-({f_m};\ z)$, respectively
holomorphic in $\ZZ_{n+}$ and $\ZZ_{n-} = \ZZ_{n+}^*$,
having boundary values
$F_{\pm}^{(b)}({f_m};\ x)$ on $\Xdn$ (resp. $\wXdn$)
in the sense of distributions, such that
for every $g \in \BB_n$ with compact support,
\begin{eqnarray}
&\displaystyle\int_{\Xdn}&
F_+^{(b)}({f_m};\ x_1,\ldots,\ x_n)\,{g_n}(x_1,\ldots,\ x_n)\,
d\sigma(x_1)\ldots d\sigma(x_n) =\nonumber\\
&\displaystyle\int_{X_d^{m+n}}&
\WW_{m+n}(w_1,\ldots,\ w_m,\ x_1,\ldots,\ x_n)
{f_m}(w_1,\ldots,\ w_m)\,{g_n}(x_1,\ldots,\ x_n)\nonumber\\
&& d\sigma(w_1)\ldots d\sigma(w_m)d\sigma(x_1)\ldots d\sigma(x_n)\ ,
\label{5.12}\end{eqnarray}
\begin{eqnarray}
&\displaystyle\int_{\Xdn}&
F_-^{(b)}({f_m};\ x_1,\ldots,\ x_n)\,{g_n}(x_1,\ldots,\ x_n)\,
d\sigma(x_1)\ldots d\sigma(x_n) =\nonumber\\
&\displaystyle\int_{X_d^{m+n}}&
\WW_{m+n}(x_n,\ldots,\ x_1,\ w_1,\ldots,\ w_m)\,
{f_m}(w_1,\ldots,\ w_m)\,{g_n}(x_1,\ldots,\ x_n)\nonumber\\
&& d\sigma(w_1)\ldots d\sigma(w_m)d\sigma(x_1)\ldots d\sigma(x_n)\ ,
\label{5.13}\end{eqnarray}
(with $\wt X_d$ instead of $X_d$ when needed).
The functions $z\mapsto F'_+(f_m;\ z) = F_-(f_m;\ z_\leftarrow)$
and $z\mapsto F'_-(f_m;\ z) = F_+(f_m;\ z_\leftarrow)$
are respectively holomorphic in $\ZZ'_{n_+}$ and $\ZZ'_{n_-}$.
Their boundary values at real points, in the sense of distributions,
are ${F'_+}^{(b)}(f_m;\ x) = F_-^{(b)}(f_m;\ x_\leftarrow)$
and ${F'_-}^{(b)}(f_m;\ x) = F_+^{(b)}(f_m;\ x_\leftarrow)$.

In the sense of distributions,
$F_+^{(b)}({f_m};\ x)$ and $F_-^{(b)}({f_m};\ x)$ coincide for
$x \in \KK_{n-}$ by virtue of local commutativity.
Hence, by the edge-of-the-wedge theorem, $z\mapsto F_+({f_m};\ z)$
and $z\mapsto F_-({f_m};\ z)$ have a common holomorphic extension
$z\mapsto F({f_m};\ z)$ in
$\Delta_n = \ZZ_{n+} \cup \ZZ_{n-} \cup \VV_n$, where
$\VV_n$ is a complex neighborhood of $\KK_{n-}$ such that
$[\lambda]\VV_n = \VV_n$ for all $\lambda >0$. Let $a \in \JJ_{n0}$.
As we have noted $[e^{is}]a \in \ZZ_{n+}$ for all $s \in (0,\ \pi)$.
Moreover $[e^{i\pi}]a$ is in $\VV_n$. Hence there is an $\veps >0$
(depending on $a$) such that $[e^{is}]a$ is in $\Delta_n$ for all
$s \in (0,\ \pi+\veps)$. This also means that, denoting
$a' = [e^{i\veps/2}]a \in \ZZ_{n+}$, all points of the compact set
$\{z = [e^{is}]a'\ :\ 0 \le s \le \pi\}$ belong to $\Delta_n$.
Since the latter is open, there exists a $\rho_a >0$ such that
the open ball $B_{a'}(\rho_a) =\{z\in \Xcdn\ :\ ||z-a'|| < \rho_a\}$
is contained in $\ZZ_{n+}$, and
$[e^{is}]B_{a'}(\rho_a) \subset \Delta_n$ for all $s\in [0,\ \pi]$
hence $[\lambda] B_{a'}(\rho_a) \subset \Delta_n$ for all
$\lambda \in \ovl{\bC_+}\setminus \{0\}$. The function
$(z,\ \lambda) \mapsto G(f_m;\ z,\ \lambda) = F(f_m;\ [\lambda]z)$
is holomorphic in
\begin{equation}
\{(z,\ \rho e^{i\theta}) \in \ZZ_{n+}\times \bC\ :\
\rho > 0,\ \ \ |\sin \theta| < \alpha(z)\},
\label{5.14}\end{equation}
where $\alpha(z) > 0$ for all $z \in \ZZ_{n+}$. Moreover
$G$ is also holomorphic in $B_{a'}(\rho_a) \times \bC_+$.
By Lemma 3 of \cite{BEM} (Appendix A), $G$ extends to a function
holomorphic in $\ZZ_{n+}\times \bC_+$. However we wish to
prove that the boundary values of $G$ as $z$ tends to the reals
(in the sense of distributions) are also analytic in $\lambda$
in $\bC_+$. As a consequence of the temperedness assumption,
for real $\lambda$, $G(f_m;\ z,\ \lambda)$ defines a holomorphic function
of tempered behavior in $\ZZ_{n+}$ with values in the functions
of $\lambda$ bounded by some power of $(|\lambda| +|\lambda|^{-1})$.
If $z \in B_{a'}(\rho_a)$, $\lambda \mapsto G(f_m;\ z,\ \lambda)$ extends to
a function analytic in $\bC_+$ and also bounded there by
some power of $(|\lambda| +|\lambda|^{-1})$. Let
$\vhi(t) = \int \wt \vhi(p)\,e^{-itp}\,dp$, where
$\wt \vhi$ is a $\CC^\infty$ function with compact support contained in
$(-\infty,\ 0)$. For $z \in B_{a'}(\rho_a)$, and sufficiently
large $N \in \bN$,
\begin{equation}
\int_\bR \vhi(\lambda)\,\lambda^N\,G(f_m;\ z,\ \lambda)\,d\lambda = 0\ .
\label{5.15}\end{equation}
The l.h.s of this equation is holomorphic and of tempered behavior
in $\ZZ_{n+}$, hence it vanishes together with its boundary values.
Therefore the boundary value $G^{(b)}(f_m;\ x,\ \lambda)$ extends to
a function of $\lambda$ holomorphic in $\bC_+$. We note that
$G^{(b)}(f_m;\ x,\ \lambda) = F_+^{(b)}({f_m};\linebreak[0]\ [\lambda]x)$ for
$\lambda > 0$ and $G^{(b)}(f_m;\ x,\ \lambda) = F_-^{(b)}({f_m};\ [\lambda]x)$
for$\lambda < 0$.

In the same way $F'_+(f_m;\ z)$ and $F'_-(f_m;\ z)$ have a common
extension $F'(f_m;\ z)$ holomorphic in
$\Delta'_n = \ZZ'_{n_+} \cup \ZZ'_{n_-} \cup \VV'_n$ with
$\VV'_n = \{x\ :\ x_\leftarrow \in \VV_n\}$. Note that the domain
$\Delta'_n$ is equal to $\{z\ :\ z_\leftarrow^* \in \Delta_n\}$.
Hence $G'(f_m;\ z,\ \lambda) = F'(f_m;\ [\lambda]z)$ extends to
a function holomorphic in $\ZZ'_{n_+} \times \bC_-$, and its
boundary value ${G'}^{(b)}(f_m;\ x,\ \lambda)$
has properties that mirror those of
$G^{(b)}(f_m;\ x,\ \lambda)$. Finally suppose that
$x \in W_R(e_d)^n$. Then, by local commutativity, for $\lambda <0$,
$F_+^{(b)}(f_m;\ [\lambda]x)$ coincides with
$F_-^{(b)}(f_m;\ [\lambda]x_\leftarrow) = {F'}^{(b)}_+(f_m;\ [\lambda]x)$.
Hence if $x \in W_R(e_d)^n$, $\lambda \mapsto G^{(b)}(f_m;\ x,\ \lambda)$
and $\lambda \mapsto {G'}^{(b)}(f_m;\ x,\ \lambda)$ have a common
analytic continuation in $\bC \setminus \bR_+$.
This ends the proof of Theorem \ref{bw}.

\begin{remark}
\label{bwrem}\rm
The tuboids $\ZZ_{n\pm}$ and $\ZZ'_{n\pm}$ can be replaced
in the above discussion
by $\TT_{n\pm}$ and $\TT'_{n\pm}$ (similarly defined).
\end{remark}

\subsection{``CTP''}
In the proof of the preceding subsection, if we assume that
$f_m$ has support in the left, instead of the right wedge,
we find that $(z,\ \lambda) \mapsto G(f_m;\ z,\ \lambda)$
extends to a function holomorphic in $\TT_{n+} \times \bC_-$
instead of $\TT_{n+} \times \bC_+$. Thus in case $m=0$ it is easy
to obtain the following special case of the Glaser-Streater theorem:

\begin{lemma}
\label{bwm=0}
If a set of Wightman functions satisfies the locality and tempered
spectral conditions, then for all integer $n \ge 1$, there
exists a function $(z,\ \lambda) \mapsto G_n(z,\ \lambda)$,
holomorphic and of tempered growth in
$\TT_{n+} \times (\bC \setminus \{0\})$, such that
\begin{eqnarray}
G_n(z_1,\ldots,\ z_n,\ \lambda) &=&
\WW_n([\lambda]z_1,\ldots,\ [\lambda]z_n) \ \ {\rm for}\
\lambda > 0, \nonumber\\
G_n(z_1,\ldots,\ z_n,\ \lambda) &=&
\WW_n([\lambda]z_n,\ldots,\ [\lambda]z_1) \ \ {\rm for}\
\lambda < 0.
\label{5.16}\end{eqnarray}
\end{lemma}
If we now assume the covariance condition (\ref{4.3}) holds,
then $G_n$ is actually independent of $\lambda$ and we obtain

\begin{lemma}
\label{ctp}
If a set of Wightman functions satisfies the locality, covariance,
and tempered spectral conditions, then for all integer $n \ge 1$,
and all $z \in \TT_{n+}$,
\begin{equation}
\WW_n(z_1,\ldots,\ z_n) =
\WW_n([-1]z_n,\ldots,\ [-1]z_1).
\label{5.17}\end{equation}
\end{lemma}
If positivity holds this implies, as usual, the existence of an
anti-unitary operator $\theta$ such that
$\theta \phi(x) \theta^{-1} = \phi([-1]x)^*$.
In \cite{BFS},
the existence of this operator is a non-trivial
step in the derivation of commutativity for opposite wedges.
Note that the above proof of
Lemma \ref{ctp}, also valid in Minkowski
space, does not require the Bargmann-Hall-Wightman Lemma.

\section{Wick rotations and Osterwalder-Schrader reconstruction
on the covering of AdS}

\label{WR}

\def\param{{\wchi}}
In this section, we consider QFT's on the {\sl covering} of AdS
which satisfy the {\sl tempered spectral condition} and we wish
to show that such theories can be formulated equivalently in terms of
theories on the ``Euclidean'' AdS spacetime $X_d^{(\cal E)}$
in a way which is reminiscent of the Wick rotation in complex Minkowski
space. In fact, the simple geometrical fact which allows the
latter to hold is the property of the tuboids $\wt \ZZ_{n+}$
described in lemma \ref{tubeucl} and obtained by
making use of the complexification
in $\tau$ of the map
\begin{equation}
(\tau,\ \vec{x}) \mapsto \param(\tau,\ \vec{x}) =
\exp(\tau M_{0d})(0,\ \vec{x},\ \sqrt{\vec{x}^2 +1})
\label{7.1}\end{equation}
of $\bR \times \bR^{d-1}$ into $\wt X_d$.
As a matter of fact,
making $\tau$ and $\vec{x}$ complex yield charts for
a certain complexification denoted $ [\wt X_d]^{(c)}$ of $\wt X_d$:
it is defined as the image of the extension of $\wchi$
to the set
$\{\tau \in \bC\} \times\{\vec{z} \in \bC^{d-2}\ :
\ \vec{z}^2 + 1 \not\in \bR_-\} $. (Of course this set
$ [\wt X_d]^{(c)}$ is the covering of a complex region which is not the full
space $X_d^{(c)}$, since
$ \wt X_d^{(c)} \equiv  X_d^{(c)}$).

In the sequel we omit the symbol $\param$ and identify
a point $(\tau,\ \vec{x})$ with its image $\param(\tau,\ \vec{x})$.
According to lemma \ref{tubeucl}, all points
$z = ((\tau_1,\ \vec{x}_1),\ldots,\ (\tau_n,\ \vec{x}_n)) \in
\left( [\wt X_d]^{(c)}\right)^n$
such that $\vec{x}_j$ is real and
$0 < \Im \tau_1 < \cdots <  \Im \tau_n$, are in $\wt\ZZ_{n+}$.
Therefore (using the global invariance under $\wt G_0$)
\begin{equation}
\WW_n((\tau_1,\ \vec{x}_1),\ldots,\ (\tau_n,\ \vec{x}_n))
\label{7.3}\end{equation}
extends to a function of $(\tau_1,\ldots,\tau_n)$, holomorphic
in the tube
$\{(\tau_1,\ldots,\tau_n)\ : \Im \tau_1 < \ldots < \Im \tau_n\}$,
of tempered growth at infinity and at the boundaries, with values in
the tempered distributions (in fact the polynomially bounded
$\CC^\infty$ functions) in $\vec{x}_1,\ldots,\vec{x}_n$. (This function
depends only on the differences of the $\tau_j$ variables, but
this does not, of course, hold for the $\vec{x}_j$.)
Together with the other permuted Wightman functions, this defines
a function of $(\tau_1,\ldots,\tau_n)$, holomorphic in
$\{(\tau_1,\ldots,\tau_n)\ : \Im (\tau_j -\tau_k) \not= 0 \ \
\forall j \not= k\}$.

We denote
\begin{equation}
S_n(s_1,\ \vec{x}_1,\ldots,\ s_n,\ \vec{x}_n) =
\WW_n((is_{\pi(1)},\ \vec{x}_{\pi(1)}),\ldots,\
(is_{\pi(n)},\ \vec{x}_{\pi(n)}))
\label{7.4}\end{equation}
for real $s_1,\ldots,s_n$ such that $s_{\pi(1)} < \cdots < s_{\pi(n)}$,
for all permutations $\pi$ of $(1,\ldots,n)$.
Standard analytic completions, using the edge-of-the-wedge theorem
and Bremermann's Continuity Theorem, show that $S_n$ is actually analytic
at all non-coinciding points of $\bR^{nd}$.
Geometrically, in view of the representation (\ref{parameucl})
of $X_d^{(\EE)}$, this means that
{\sl the functions $S_n(z_1,\ldots,z_n)$ appear
as $n$-point Schwinger functions defined at all
noncoinciding points of the ``Euclidean'' AdS spacetime
$X_d^{(\EE)}$},
namely
\begin{equation}
\{(z_1,\ldots,z_n) \in X_d^{(\EE) n};\ z_i \neq z_j
\ {\rm for }\ \ {\rm all} \ \  i,j,\   i\neq j\}
\label{noncoinc}\end{equation}

By construction,
$S_n(s_1,\ \vec{x}_1,\ldots,\ s_n,\ \vec{x}_n) =
S_n(s_{\pi(1)},\ \vec{x}_{\pi(1)},\ldots,\ s_{\pi(n)},\ \vec{x}_{\pi(n)})$
for every permutation $\pi$. In case the positive definiteness
condition holds, the sequence $\{S_n\}$ has the
Osterwalder-Schrader positivity property: if $f_0 \in \bC$ and,
for $1 \le n \le N$, $f_n \in \SS(\bR^{nd})$ has its support in
$\{(s_1,\ \vec{x}_1,\ldots,\ s_n,\ \vec{x}_n)\ :\ 0 < s_1 < \cdots < s_n\}$,
then, with the convention $S_0 = 1$,
\begin{eqnarray}
\lefteqn{
\sum_{m,n = 0}^N \int
\ovl{f_m(s'_1,\ \vec{x}'_1,\ldots,\ s'_m,\ \vec{x}'_m)}\,
f_n(s_1,\ \vec{x}_1,\ldots,\ s_n,\ \vec{x}_n)}\nonumber\\
&& S_{m+n}(-s'_m,\ \vec{x}'_m,\ldots, -s'_1,\ \vec{x}'_1,
s_1,\ \vec{x}_1,\ldots,\ s_n,\ \vec{x}_n)\nonumber\\
&&\hbox to 4cm {\hfill} ds'_1\,d\vec{x}'_1 \ldots ds_n\,d\vec{x}_n \ge 0\ .
\label{7.5}\end{eqnarray}
This follows from the existence of the vector-valued holomorphic
functions $\Phi_n$ (see Sec. \ref{HYP}). Note that the analytic completion
mentioned above applies to these vector-valued functions.

\vskip 0.2cm
Conversely, let $\{S_n\}$ be a sequence of functions defined
at all noncoinciding points of
$X_d^{(\EE)}$, namely on the images of the sets
$\{((s_1,\ \vec{x}_1),\ldots,
 (s_n,\ \vec{x}_n)) \in \bR^{nd}\ :\
s_j \not= s_k\
\vec x_j \not= \vec x_k\
\ \forall\,j\not= k\}$, symmetric, invariant under a
common translation of the variables $s_j$, and satisfying
the Osterwalder-Schrader positivity property (\ref{7.5}) when
the supports of the $\{f_n\}$ are as above. Then it is possible,
by the same method as in the ``flat'' case (\cite{OS1,OS2,G2}),
to construct a Hilbert space $\HH$, $\Omega \in \HH$, and a sequence of
functions $\{\Phi_n\}_{n \in \bN}$ with values in $\HH$ on
$\{((\tau_1,\ \vec{x}_1),\ldots,(\tau_n,\ \vec{x}_n))\ :\
0 < \Im \tau_1 < \cdots < \Im \tau_n,\ \  \vec{x}_j \in \bR^{d-1}\}$,
holomorphic in the $\tau_j$ and  $\CC^\infty$ in the $\vec{x}_j$,
such that
\begin{eqnarray}
\lefteqn{
S_{m+n}(-s'_m,\ \vec{x}'_m,\ldots -s'_1,\ \vec{x}'_1,
s_1,\ \vec{x}_1,\ldots,\ s_n,\ \vec{x}_n) = }\nonumber\\
&&(\Phi_m((is'_1,\ \vec{x}'_1),\ldots,(is'_m,\ \vec{x}'_m)),\
\Phi_n((is_1,\ \vec{x}_1),\ldots,(is_n,\ \vec{x}_n)))
\label{7.6}\end{eqnarray}
whenever $0 < s'_1 < \cdots < s'_m$ and
$0 < s_1 < \cdots < s_n$. The temperedness of the $\Phi_n$ at the
real points, hence the existence of boundary values,
can be obtained if, as we shall assume,
the sequence $\{S_n\}$ satisfies
suitable growth conditions, as in \cite{OS1,OS2,G2}.
Note that
the boundary values
of the scalar products of the $\Phi_n$, namely the distributions
$\WW_n$ thus obtained are automatically defined on the covering
space $\wXdn$ of $X_d^n$, since the analytic continuation
in the complex variables
$\tau_j \in {\bf C}$ corresponds to travelling in the covering space
$\left( [\wt X_d]^{(c)}\right)^n$ (no periodicity condition in the
variables $\tau_j$ being produced in general).

If the functions $\wt S_n$ are
invariant under the transformations of $\wt G_0^{(c)}$ which preserve
$\{z \in \wXcdn\ :\ z^0 \in i\bR^n,\ \ \vec{z} \in \bR^{n(d-1)}\}$,
then differential operators representing $\GG$ can be used as in
\cite{OS1} to show that the $\WW_n$ are invariant under $\wt G_0$.
Finally the analyticity of the $\WW_n$ shows that the operator
$\wh M_{0d}$ representing $M_{0d}$ in $\HH$ has positive spectrum
and so does the representative $\wh M$ of any $M \in \CC_1$ by invariance
under $\wt G_0$.

\section{Final remarks}
\label{O}

First of all, we wish to emphasize the peculiar thermal aspects of
``generic'' field theories on AdS or its covering,
which are strongly related to the
geometry of the AdS quadric, namely to the existence of
uniformly accelerated motions on three types of trajectories:
while the elliptic and parabolic observers perceive a
world at zero temperature (satisfying a condition
of energy-positivity)
in spite of the fact that the acceleration can take all values
between zero and $1\over R$, the hyperbolic observers
perceive an Unruh-type temperature effect growing from zero to
infinity when the acceleration grows from $1\over R$
to infinity.

Our subsequent remarks
concern some important peculiarities of QFT's on the {\sl pure} AdS
spacetime with respect to those on its covering, which appeared
in the study of general two-point functions
(see Section \ref{2pt}), namely:

i) The vanishing of the commutator vacuum expectation value in the region of
spacelike separation $(x_1 -x_2)^2 < 0$
implies its vanishing in the region of time-like separation
$(x_1 +x_2)^2 < 0$, which we called the ``region of
exotic causal separation'',

ii) the periodicity condition on time-like geodesics implies that
the two-point function is a holomorphic function of
$(z_1,z_2)$ in the domain ${\bf C} \setminus [-1,+1]$ itself
instead of its covering. In the free-field case, this selects
the Legendre functions $Q^{(d+1)}_{\lambda}((z_1,z_2))$ for which
$\lambda = \ell$ integer, which results in a quantization of the
mass parameter $m^2 = \ell(\ell +d-1)$.  More generally
all the fields on the pure AdS spacetime have
two-point functions which are (positive) superpositions
of the
free-field functions
$Q^{(d+1)}_{\ell}((z_1,z_2)) $ with integer indices $\ell$.

These peculiarities suggest respectively the following questions
which may be linked together and deserve to be studied in
further works.

a) Does local commutativity, formulated as the vanishing of the
commutator in the region of spacelike separation implies its
vanishing in the region of exotic causal separation (or
``exotic causality'') ?

b) Do the axioms of Section \ref{HYP} applied to the pure AdS
spacetime
exclude the existence of non-trivial interacting QFT ?

It has in fact been noted in \cite{BFS}
that the region of spacelike separation does not
really deserve that name in the pure AdS case, since
pairs of points in that region can also be connected
by classes of {\sl non-geodesic} timelike paths;
as it is stated, the condition of local commutativity therefore
appears as a strong constraint which would force the
interactions to respect the time-periodicity of the geometry.
One can then also say that the condition of exotic locality
would represent a constraint of similar nature acting at
odd number of half-periods of time, and that is why
one could possibly expect it to be a consequence
of local commutativity.

We know that, besides generalized free fields,
their Wick polynomials
are allowed to exist, since their $n$-point functions,
which are combinations of
products of two-point functions, do satisfy the required
(periodic) analyticity properties in the relevant tubes
${\cal T}_{n\pm}$
of the pure AdS spacetime. However, if we think of
nontrivial interacting fields in terms of perturbation theory
and consider Feynman-type convolution products of free two-point
functions involving not only products but integrals on the AdS
spacetime, one can show that the constraints of periodic analyticity
cannot be maintained in general: in particular the action of
retarded propagators can be seen to propagate the interaction
in the covering tubes
$\wt{\cal T}_{n\pm}$.
So one could formulate as a conjecture, that there might be
no other QFT's on the pure AdS spacetime than the ones previously
mentioned, which would then also entail (as a byproduct) a positive
answer to our question a). However, there is at the moment
no genuine proof relying on the axioms of Section \ref{HYP}
that such a conjecture is valid.

We end these remarks
with a brief comparison of the formalism of this paper
with that proposed by \cite{BFS}. A difficulty is of course the usual gap
between a formulation based on fields and one based on local
algebras. This difficulty is compounded in the AdS case by the
lack of a translation group. In \cite{BFS} the pure AdS, and
not its covering space, is considered. From the principle of
passivity of the vacuum, the authors succeed in deriving the positivity
of the energy, the existence of a CTP operator (both in the same sense
as here), and the commutativity of operators localized in opposite
wedges. If we suppose that, in terms of fields, these results
correspond to the tempered spectral, covariance, and
positivity conditions, and a portion of
local commutativity, then, by Remark \ref{extloc} (Section~\ref{HYP}),
this implies
full local commutativity, and thus all our axioms.

\def\varcm{truecm}
\appendix
\section{Appendix. More on two-point functions}
\label{ap2pt}

\subsection{The case of $X_d$}
\label{x_d2pt}
We begin by considering a complex function $w_+$ holomorphic and of
polynomial growth on $\TT_{1-}\times \TT_{1+}$. The function $w_-$
defined by $w_-(z_1,\ z_2) = w_+(z_2,\ z_1)$ is holomorphic in
$\TT_{1+}\times \TT_{1-}$. We denote $w^{(b)}_\pm$ the boundary
value of $w_\pm$ on $X_d$ in the sense of tempered distributions.
We suppose that $w^{(b)}_+$ and $w^{(b)}_-$ coincide on the real
open subset $\RR_2$ of $X_d^2$ defined by
\begin{equation}
\RR_2 = \{x \in X_d^2\ :\ (x_1 - x_2,\ x_1 - x_2) < 0\}
\ \ = \{x \in X_d^2\ :\ (x_1,x_2) > 1\}\ .
\label{a2pt.1}\end{equation}
Since we are ultimately interested in the case when
$w_\pm(z_1,\ z_2) = w_\pm(\Lambda z_1,\ \Lambda z_2)$ for all
$\Lambda \in G_0$, we make the simplifying assumption that
$w_\pm(x_1+iy_1,\ x_2+iy_2)$ has a boundary value which is
$\CC^\infty$ in $x_1$ and of tempered growth in $z_2 = x_2+iy_2$
when $y_1$ tends to 0 while $z_2 \in \TT_{1\pm}$. Actually the
general case can be reduced to the simplified one by considering
$\int_{G_0} \vhi(\Lambda) w_\pm(\Lambda z_1,\ \Lambda z_2)\,d\Lambda$
for suitable test-functions $\vhi$. The functions $f_\pm$ defined by
$f_\pm(z) = w_\pm(e_d,\ z)$ are respectively holomorphic and of tempered
growth in $\TT_{1\pm}$, and have boundary values $f^{(b)}$ on $X_d$
in the sense of tempered distributions. These boundary values
coincide in the real open subset $\RR$ of $X_d$ defined by
\begin{equation}
\RR = \{x \in X_d\ :\ (x-e_d)^2 < 0\}\ \ \ =
\{x \in X_d\ :\ x^d > 1\}\ .
\label{a2pt.2}\end{equation}
We also define
\begin{equation}
\RR' = \{x \in X_d\ :\ (x+e_d)^2 < 0\}\ \ \ =
\{x \in X_d\ :\ x^d < -1\}\ .
\label{a2pt.2.1}\end{equation}
Let $H$ (resp. $H^{(c)}$) denote the subgroup of $G_0$ (resp. $G_0^{(c)}$)
which leaves $e_d$ unchanged. This is just the connected real
(resp. complex) Lorentz group for the $d$-dimensional Minkowski space
$\Pi_0 = \{x \in \amb\ :\ x^d = 0\}$
(resp. $\Pi_0^{(c)} = \{z \in \ambc\ :\ z^d = 0\}$ ). The properties
mentioned above for $f_\pm$ are invariant under $H$. Let
\begin{equation}
\TT'_{1,d} = H^{(c)}\,\TT_{1+} = H^{(c)}\,\TT_{1-}\ .
\label{a2pt.3}\end{equation}
The last equality is due to the fact that $I_{01} \in H^{(c)}$,
where $I_{01}e_0 = -e_0$, $I_{01}e_1 = -e_1$, $I_{01}e_\mu = e_\mu$
for $1< \mu\le d$, and that $I_{01}$ is a bijection of
$\TT_{1+}$ onto $\TT_{1-}$. Obviously $\TT'_{1,d} = H^{(c)}\,\TT'_{1,d}$.
We also denote:
\begin{equation}
\Pi_\pm = \{x\in \amb\ :\ x^d = \pm 1\},\ \ \ \
\Pi^{(c)}_\pm = \{z\in \ambc\ :\ z^d = \pm 1\}\ ,
\label{a2pt.3.1}\end{equation}
\begin{equation}
Q_\pm = \{z\in \ambc\ :\ (z \mp e_d)^2 = 0\}\ ,
\label{a2pt.3.2}\end{equation}
\begin{equation}
Q_0 = Q_\pm \cap \Pi_0^{(c)} =
\{Z \in \Pi_0^{(c)}\ :\ (Z,Z) = -1\}\ .
\label{a2pt.3.3}\end{equation}
The intersection of $X_d$ with the light-cone with apex at $-e_d$
is contained in the hyperplane $\Pi_-$, in fact
\begin{equation}
\Xcd \cap Q_- = \Xcd \cap \Pi_-^{(c)}\ .
\label{a2pt.3.4}\end{equation}
Note also that $\TT_{1\pm}\cap \Pi^{(c)}_\pm = \emptyset$. Indeed we
know that the image of $\TT_{1\pm}$ under the map
$z \mapsto z^d$ is $\bC \setminus [-1,\ 1]$. As a consequence
$\TT'_{1,d} \cap \Pi^{(c)}_\pm = \emptyset$. Indeed the image of
$\TT'_{1,d}$ under $z \mapsto z^d = (e_d,z)$ is the same as that of
$\TT_{1\pm}$ since $H^{(c)}e_d = \{e_d\}$ by definition.
We will prove:

\begin{lemma}
\label{GS}
Let $f_\pm$ be functions respectively holomorphic and of tempered
growth in $\TT_{1\pm}$, $f_\pm^{(b)}$ their boundary values,
in the sense of tempered distributions, on $X_d$. We suppose
that $f_+^{(b)}$ and $f_-^{(b)}$ coincide in $\RR$. Then:\HB
(i) There exists a function
$f$ holomorphic on $\TT'_{1,d}$ whose restriction to
$\TT_{1\pm}$ is $f_\pm$.\HB
(ii) $\TT'_{1,d} =
\{z \in \Xcd\ :\ z^d \in \Delta = \bC \setminus [-1,\ 1]$\}.
In particular $\TT'_{1,d}$ contains $\RR$ and $\RR'$.\HB
(iii) If $f^{(b)}_\pm$ is invariant under $H$, i.e. if
$f^{(b)}_\pm(x) = f^{(b)}_\pm(\Lambda\,x)$ in the sense of
distributions for all $\Lambda \in H$, then there exists
a function $h$ holomorphic on $\Delta$ such that
$f(z) = h(z^d)$ for all $z \in \TT'_{1,d}$.
\end{lemma}

\begin{figure}
\newdimen\fixhoffset
\fixhoffset=\hsize \relax
\advance\fixhoffset by -12.000\varcm
\divide \fixhoffset by 2
\hbox{\kern\fixhoffset\vbox to 10.00 \varcm{\offinterlineskip
\def\point#1 #2 #3 {\rlap{\kern #1 \varcm
\raise #2 \varcm \hbox{#3}}}
\def\spot{{\kern -0.2em\lower.55ex\hbox{$\bullet$}}}
\vfill\includegraphics{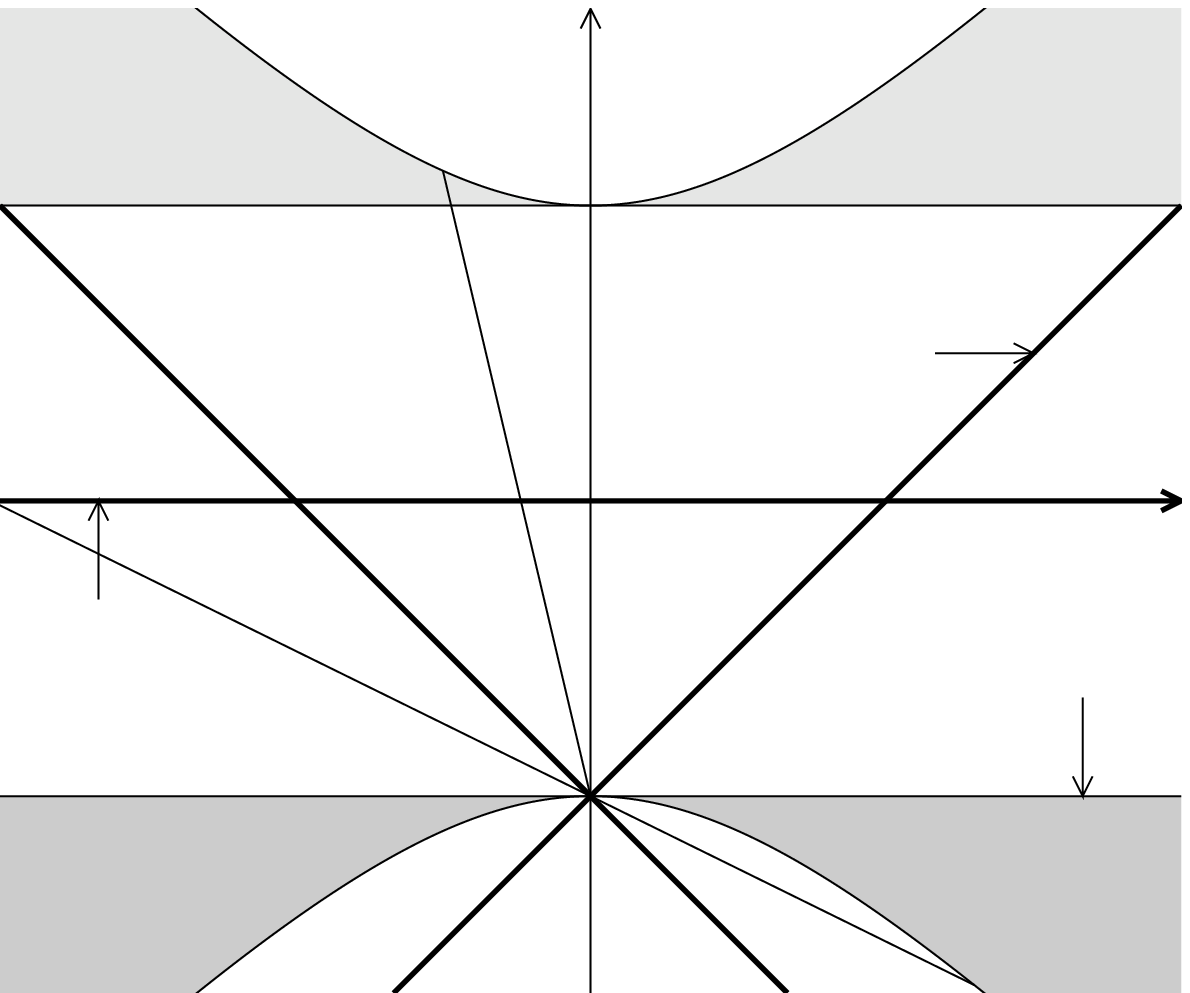}
\smash{\hbox to \hsize{%
\point 6.10 5.10 {0}
\point 11.50 4.50 {$x_1$}
\point 6.20 9.50 {$x_d$}
\point 6.50 2.10 {$-e_d$}
\point 6.20 7.60 {$e_d$}
\point 1.00 3.50 {$\Pi_0$}
\point 11.00 3.20 {$\Pi_-$}
\point 8.90 6.40 {$Q_-$}
\point 4.50 8.35 {$\spot$}
\point 4.60 8.45 {$z$}
\point 5.29 5.00 {$\spot$}
\point 4.29 5.20 {$\vhi_+(z)$}
\point 9.90 0.08 {$\spot$}
\point 10.10 0.28 {$z'$}
\point -0.09 5.00 {$\spot$}
\point 0.01 5.20 {$\vhi_+(z')$}
\point 1.00 9.00 {$\RR$}
\point 1.00 1.00 {$\RR'$}
\hfill}}}\hfill}

\caption{Schematic representation of the map $\vhi_+$
in projection onto the $(1,d)$ plane.}
\newdimen\fixhoffset
\fixhoffset=\hsize \relax
\advance\fixhoffset by -12.000\varcm
\divide \fixhoffset by 2
\hbox{\kern\fixhoffset\vbox to 8.00 \varcm{\offinterlineskip
\def\point#1 #2 #3 {\rlap{\kern #1 \varcm
\raise #2 \varcm \hbox{#3}}}
\def\spot{{\kern -0.2em\lower.55ex\hbox{$\bullet$}}}
\vfill\includegraphics{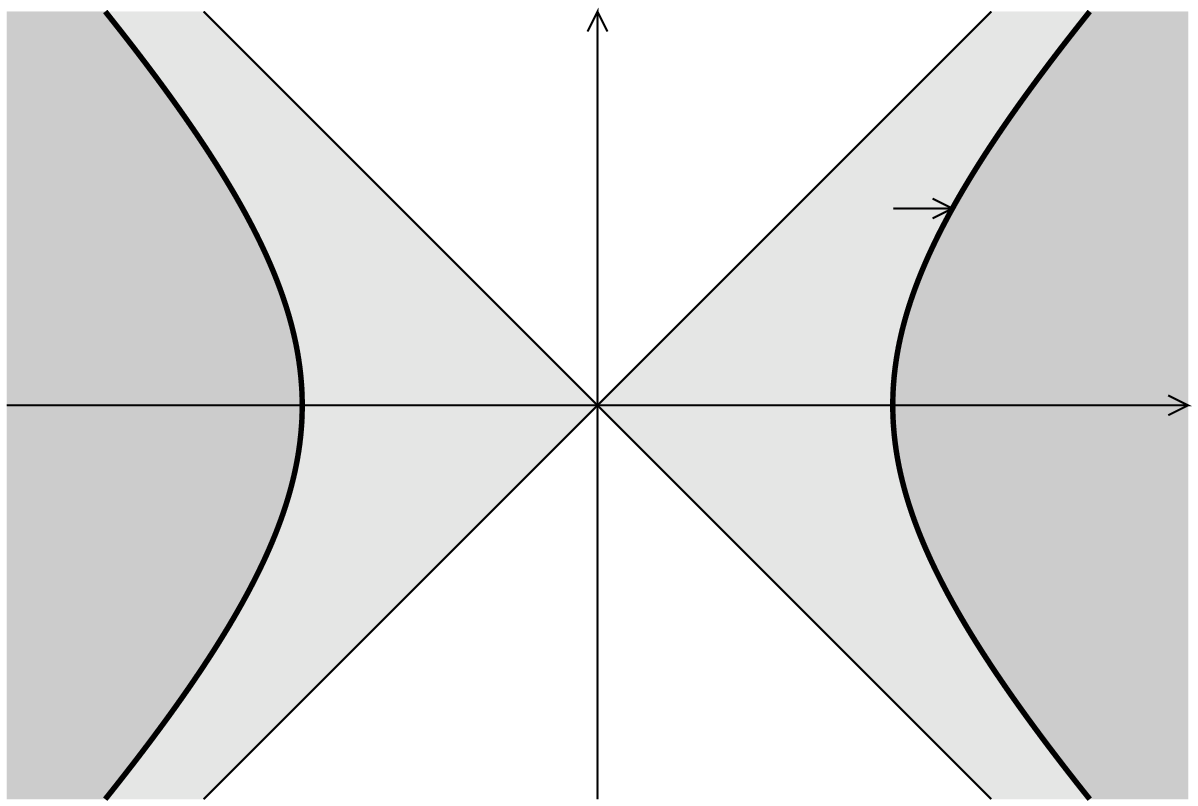}
\smash{\hbox to \hsize{%
\point 8.51 6.00 {$Q_0$}
\point 6.10 3.50 {$0$}
\point 3.10 3.50 {$-1$}
\point 8.60 3.50 {$1$}
\point 11.50 3.50 {$X^1$}
\point 6.10 7.50 {$X^0$}
\point 7.50 4.50 {$\vhi_+({\cal R})$}
\point 10.00 4.50 {$\vhi_+({\cal R'})$}
\point 3.60 4.50 {$\vhi_+({\cal R})$}
\point 0.80 4.50 {$\vhi_+({\cal R'})$}
\hfill}}}\hfill}

\caption{In the subspace $\Pi_0$ the region $\vhi_+(\RR)$
(light gray) and the region $\vhi_+(\RR')$ (dark gray) in the case $d=2$.}
\end{figure}

\noindent{\it Proof}\HB
(i) We will reduce this statement to a simple case of the
Glaser-Streater Theorem \cite{Sr,J} by using the inversion
(stereographic projection)
\begin{equation}
 z \mapsto \vhi_+(z) = Z,\ \ \
Z+e_d = 2\, {z + e_d \over (z + e_d)^2},\ \ \ \
z+e_d = 2\, {Z + e_d \over (Z + e_d)^2}\ ,
\label{a2pt.4}\end{equation}
\begin{equation}
(Z+e_d)^2 = {4 \over(z+e_d)^2}\ .
\label{a2pt.5}\end{equation}
$\vhi_+$ is a holomorphic involution of $\ambc \setminus Q_-$ onto
itself and commutes with
every element of $H^{(c)}$. If
$z = x+iy \in \Xcd \setminus Q_- = \Xcd \setminus \Pi_-^{(c)}$,
and $Z = X+iY = \vhi_+(z)$, we find
\begin{equation}
Z+e_d = 2\,{z+e_d \over (z+e_d)^2} = \,{z+e_d \over z^d + 1},
\label{a2pt.6}\end{equation}
which imply $Z^d = 0$. Conversely if $Z \in \Pi_0^{(c)} \setminus Q_0$,
it follows from (\ref{a2pt.4}) that $(z,z) = 1$.
Therefore $\vhi_+$ restricted to
$\Xcd \setminus \Pi_-^{(c)}$ is a holomorphic bijection
onto $\Pi_0^{(c)} \setminus Q_0$. Moreover
(if $z \in \Xcd \setminus \Pi_-^{(c)}$)
\begin{equation}
(Z+e_d)^2 = (Z,Z)+1 = {2\over z^d +1},\ \ \
z+e_d = 2\,{Z+e_d \over (Z,Z)+1}\ ,
\label{a2pt.7}\end{equation}
\begin{equation}
Y = \,{(x^d + 1)y - y^d(x+e_d) \over (x^d +1)^2 + y_d^2},\ \ \
(Y,Y) = {(y,y) \over (x^d +1)^2 + y_d^2}\ ,
\label{a2pt.8}\end{equation}
\begin{equation}
Y^0 = \,{y^0(x^d+1) - y^d x^0 \over (x^d +1)^2 + y_d^2}\ .
\label{a2pt.9}\end{equation}
Assume that $z \in \TT_{1+}$. Then $(y,y) > 0$, hence $(Y,Y) >0$.
Moreover $(x,x) = (y,y)+1 >1$, $(x,y) = 0$, $y^0 x^d - y^d x^0 > 0$, and
\begin{eqnarray}
(y^0 x^d - y^d x^0)^2 &=& ((y,y)+\vec{y}^2)((x,x)+\vec{x}^2) -
(\vec{x}\cdot\vec{y})^2\nonumber\\
&\ge& ((y,y)+\vec{y}^2)(x,x) > (y^0)^2,
\label{a2pt.10}\end{eqnarray}
hence $Y^0 > 0$. Thus $Z$ belongs to $\TT_+$,
the future tube in the complex Minkowski space $\Pi_0^{(c)}$,
and in fact $Z \in \TT_+ \setminus Q_0$.
Similarly $\vhi_+$ maps $\TT_{1-}$ into the past
tube of $\Pi_0^{(c)}$. Conversely suppose that
$Z = X+iY \in \TT_+ \setminus Q_0$. Then $Z = \vhi_+(z)$ where
$z = x+iy$ is given by the last equality in (\ref{a2pt.4}). It satisfies
$(y,y) >0$ by (\ref{a2pt.8}), and $y^0x^d - y^d x^0 >0$ since otherwise
we would have $Y^0 < 0$ by (\ref{a2pt.9}).

For real $x \in X_d \setminus \Pi_-$ and
$X = \vhi_+(x) \in \Pi_0 \setminus Q_0$,
the preceding formulae specialize to
\begin{equation}
(X,X) = {1-x_d \over 1+x_d}\ .
\label{a2pt.11}\end{equation}
As a consequence
\begin{eqnarray}
\vhi_+(\RR) &=& \{X \in \Pi_0\ :\ -1 < (X,X) <0\},
\label{a2pt.12}\\
\vhi_+(\RR') &=& \{X \in \Pi_0\ :\ (X,X) <-1\}.
\label{a2pt.13}\end{eqnarray}
The extended tube $\TT'$ in $\Pi_0^{(c)}$ is defined as usual
as $\TT' = H^{(c)}\TT_+ = H^{(c)}\TT_-$. Obviously
$\vhi_+(\TT'_{1,d}) = \TT' \setminus Q_0$
(recall that $\TT'_{1,d}\cap \Pi_-^{(c)} = \emptyset$).
As it is well-known $\vhi_+(\RR) \subset \TT'$ hence
$\vhi_+(\RR) \subset \TT' \setminus Q_0$, and also
$\vhi_+(\RR') \subset \TT' \setminus Q_0$. Furthermore
\begin{equation}
\TT' = \{Z \in \Pi_0^{(c)}\ :\ (Z,Z) \notin \bR_+\}\ ,
\label{a2pt.15}\end{equation}

Let $g_\pm = f_\pm \circ \vhi_+^{-1}$. Then $g_\pm$ is holomorphic and
of tempered growth in $\TT_\pm \setminus Q_0$, where
$g_\pm(Z) = f_\pm(z)$, $z \in \TT_{1\pm} \setminus \Pi_-^{(c)}$
being given by (\ref{a2pt.7}). The boundary values if $g_\pm$
coincide in $\vhi_+(\RR)$ so that, by the edge-of-the-wedge theorem,
$g_\pm$ have a common holomorphic extension in
$(\TT_+\cup\TT_- \setminus Q_0)\cup \NN$, where $\NN$ is a complex
open neighborhood of $\vhi+(\RR)$ invariant under $H$.
By following the steps of the proof of the Glaser-Streater Theorem
as presented e.g. in \cite{BEG} it is immediate to see
that all the arguments apply to the present case, owing to the fact
that $Q_0$ is invariant under $H^{(c)}$.
We therefore conclude that $g_\pm$ have a common extension $g$
holomorphic in $\TT'\setminus Q_0$. Note in particular that $g$
is holomorphic in a neighborhood of $\vhi_+(\RR')$, hence that
the boundary values of $g_\pm$  coincide there. Since $\vhi_+$
is a bijection of $\TT'_{1,d}$ onto $\TT' \setminus Q_0$,
setting $f(z) = g(\vhi_+(z))$ proves part(i).

\noindent(ii) follows from (\ref{a2pt.15}) and from the fact that
for $z \in \Xcd \setminus \Pi_-^{(c)}$, $Z = \vhi_+(z)$,
\begin{equation}
(Z,Z) = {1-z^d \over 1+z^d},\ \ \ \
z^d = {1 -(Z,Z)\over 1+(Z,Z)}\ .
\label{a2pt.16}\end{equation}
This implies that $\RR = \{x \in X_d\ :\ x^d > 1\} \subset \TT'_{1,d}$,
and similarly $\RR' \subset \TT'_{1,d}$.

\noindent (iii) In addition to the preceding hypotheses we
now assume that $f_\pm$, and therefore
$g_\pm$, are invariant under the real Lorentz group $H$.
Applying the Bargmann-Hall-Wightman Theorem \cite{SW,J}, we find
that there exists a function $\wh g$, holomorphic in
$\bC \setminus \bR_+ \setminus \{-1\}$ such that
$g(Z) = \wh g(Z^2)$ for all $Z \in \TT'\setminus Q_0$.
Setting $h(\zeta) = \wh g((1-\zeta)/(1+\zeta))$ we obtain
part (iii). \endprf

\vskip 0.5 truecm
\begin{remark}
\label{remA1}\rm
Another proof of part (i) can be given by using the temperedness of $f_\pm$.
Indeed there is an integer $M > 0$ such that the functions
$Z \mapsto g'_\pm(Z) = ((Z,Z)+1)^M g_\pm(Z)$ are holomorphic
in $\TT_\pm$ (respectively) and coincide on $\vhi_+(\RR)$.
By the ``dou\-ble-cone theorem'' \cite[p. 68]{Bo2}, their region of coincidence
extends to $\{X \in \Pi_0\ :\ (X,X)<0\}$ and the standard
Glaser-Streater Theorem can be applied.
\end{remark}

\vskip 1 truecm
We return to the functions $w_\pm$, which we now assume
invariant under $G_0$, and apply Lemma \ref{GS}
to $f_\pm(z) = w_\pm(e_d,\ z)$: these functions have a common
holomorphic extension to $\TT'_{1,d}$, and
there exists a function $h$,
holomorphic on $\Delta$, such that $w_\pm(e_d,\ z) = h(z^d)$.
For $z_1 \in \TT_{1-}$ and $z_2 \in \TT_{1+}$, let
$w'_+(z_1,\ z_2) = h((z_1,z_2))$.
For real $x_1 \in X_d$ and $z_2 \in \TT_{1+}$, we have
$w'_+(x_1,\ z_2) = w_+(x_1,\ z_2)$. Indeed we can write
$x_1 = \Lambda e_d$, $z_2 =  \Lambda z'$ for some $\Lambda \in G_0$
and $z' \in \TT_{1+}$, so that
$w_+(x_1,\ z_2) = w_+(e_d,\ z') = h(z^{\prime d}) = w'_+(x_1,\ z_2)$.
Therefore $w_+(z_1,\ z_2)$ coincides with $w'_+(z_1,\ z_2)$'
i.e. with $h((z_1,z_2))$
for all $z_1 \in \TT_{1-}$ and $z_2 \in \TT_{1+}$. Similarly
$w_-(z_1,\ z_2)$ coincides with $h((z_1,z_2))$ on
$\TT_{1+} \times \TT_{1-}$.

We have proved:

\begin{lemma}
\label{ads2pt}
With the preceding assumptions on $w_\pm$, and assuming
in addition that these functions are invariant under $G_0$,
there exists a function $h$, holomorphic on $\Delta$,
such that $w_\pm$ coincide with
$(z_1,\ z_2) \mapsto h((z_1,z_2))$ in their domains of definition.
\end{lemma}

\vskip 1 truecm
\begin{remark}
\label{remA2}\rm
This implies that $w_\pm^{(b)}$ coincide not only on $\RR_2$,
but also on the ``exotic region'' $\RR'_2$:
\begin{equation}
\RR'_2 = \{x \in X_d^2\ :\ (x_1 + x_2,\ x_1 + x_2) < 0\}
\ \ = \{x \in X_d^2\ :\ (x_1,x_2) < -1\}\ .
\label{a2pt.17}\end{equation}
Our proof shows that
this also holds without assuming that $w_\pm$ are invariant
under $G_0$.
\end{remark}

\subsection{The case of $\wXd$}
\label{wx_d2pt}
The topology of $\TT_{1\pm}$ (and hence of $\wTpm$) is made clear by
the holomorphic bijection $\vhi_+$, which maps $\TT_{1\pm}$ onto
$\TT_\pm \setminus Q_0$ in the complex Minkowski space $\Pi_0^{(c)}$.
To make things even clearer, one can use the map $\psi$ defined
in $\Pi_0^{(c)}$ by
\begin{equation}
\psi(Z) + e_{d-1} = -2{Z+e_{d-1} \over (Z+e_{d-1})^2}\ .
\label{w2pt.1}\end{equation}
This is a holomorphic bijection of
$\{Z \in \Pi_0^{(c)}\ :\ (Z+e_{d-1})^2 \not= 0\}$ onto itself,
which maps $\TT_+$ onto itself (i.e. in
particular $(Z+e_{d-1})^2 \not= 0$ for $Z \in \TT_\pm$). It maps
$\{Z \in \Pi_0^{(c)}\ :\ (Z,Z) = -1,\ \ \ (Z+e_{d-1})^2 \not= 0\}$
onto $\{Z \in \Pi_0^{(c)}\ :\ Z^{d-1} = 0,\ \ \ (Z+e_{d-1})^2 \not= 0\}$
and therefore it is a holomorphic bijection of
$\TT_+ \setminus Q_0$ onto
$\TT_+ \cap \{Z \in \Pi_0^{(c)}\ :\ Z^{d-1} \not= 0\}$ (and
similarly for $\TT_- \setminus Q_0$).
We shall however continue to work with $\TT_\pm \setminus Q_0$
which has the advantage of being invariant under $H$. We denote
\begin{equation}
\LL = \{z \in \Pi_0^{(c)}\ :\ (z,z) +1 \in \bR_-\}\ .
\label{w2pt.5}\end{equation}
This is an analytic hypersurface containing $Q_0$.

\begin{lemma}
\label{connec}
The open set $\TT_+ \setminus \LL$ is connected and simply connected.
\end{lemma}

\noindent{\it Proof.}
The set $A = \TT_+ \setminus \LL$ is star-shaped with respect to 0,
i.e. $\rho\,A \subset A$ for every $\rho \in (0,\ 1)$ (but
$0 \notin A$). For every $z\in \Pi_0^{(c)}$, $|(z,z)| \le ||z||^2$.
hence if $B$ denotes the open ball
$B = \{z \in \Pi_0^{(c)}\ :\ ||z||^2  < 1/2\}$, we have
$A\cap B$ = $\TT_+\cap B$ and this intersection is convex.
We can define a map $\sigma(t, z) = ((1-t) + t||2z||^{-1})z$ of
$[0,\ 1]\times A$ into $A$ such that $\sigma(0,\ z) = z$ and
$z \mapsto \sigma(1,\ z)$ sends $A$ into $A\cap B$. Hence any two
points in $A$ can be connected by a continuous arc, and every
closed curve in $A$ is homotopic to 0. \endprf

We now suppose given a pair of functions $g_\pm$ respectively
holomorphic and of tempered growth in the covering of
$\TT_\pm \setminus Q_0$. This is equivalent to giving a pair
of sequences $\{g_{n\pm}\ :\ n \in \bZ\}$, with the following
properties:\HB
(1) for each $n \in \bZ$,
$g_{n\pm}$ is holomorphic in $\TT_\pm \setminus \LL$;\HB
(2) every $z \in \LL\cap (\TT_+ \setminus Q_0)$
has an open complex neighborhood
$V_z$ such that $g_{n+}|V_z \cap \{Z\ :\ \Im (Z,Z)>0\}$ and
$g_{(n+1)+}|V_z \cap \{Z\ :\ \Im (Z,Z)<0\}$ have a common holomorphic
extension in $V_z$;\HB
(3) Similarly for $g_{n-}$ in $\TT_-$.\HB
In addition we suppose that\HB
(4) the boundary values of $g_{0\pm}$
coincide in $\vhi_+(\RR)$.

The proof of the Glaser-Streater Theorem again applies to
show that $g_{0\pm}$ have a common holomorphic extension
$g_0$ in $\TT' \setminus \LL$, in particular that,
the map $(\Lambda,\ z) \mapsto g_{0\pm}(\Lambda z)$ of
$H\times (\TT_\pm \setminus \LL)$ into $\bC$ extends to a holomorphic
function on $H^{(c)}\times (\TT_\pm \setminus \LL)$. From this it follows
that $(\Lambda,\ z) \mapsto g_{n\pm}(\Lambda z)$ also
extends to a holomorphic function on $H^{(c)}\times (\TT_\pm \setminus \LL)$.
The Bargmann-Hall-Wightman Lemma again shows that each of the
functions $g_{n\pm}$ extends to a function holomorphic in
$\TT' \setminus \LL$. Moreover these two functions coincide, and we
denote $g_n = g_{n\pm}$. This can be seen, for $n \ge 0$,
by induction on $n$. Supposing it to hold up to $n-1$,
we consider points of the form $iy$ for $y \in V_+$ and
$(y,y) > 1$, which belong to $\TT_+\cap \LL$. At such a point
\begin{eqnarray}
\lim_{\veps \downarrow 0} g_{n+}((i-\veps)y) &=&
\lim_{\veps \downarrow 0} g_{(n-1)+}((i+\veps)y)
\nonumber \\
= \lim_{\veps \downarrow 0} g_{(n-1)-}((i+\veps)y) &=&
\lim_{\veps \downarrow 0} g_{n-}((i-\veps)y)\ .
\label{w2pt.6}\end{eqnarray}
Note however that if in addition the ``hermiticity
condition'' $g_{0-}(z) = g_{0+}(z^*)^*$ holds, this extends to
$g_{-n-}(z) = g_{n+}(z^*)^*$ for all $n \in \bZ$.
There is no general reason for
$g_{n-}(z)$ and $g_{n+}(z^*)^*$ to coincide for
$n \not= 0$.

If we suppose that $g_{0\pm}$ are invariant under $H$, then
for each $n \in \bZ$ the functions $g_n$ are locally Lorentz invariant
and, again by the Bargmann-Hall-Wightman Lemma, there is a
function $h_n$, holomorphic in
$\bC \setminus  R_+  \setminus  (-1+\bR_-)$,
such that $g_n(z) = h_n((z,z))$. Moreover
$h_{n+1}(t-i0) = h_n(t+i0)$ for all $t \in (-\infty,\ -1)$
and all $n \in \bZ$. Therefore there exists a function
$\wt h$, holomorphic on
$\bC \setminus \bigcup_{n \in \bZ} (2i\pi n + \bR_+)$
such that, for each $n \in \bZ$, $h_n(e^w -1) = \wt h(w+2in\pi)$
in $\{w\in \bC\setminus \bR_+\ :\ -i\pi < \Im w < i\pi\}$

Let now $w_\pm$ be functions holomorphic and of tempered
growth on $\wt \TT_{1-}\times\wt \TT_{1+}$ and
$\wt \TT_{1+}\times\wt \TT_{1-}$, respectively.
We suppose that
$w_\pm(\Lambda z_1,\ \Lambda z_2) = w_\pm(z_1,\ z_2)$
for all $\Lambda \in \wt G_0$ and all $z$ in the relevant
domain, and that the boundary values of $w_\pm$ on $\wXd$
coincide at space-like separated arguments.
As in the case
of $X_d$, we assume that the boundary values can be extended
to $\CC^\infty$ functions of a real first argument, holomorphic
and of tempered growth in the second argument in
$\wt \TT_{1\pm}$ (respectively) and define
$f_\pm(z) = w_\pm(e_d,\ z)$, $g_\pm(Z) = f_\pm(\vhi_+^{-1}(Z))$.
Then $g_\pm$ satisfy the conditions (1)-(4) mentioned above,
$g_{n\pm}$ are $H$ invariant, and the preceding remarks
provide the functions $g_n$, $h_n$ and $\wt h$.
We can transport back these properties to the functions $w_\pm$.

\begin{remark}
\label{remA3}\rm
It is worth noting the following formula,
where $z_1,\ z_2 \in \Xcd \setminus \Pi_-^{(c)}$ and
$Z_1 = \vhi_+(z_1)$, $Z_2 = \vhi_+(z_2)$\ :
\begin{equation}
(z_1,z_2) -1 =
{-2(Z_1-Z_2)^2 \over (Z_1^2+1)(Z_2^2+1)}\ ,
\label{rm.1}\end{equation}
\end{remark}

\section{Appendix. Proof of Lemma \ref{g+open}}

\label{g+pf}
We start with
\begin{lemma}
\label{basis}
In any neighborhood of $M_{0d}$ in $\GG$, one can find a basis
$\{h_{\mu,\nu} \in \CC_+\ :\ 0 \le \mu < \nu \le d\}$ of $\GG$
such that
$M_{0d} = 2\sum_{0 \le \mu < \nu \le d} h_{\mu,\nu}/d(d+1)$.
\end{lemma}

\noindent{\it Proof.}
Let $\kappa > 0$ be sufficiently small. We denote $u_j = \kappa e_j$
for all $j = 1,\ldots,d-1$. For $0 \le \mu < \nu \le d$, we define
$f_{\mu,\nu} \in \bigwedge_2(E_{d+1})$ as follows ($j$ and $k$ are
integers in $[1,\ d-1]$):
\begin{eqnarray}
f_{0,d} &=& \left ( e_0 + \sum_{j=1}^{d-1} u_j \right ) \wedge
\left ( e_d + \sum_{k=1}^{d-1} k\,u_k \right )\nonumber\\
&=& e_0 \wedge e_d +\sum_{k=1}^{d-1} k\,e_0 \wedge u_k
+ \sum_{j=1}^{d-1} u_j \wedge e_d
+\sum_{1 \le j < k \le d-1} (k-j)\,u_j \wedge u_k\ ,\nonumber\\
f_{j,k} &=& (e_0 + (k-j)\,u_j) \wedge (e_d - u_k)\ \ {\rm for}\
1 \le j < k \le d-1 \nonumber\\
&=& e_0 \wedge e_d - e_0 \wedge u_k +(k-j)\,u_j \wedge e_d
- (k-j)\,u_j \wedge u_k\ ,\nonumber\\
f_{0,j} &=& e_0 \wedge (e_d -a_j u_j),\ \ \
f_{j,d} = (e_0 - b_j u_j) \wedge e_d\ \
{\rm for}\ 1 \le j \le d-1\ .
\end{eqnarray}
This gives
\begin{eqnarray}
\sum_{0 \le \mu < \nu \le d} f_{\mu,\nu} &=&
{(d+1)d \over 2} e_0 \wedge e_d \nonumber\\
&+& \sum_{k=1}^{d-1} \left (k - \sum_{1 \le j <k} 1 -a_k \right)
e_0 \wedge u_k \nonumber\\
&+& \sum_{j=1}^{d-1} \left (1 + \sum_{j < k \le d-1}(k-j) - b_j \right)
u_j \wedge e_d\ .
\end{eqnarray}
We adjust $a_j$ and $b_j$ so that only the first term survives in the
rhs of this identity, i.e
\begin{equation}
a_j = 1,\ \ \ b_j = 1 + (d-1-j)(d-j)/2\ \ \
{\rm for}\ 1 \le j \le d-1\ .
\end{equation}
With these values, $e_0\wedge u_j$ and $u_j\wedge e_d$ can be
recovered from $f_{0,j}$ and $f_{d,j}$, respectively, then
$u_j \wedge u_k$ can be recovered from $f_{j,k}$. Thus the
$\{f_{\mu,\nu}\}_{0 \le \mu < \nu \le d}$ form a basis of
$\bigwedge_2(E_{d+1})$. We obtain a basis
$\{h_{\mu,\nu}\}_{0 \le \mu < \nu \le d}$ of $\GG$ by setting
$h_{\mu,\nu} = \ell(f_{\mu,\nu})$. Clearly, given a neigborhood $V$
of $M_{0d} = \ell(e_0 \wedge e_d)$, $\kappa$ can be chosen so
small that $h_{\mu,\nu} \in V$ for all $\mu$ and $\nu$.
\endprf

\begin{corollary}
\label{whc+open}
The convex cone $\wh\CC_+$ generated by $\CC_1$ is open. Also
\begin{equation}
G_0 = \{\exp(s_1 M_1)\ldots\exp(s_N M_N)\ :\
N \in \bN,\ \ s_j \in \bR,\ \ M_j \in \CC_1\ \ \forall j\}.
\end{equation}
\end{corollary}

\begin{corollary}
\label{g+open2}
$G_0^+$ is open in $G_0^{(c)}$.
\end{corollary}

\noindent{\it Proof.}
We first show that, if $\tau \in \bC_+$ has sufficiently small
modulus, $\exp(\tau M_{0d})$ is an interior point of $G_0^+$.
Denoting $L = d(d+1)/2$, let
$(LA_1,\ldots,LA_L) = (h_{0,1},\ldots, \ \linebreak[0]h_{d-1,d})$,
be the basis of $\GG$ constructed in
the proof of Lemma \ref{basis}. In particular
$M_{0d}= (A_1 +\cdots + A_L)$.
We consider the two holomorphic maps of $\bC^L$ into $G_0^{(c)}$
\begin{eqnarray}
h_1(z_1,\ldots,z_L) &=& \exp(z_1 A_1) \cdots \exp(z_L A_L),\nonumber\\
h_2(z_1,\ldots,z_L) &=& \exp(z_1 A_1 + \cdots + z_L A_L)\ .
\end{eqnarray}
These maps are tangent at 0, and, for sufficiently small $\veps >0$,
both are biholomorphic maps of the polycylinder
$\{z \in \bC^L\ :\ |z_j| < \veps\ \forall j\}$ into $G_0^{(c)}$.
In particular the subset
$V_1 = h_1(\{z \in \bC^L\ :\ |z_j| < \veps,\ \ \Im z_j > 0\ \forall j\})$
of $G_0^{(c)}$ is open and contained in $G_0^+$. For sufficiently
small $|\tau|$ the curve
$\tau \mapsto h_1^{-1}\circ h_2(\tau,\ldots,\tau)$ exists and is
tangent to $\tau \mapsto (\tau,\ldots,\tau)$. Therefore there exists an
$\eta > 0$ such that $|\tau| < \eta$ and $\Im \tau > 0$ imply
$h_2(\tau,\ldots,\tau) \in V_1$, i.e.
$\exp (\tau M_{0d}) \in V_1$. We now consider a point $\Lambda \in G_0^+$
of the form $\exp(\tau_1 M_1)\cdots \exp(\tau_N M_N)\exp \tau M_0d$,
where $\tau_1,\ldots,\ \tau_N,\ \tau \in \bC_+$ and
$M_1,\ldots,M_N \in \CC_1$. This point can be rewritten as
$\Lambda = \Lambda_\theta \exp(\theta\tau M_{0d})$ where $\theta$ is
arbitrary in $(0,\ 1)$ and $\Lambda_\theta \in G_0^+$. For sufficiently small
$\theta$, $\exp(\theta\tau M_{0d}) \in V_1$ hence
$\Lambda \in \Lambda_\theta V_1$, an open set contained in $G_0^+$.
Since $G_0^+$ is invariant under conjugations from $G_0$, this proves
the corollary. \endprf

\begin{corollary}
\label{g0g+}
\begin{equation}
G_0^+ = G_0\,G_0^+  = G_0^+\,G_0\ .
\end{equation}
\end{corollary}

\noindent{\it Proof.}
Let $\Lambda \in G_0^+$ and $M \in \CC_+$. For any real $s$ and $t$,
\begin{equation}
e^{sM}\,\Lambda = e^{(s+it)M}(e^{-it M}\Lambda)\ .
\end{equation}
Since $G_0^+$ is open, for sufficiently small $t > 0$,
$e^{-it M}\Lambda \in G_0^+$ and $e^{(s+it)M} \in G_0^+$ hence
$e^{sM}\,\Lambda \in G_0^+$. Any $S \in G_0$ can be written
as a finite product $\exp(s_1 M_1)\ldots\linebreak[0]\exp(s_N M_N)$ where
$s_j \in \bR$ and $M_j \in \CC_+$, hence $S\Lambda \in G_0^+$
for any $\Lambda \in G_0^+$. Thus $G_0^+ = G_0\,G_0^+$, and since
$SG_0^+S^{-1} = G_0^+$ for all $S \in G_0$, this implies
$G_0^+ = G_0^+\,G_0$. \endprf

\end{document}